\documentclass[twoside]{articlek}

\usepackage[normal]{caption}

\renewcommand{\refname}{REFERENCES}

\def\BibTeX{{\rm B\kern-.05em{\sc i\kern-.025em b}\kern-.08em
		T\kern-.1667em\lower.7ex\hbox{E}\kern-.125emX}}

\usepackage{graphicx}

\renewcommand{\refname}{REFERENCES}

\def\BibTeX{{\rm B\kern-.05em{\sc i\kern-.025em b}\kern-.08em
		T\kern-.1667em\lower.7ex\hbox{E}\kern-.125emX}}

\usepackage{graphicx}

\begin{document}

\begin{flushleft}
	{\large\bf K  Isomers in Transuranium Nuclei 
		}
	\vspace*{25pt}

	{\bf Fritz Peter He\ss berger$^{1,2,}\footnote{E-mail: \texttt{f.p.hessberger@gsi.de}}$} \\
	\vspace{5pt}
	{$^1$GSI - Helmholtzzentrum f\"ur Schwerionenforschung GmbH, Planckstra\ss e 1, 64291 Darmstadt, Germany\\
		$^2$Helmholtz Institut Mainz, Staudingerweg 18, 55128 Mainz, Germany}\\

\end{flushleft}
          
   Version: October, 4, 2023\\

\abstract{K isomers in transuranium nuclei have become a most interesting subject in nuclear structure
	investigation in many laboratories within the recent twenty years. In this paper an overview on the present day situation will be given. It will focus on the conditions
	for occuring of this kind of isomers, their decay properties and systematics in their properties as far as experimental data are available.
	}

\section{1. Introduction}
Usually isomers are defined as 'longlived' excited levels of an atomic nucleus, whereas the term 'longlived' is not specified initially. At the time of discovery 
of isomeric states it was assumed that nuclear levels had lifetimes $<$1$\times$10$^{-13}$s \cite{DrW16}. It thus seemed justified to denote nuclear levels 
with life-times $\tau$$>$10$^{-13}$ s as 'long-lived' or 'isomeric'. Recently G.D. Dracoulis et al. \cite{DrW16} have expressed it in the following way:
'on the technical side any state which has a directly measurable lifetime in the sense of an electronic measurement, even down to the subnanosecond  region could be
termed an isomer, or more properly a metastable state'. We will follow this wording in the present paper.\\
The condition for emergence of an isomeric state is due to the fact that decay into lower lying excited nuclear levels or the ground state does not occur promptly but
is delayed or hindered. P.M. Walker and G.D. Dracoulis here distinguish three cases \cite{WaD99}:\\
a) Shape isomers: these occur, if a second potential minimum exists at large elongation of the nucleus. Decay into levels with lower excitation energies or the ground state is thus connected with a large change of the nuclear shape. Typical examples are the 'fission isomers' in 
transuranium nuclei, e.g. $^{242m}$Am. But also internal transitions may compete with spontaneous fission.\\
b) Spin - isomers: these isomers occur, if the spin differences between the isomeric state and lower lying levels are large and transitions between them 
are only possible by
emission of electromagnetic radiation of high multipolarity. Depending on the structure of the individual nuclei in these cases also 
$\beta^{+}$\,-, electron capture (EC)\,-, $\beta^{-}$\,-, $\alpha$\,- decay or even spontaneous fission may compete with internal transitions
($\gamma$\,-emission, internal conversion (IC)).\\
c) K isomers: these isomers are specific cases of spin isomers for which not only the spin - difference, but also the orientation of the 
spin vector plays a role. These isomers accur in axially symmetric deformed nuclei. The quantum number {\it{K}} denotes the projection of the
total angular momentum (total 'spin') onto the symmetry axis.\\
The situation is illustrated in fig. 1. Figs. 1a and 1b show the relations in a deformed nucleus with one unpaired nucleon. The total angular momentum
(total spin)  {\it{$\vec{\jmath}$}} is given by the vector sum of the orbital momentum {\it{$\vec{l}$}} and the spin
{\it{$\vec{s}$}} of the unpaired nucleon. Here two cases have to be distinguished: a) orbital angular momentum {\it{$\vec{l}$}} and
and spin {\it{$\vec{s}$}} vectors are 'parallel' (fig. 1a) and b) orbital angular momentum {\it{$\vec{l}$}} and
and spin {\it{$\vec{s}$}} vectors are 'anti-parallel' (fig. 1b). In case a) the projection {\it{$\Lambda$}} of the orbital angular momentum {\it{$\vec{l}$}} onto the symmetry axis is lower than {\it{$\Omega$}}, the projection of the total angular momentum onto the symmetry axis, 
i.e. {\it{$\Omega$}} $>$ {\it{$\Lambda$}}, while in case b) the projection {\it{$\Lambda$}} of the orbital angular momentum onto the
symmetry axis is larger than  {\it{$\Omega$}},
i.e. {\it{$\Omega$}} $<$ {\it{$\Lambda$}}.\\
Using the usual notation for Nilsson - levels {\it{{$\Omega$$^{\pi}$$[$N,n$_{z}$,$\Lambda$$]$}}}, where $\pi$ denotes the parity of the level, N
denotes the total number of oszillator quanta and n$_{z}$ the number of oszillator quanta along the symmetry axis, the case {\it{$\Omega$}} $>$ {\it{$\Lambda$}} is denoted by an 'uparrow' $\uparrow$, i.e. {\it{{$\Omega$$^{\pi}$$[$N,n$_{z}$,$\Lambda$$]$$\uparrow$}}}, ('spin-up state'),
while the case {\it{$\Omega$}} $<$ {\it{$\Lambda$}} is denoted by a 'downarrow' $\downarrow$, i.e. {\it{{$\Omega$$^{\pi}$$[$N,n$_{z}$,$\Lambda$$]$$\downarrow$}}}, ('spin-down state').\\
Fig. 1c shows the situation in case of two unpaired nucleons where the total spins of the individual nucleons
{\it{$\vec{\jmath_{1}}$}} and  {\it{$\vec{\jmath_{2}}$}} add up to the total spin  {\it{$\vec{\jmath}$}}, the projection of which onto the 
symmetry axis is denoted as {\it{K}}. Comparing figs. 1a and 1b with fig. 1c it is evident that formally for a single unpaired nucleon the
quantum number {\it{$\Omega$}} is identical to {\it{K}} for the case of two unpaired nucleons taking $\Omega_{2}$\,=\,0, or formally {\it{$\Omega$=K}}. Therefore it
meanwhile became common (see e.g. \cite{HaL22}) also to denote in specific cases single particle 'spin isomers' as 'K isomers'. 
This feature will be discussed in detail in section 5.\\

 \begin{figure*}
	\resizebox{0.99\textwidth}{!}{
		\includegraphics{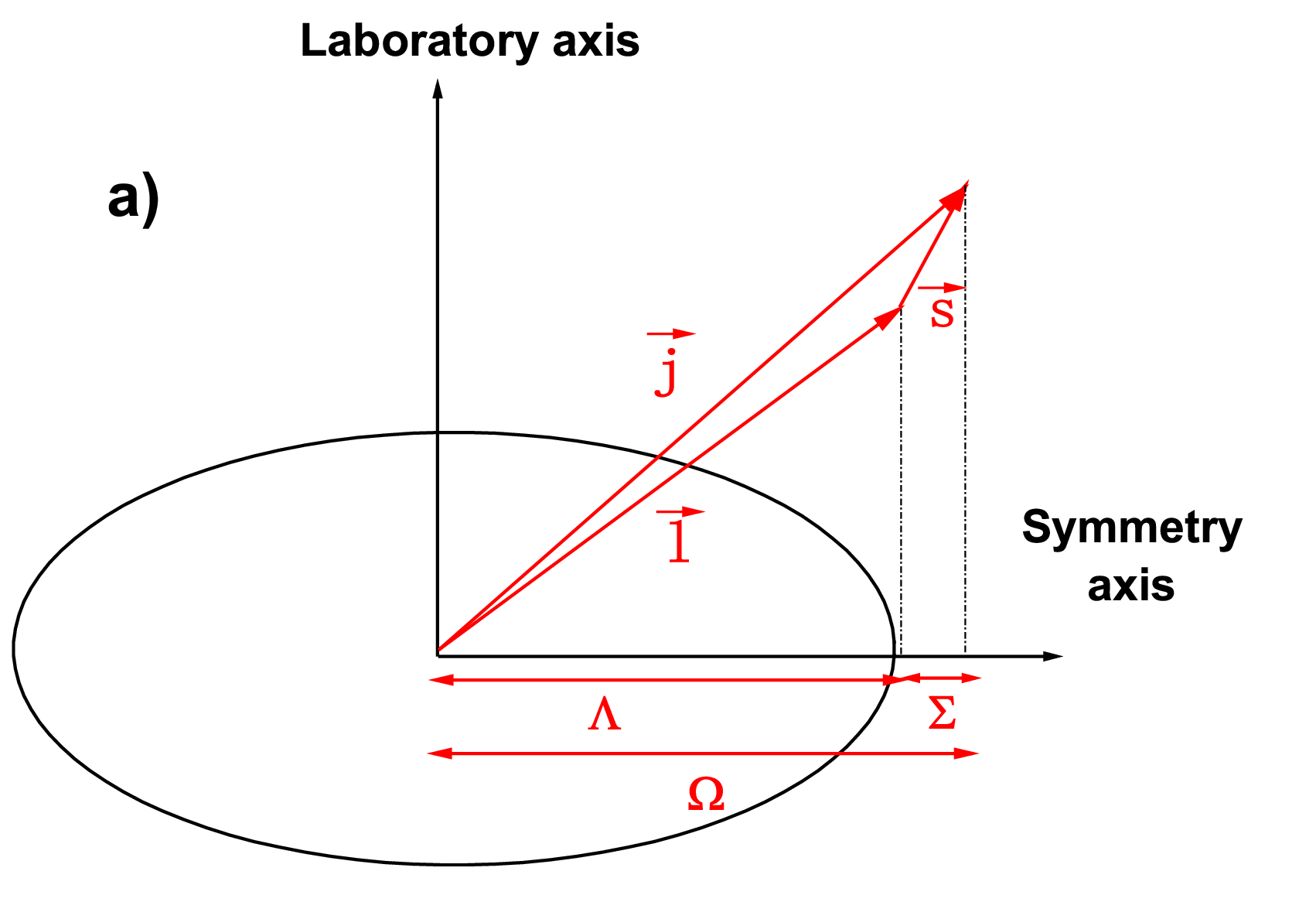}
		\includegraphics{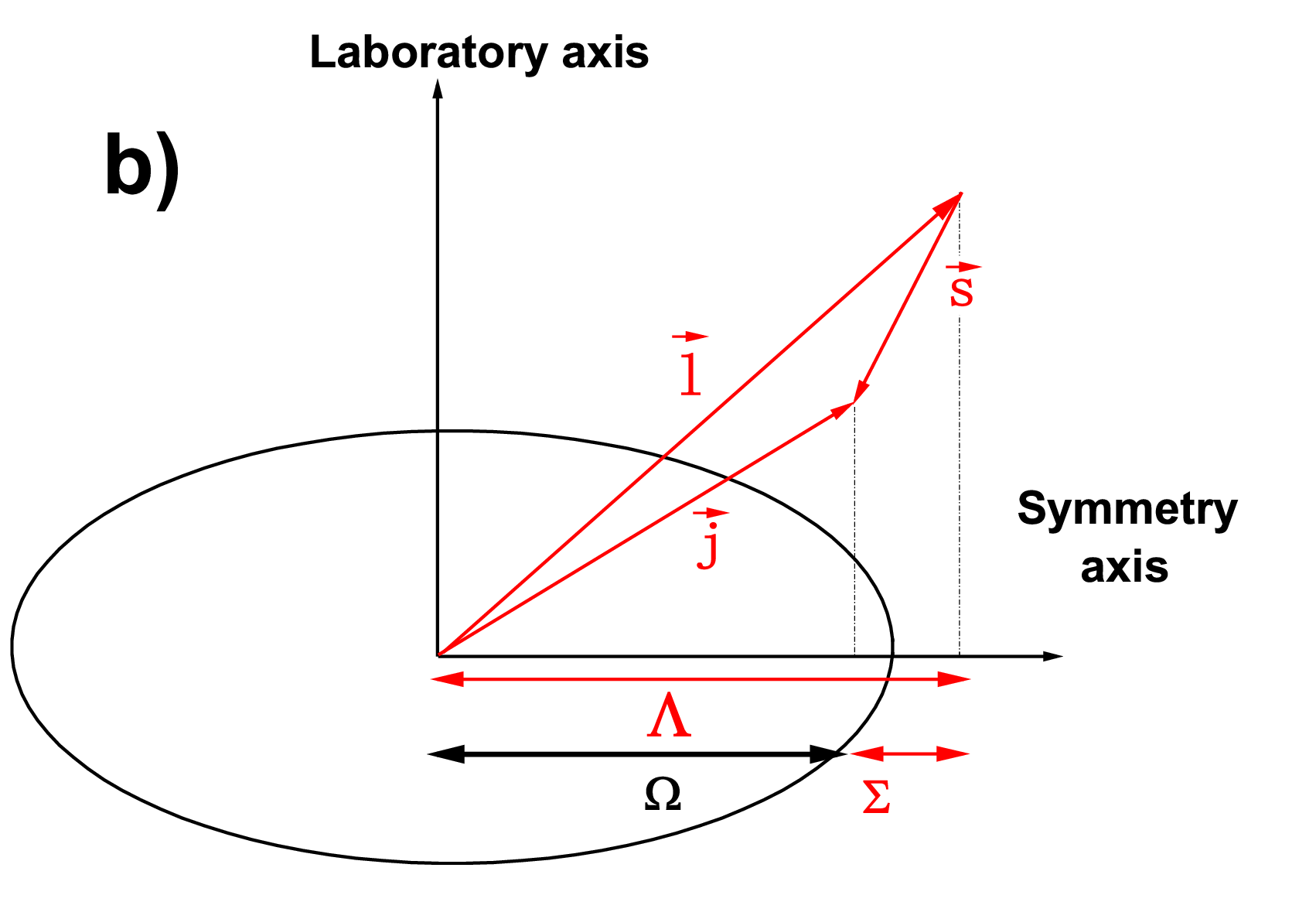}
		\includegraphics{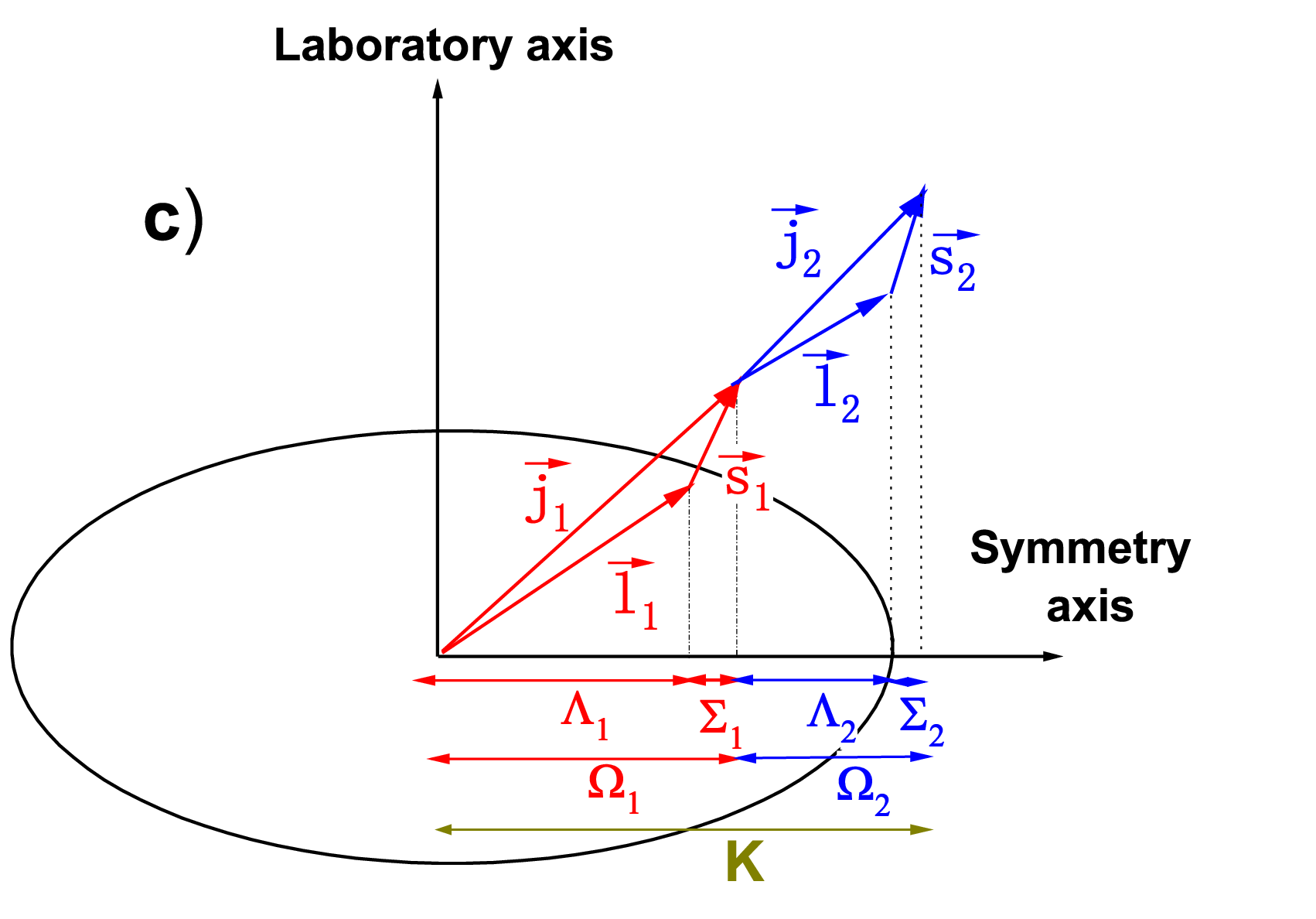}
	}
	\caption{Angular momentum coupling schemes; a) single unpaired nucleon with nucleon orbital angular momentum and spin vectors in 'parallel';
		b) single unpaired nucleon with nucleon orbital angular momentum and spin vectors 'anti-parallel'; c) coupling scheme for two unpaired 
		nucleons (for nonrotating nuclei). }
	\label{fig:1}       
\end{figure*}

\section{2. Angular Momentum Coupling}
As the K isomers we are dealing with mostly result from 2-quasi-particle or 4-quasi-particles states, let us first discuss the conditions
of their occurence which is strongly related to the coupling of angular momenta of two or more unpaired nucleons.\\
In the previous section schematic coupling of angular momenta for two unpaired nucleons was discussed and also shown in fig. 1c.
In that specific case orbital momentum and spin of the individual nucleons add up to the total angular momenta of each nucleon, which then
add to the total angular momentum (or 'total' spin) of the nucleus. However, this is only one 'extreme' case of angular momentum coupling
denoted as 'jj-coupling'. An alternative case is coupling of the orbital angular momenta of both nucleons to a total orbital
angular momentum of the nucleus  {\it{$\vec{L}$}}, and coupling of the spins of both nucleons to the total spin of the nucleus  {\it{$\vec{S}$}}, which then couple to the total angular momentum of the nucleus {\it{$\vec{J}$}}. This kind of coupling is known as LS - coupling or 
Russel-Saunders coupling. In the following both schemes will be briefly introduced.

\subsection{\bf{2.1 LS coupling (Russel-Saunders - coupling)}}
In this coupling scheme it is assumed that there is only a negligibly weak coupling between the orbital angular momentum {\it{$\vec{l}$}} and the spin
{\it{$\vec{s}$}} of an individual nucleon, but that there is a strong coupling between the orbital angular momenta {\it{$\vec{l}$}} of the different nucleons and also
a strong coupling between the spins {\it{$\vec{s}$}} of the different nucleons (see e.g. \cite{Evans55}). Consequently the individual orbital angular momenta will 
couple to a total orbital angular momentum {\it{$\vec{L}$}}  of the nucleus, i.e. 
{\it{$\vec{L}$}} = {\it{$\sum_{i=1}^{n} \vec{l}$}}, while the individual spins will couple to a total spin {\it{$\vec{S}$}} of the nucleus, i.e
{\it{$\vec{S}$}} = {\it{$\sum_{i=1}^{n} \vec{s}$}}, where  {\it{n}} denotes the number of involved nucleons. The total orbital angular momentum and the total
spin then couple to a total angular momentum  {\it{$\vec{J}$}} of the nucleus,  {\it{$\vec{J}$}} = {\it{$\vec{L}$}} + {\it{$\vec{S}$}}. In this coupling
scheme it is also presumed that states of different {\it{$\vec{L}$}} have quite different energies and states of different {\it{$\vec{S}$}} having the same 
{\it{$\vec{L}$}} value correspond to quite different energy levels, so-called spin multiplets \cite{Evans55}.\\
LS - coupling conditions are quite satisfactorily fulfilled in light nuclei.

\subsection{\bf{2.2 jj coupling}}
In the other extreme case of angular momentum coupling it is presumed that interactions between the orbital angular momentum  {\it{$\vec{l}$}} and the
spin {\it{$\vec{s}$}} of the individual nucleons dominate, while the coupling between the angular momenta and, respectively the spins, of the individual nucleon 
is small (see e.g. \cite{Evans55}), or as sometimes denoted, interaction between orbital angular momentum und spin is larger than the 
'residual interactions'. Consequently orbital angular momentum {\it{$\vec{l}$}} and spin {\it{$\vec{s}$}} of a nucleon couple to a total angular momentum 
{\it{$\vec{j}$}} = {\it{$\vec{l}$}} {\it{$\pm$}} {\it{$\vec{s}$}}, which then couple to a total angular momentum  {\it{$\vec{J}$}} of the nucleus, i.e.
{\it{$\vec{J}$}} = {\it{$\sum_{i=1}^{n} \vec{j}$}}, where {\it{n}} denotes the number of the involved nucleons.\\
jj - coupling conditions are quite satisfactorily fulfilled in heavy nuclei.

\subsection{\bf{2.3 jj-coupling in odd-odd nuclei -- 'Nordheim rule'}}
The jj - coupling discussed above states that angular momenta and spins of the individual particles
first couple to the total angular momentum {\it{$\vec{j}$}} of the individual nucleon, which then couple to a total spin of the nucleus {\it{$\vec{J}$}}.
According to the rules of quantum mechanics the total spin {\it{J}} can access the values J\,=\,{\it{j$_{1}$}}+{\it{j$_{2}$}},
{\it{j$_{1}$}}+{\it{j$_{2}$}}-{\it{1}} .... $\mid${\it{j$_{1}$}}-{\it{j$_{2}$}}$\mid$. It thus is of high interest, which of the possible
values of the total spin {\it{J}} refers to the lowest excitation energy. A first rule was suggested by L.W. Nordheim \cite{Nord50,Nord51} on the basis
of the hypothesis, that the individual configurations of neutrons and protons in odd-odd nuclei are the same as in odd A nuclei with the same number of nucleons in 
the odd particle group \cite{Nord51}:\\
\\
a) If the odd neutron and odd proton belong to different 'Schmidt groups', i.e. if {\it{j$_{1}$}}\,=\,{\it{l$_{1}$$\pm$1/2}} 
and {\it{j$_{2}$}}\,=\,{\it{l$_{2}$$\mp$1/2}},
then the result spin (for the lowest excitation energy) is {\it{J}}\,=\,$\mid${\it{j$_{1}$}}-{\it{j$_{2}$}}$\mid$  ('strong rule').\\
\\
b) If the odd neutron and odd proton belong to the same 'Schmidt group', i.e. if {\it{j$_{1}$}}\,=\,{\it{l$_{1}$$\pm$1/2}} and {\it{j$_{2}$}}\,=\,{\it{l$_{2}$$\pm$1/2}},
then the result spin  (for the lowest excitation energy) is {\it{J}}\,$>$\,$\mid${\it{j$_{1}$}}-{\it{j$_{2}$}}$\mid$ or
sometimes also written as $\mid${\it{j$_{1}$}}-{\it{j$_{2}$}}$\mid$ $<$ {\it{J}} $\le$ {\it{j$_{1}$}}+{\it{j$_{2}$}} (see e.g. \cite{GaM58}) ('weak rule').
It is, however, remarked that there is a tendency to couple to J\,=\,j$_{1}$+j$_{2}$, but this not fulfilled generally, as already stated by L.W. Nordheim 
\cite{Nord51} that these rules 'seem to hold for a great majority of cases, but not without exceptions'.

\subsection{\bf{2.4 jj-coupling in odd-odd nuclei -- 'Brennan rule'}}
The 'Nordheim rules' have been modified a couple of years later by M.H. Brennan and A.M. Bernstein \cite{BrB60}. Rule a) remained unchanged (in \cite{BrB60} denoted as'R1');\\
\\
R1: {\it{J}}\,=\,$\mid${\it{j$_{1}$}}-{\it{j$_{2}$}}$\mid$ for {\it{j$_{1}$}}\,=\,{\it{l$_{1}$$\pm$1/2}} 
and {\it{j$_{2}$}}\,=\,{\it{l$_{2}$$\mp$1/2}}\\
\\
For rule b) two cases where
distinguished: \\
R2-b1) {\it{j$_{1}$}}\,=\,{\it{l$_{1}$+1/2}} ($\uparrow$) and {\it{j$_{2}$}}\,=\,{\it{l$_{2}$+1/2}}  ($\uparrow$); in this case both nucleons couple their spins preferrably to
{\it{J}}\,$=$\,$\mid${\it{j$_{1}$}}+{\it{j$_{2}$}}$\mid$;\\
R2-b2) {\it{j$_{1}$}}\,=\,{\it{l$_{1}$-1/2}}  ($\downarrow$) and {\it{j$_{2}$}}\,=\,{\it{l$_{2}$-1/2}}  ($\downarrow$); in this case both nucleons couple their spins preferrably to
{\it{J}}\,$>$\,$\mid${\it{j$_{1}$}}-{\it{j$_{2}$}}$\mid$;\\
for the specific case of {\it{j$_{1}$}} = {\it{j$_{2}$}} = 1/2, the resulting spin is determined as {\it{J}} = {\it{j$_{1}$}} + {\it{j$_{2}$}}.\\ 
\\
The third rule concerns the case, when the configuration is a mixture of a particle and a hole state. In this case the resulting spin  {\it{J}}
is given as:\\
R3:  {\it{J}} = {\it{j$_{1}$}} + {\it{j$_{2}$}} - 1 \\
It is, however, stated by the authors, that this coupling rule is quite uncertain, and can be regared only as a tendency. \\
\\
M.H. Brennan and A.M. Bernstein investigated about 75 cases; they obtained applying R1 a correct assigment in all cases and wrong assignements
for R2 in 7.3$\%$ and R3 in 54$\%$ of the cases. It should be noted that they did not apply the rules consequently. In some cases also for particle - hole - configurations
the rules R1 and R2 were applied. Also for two - hole states R2 seems inverted, i.e. maximum spin for $\downarrow$ cases and minmum spin for $\uparrow$ cases.

\subsection{\bf{2.5 Strong coupling in two-quasi-particle states in even-even nuclei  -- 'Gallagher rule'}}
Another set of coupling rules was suggested by J.C. Gallagher \cite{Gall62}, which were applied to two particle-states in deformed even-even nuclei. They are of 
relevance for the formation of 2-quasi-particle K isomers. Using the Nilsson notation as defined above, four cases can be distinguished:\\
\\
a) $\Omega_{1}$[N$_{1}$n$_{z1}$$\Lambda_{1}$]($\uparrow$) $\otimes$  $\Omega_{2}$[N$_{2}$n$_{z2}$$\Lambda_{2}$]($\uparrow$) \\
b) $\Omega_{1}$[N$_{1}$n$_{z1}$$\Lambda_{1}$]($\downarrow$) $\otimes$  $\Omega_{2}$[N$_{2}$n$_{z2}$$\Lambda_{2}$]($\downarrow$) \\
c) $\Omega_{1}$[N$_{1}$n$_{z1}$$\Lambda_{1}$]($\downarrow$) $\otimes$  $\Omega_{2}$[N$_{2}$n$_{z2}$$\Lambda_{2}$]($\uparrow$) \\
d) $\Omega_{1}$[N$_{1}$n$_{z1}$$\Lambda_{1}$]($\uparrow$) $\otimes$  $\Omega_{2}$[N$_{2}$n$_{z2}$$\Lambda_{2}$]($\downarrow$) \\
\\
It is pointed out by J.C Gallagher that at a 'strong  coupling' states of $\Omega$ = $\mid$$\Omega_{1}$$\pm$$\Omega_{1}$$\mid$ are degenerated. The
degenration is cancelled by the residual interaction between the neutrons or protons, respectively. Further it is argued that at 'strong coupling'
the relation {\it{$\Omega$ = $\Lambda$}} + {\it{$\Sigma$}} with {\it{$\Lambda$ = $\mid$$\Lambda_{1}$$\pm$$\Lambda_{2}$$\mid$}} and 
{\it{$\Sigma$ = $\mid$$\Sigma_{1}$$\pm$$\Sigma_{2}$$\mid$}} is fulfilled. On this basis it is postulated, that for the lowest lying level {\it{$\Sigma$ = 0}} is valid. 
Thus it follows:\\
\\
a) {\it{$\Omega$(E$^{*}_{min}$) = $\mid$$\Omega_{1}$-$\Omega_{2}$$\mid$}} if {\it{$\Omega_{1}$\,=\,$\Lambda_{1}$$\pm$1/2}} and 
{\it{$\Omega_{2}$\,=\,$\Lambda_{2}$$\pm$1/2}}, i.e. for cases with parallel ($\uparrow\uparrow$) or ($\downarrow\downarrow$) spin projections \\
b) {\it{$\Omega$(E$^{*}_{min}$) = $\Omega_{1}$+$\Omega_{2}$}} if {\it{$\Omega_{1}$\,=\,$\Lambda_{1}$$\pm$1/2}} and 
{\it{$\Omega_{2}$\,=\,$\Lambda_{2}$$\mp$1/2}}, i.e. for cases with antiparallel ($\uparrow\downarrow$) or ($\downarrow\uparrow$) spin projections. \\

\begin{figure*}
	\centering
	\resizebox{0.75\textwidth}{!}{
		\includegraphics{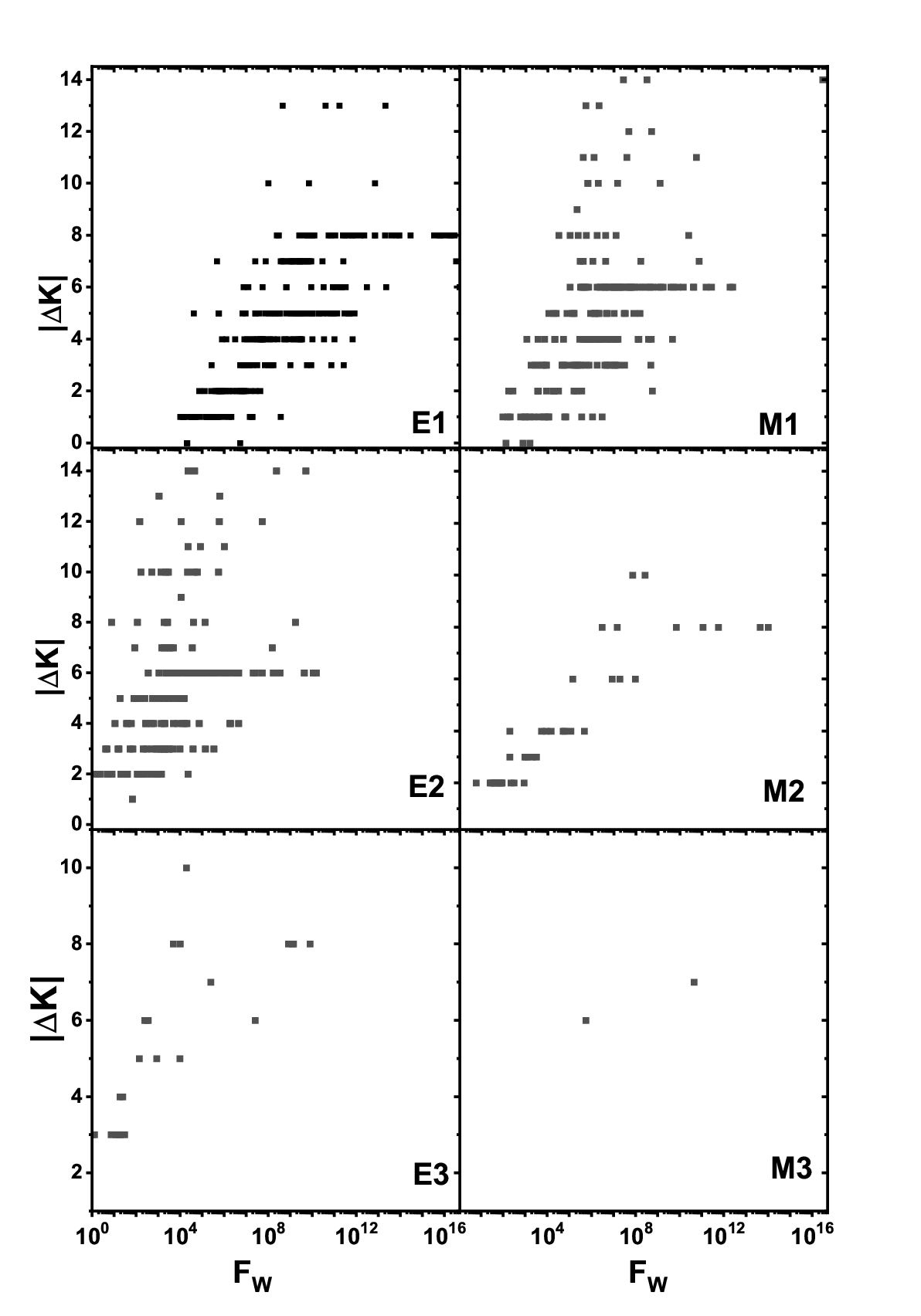}
	} 
	\caption{Hindrance factors F$_{W}$ for the decay of K isomers in dependence of the K difference {\it$\mid$$\Delta$K$\mid$} of initial and final
		states for transitions of multipolarities E1, E2, E3, M1, M2, M3. Data are taken from \cite{Kondev15}.
	}
	\label{fig:2}       
\end{figure*}

\vspace*{0.8cm} 

\section{3. K-hindrance of internal transitions}
Excited nuclear states predominantly decay by internal transitions, $\gamma$ emission, internal conversion. At a transition energy difference 
$\Delta$E $>$ 1022 keV (double rest mass energy of electrons) also decay via electron-positron pair creation is possible.
The transition probability usually depends on the angular momentum difference between the initial and final states, which also defines the multipolarity
of the transition. In general one can state, the higher the multipolarity of the transition is, the lower is the transition probability or, vice versa,
the longer is the life-time of the initial state. At long life-times also other decay modes ($\beta^{+}$-, $\beta^{-}$-, EC-, $\alpha$ - decay or spontaneous fission)
may compete with internal transitions.\\
In case of internal transition from a state {\it{i}} characterized by the quantum numbers (K$_{i}$ J$_{i}$ $\pi_{i}$) to a state {\it{f}} characterized
by (K$_{f}$ J$_{f}$ $\pi_{f}$) some selection rules have to be fulfilled \cite{Alaga55}:\\
\\
a) {\it{$\mid$$J_{i}-J_{f}$\,$\le$$\lambda$\,$\le$J$_{i}$+J$_{f}$}}\\

b) $\pi$\,=\,$\pi_{i}$$\pi_{f}$\\

c) {\it{$\mid$$K_{i}-K_{f}$$\mid$\,=\,$\Delta$K\,$\le$\,$\lambda$}}\\	
\\
where $\lambda$ denotes the multipolarity of the transition.\\

Rules a) and b) are strict, rule c) has to be discussed.

Since, generally spoken, the K quantum number is given by the intrinsic structure of a nucleus
changes of the K quantum number at internal nuclear transitions go along with changes in the
intrinsic structure of the nucleus. It is thus understandable, that changes of the K quantum
number at internal transitions will have some reaction on the transition probabilities. This 
was already demonstrated in early calculations by G. Alaga et al. \cite{Alaga55} and C.J. Gallagher
\cite{Gall60} and also discussed by A. Bohr and B.R. Mottelson \cite{Bohr98}.
They showed that the transition matrix element vanishes if $\Delta$K = $\mid$K$_{i}$-K$_{f}$$\mid$ $>$ $\lambda$,
with $\lambda$ representing the multipolarity of the radiation of the transtion between the initial and the final states.
In other words, transitions  $\Delta$K$>$ $\lambda$ are forbidden. However, this transition rule is only valid
for cases when initial or final states (or both) are 'pure' states. Small pertubative admixtures in the wave function
diminish the 'K forbiddeness'. In other words, 'K forbiddeness' will be replaced by 'K hindrance' or 'K retardation'.
Such admixtures may be 'K mixing', which 'might arise from dynamical effects such as rotation or vibration, 
random interactions with one or more close lying states, or configuration mixing with the underlying nuclear potential' \cite{Kondev15}.
See also \cite{Kondev15}
for more exhaustive discussion. \\

It should be clarified at this point, that the K quantum number was defined  as the projection
of the total angular momentum {\it{$\vec{I}$}} onto the symmetrx axis of the nucleus (see e.g. \cite{Nilsson55}).
For non rotating nuclei K\,=\,$\Omega$. At rotation the total angular momentum {\it{$\vec{I}$}} is the vector sum
{\it{$\vec{I}$}} = {\it{$\vec{J}$}} + {\it{$\vec{R}$}}, where {\it{$\vec{J}$}} is the contribution from the internal
motion of the nucleons and {\it{$\vec{R}$}} is the contribution from the collective motion, thus  K\,$>$\,$\Omega$.
Another definition is given by  G. Alaga et al. \cite{Alaga55}. Here it is argued that all the states of the rotational
band are characterized by the same intrinsic wave function {\it{$\varphi_{\tau K}$}} and thus have all the same
K quantum number, namly that of the bandhead. This is of relevance for the calculation of transition probabilities.
Thus the K hindrance (selection rule c) above) is given by the K difference of the bandheads of the initial and final states.\\

Evidently transitions violating rule c) are hindered. The degree of 
K-forbiddeness {\it{$\nu$}} can be expressed as   
{\it{$\nu$\,=\,$\Delta$K-$\lambda$}}.\\
At first glance it may seem that rules a) and c) are quasi identical. But 
definitely it is not the case for deformed nuclei. Here each single particle, 
'multi'($\ge$\,2) quasi particle or vibrational state is the 'head' of a rotational band with angular momenta I = K, K+1, K+2 ... . Thus for members
of rotational bands cases of 
{\it{$\mid$$I_{i}-I_{f}$}}$\mid$\,$\le$\,{\it{$\mid$$K_{i}-K_{f}$$\mid$}} are
possible, in other words transitions of multipolarities
{\it{$\lambda$($\Delta$I)}}\,$<$\,{\it{$\lambda$($\Delta$K)}} are possible, but
those transitions do not have transition probabilities of single particle
transitions, but are 'K-hindered'. This item will be discussed in more detail in sect. 5.\\
Quantitatively the degree K-hindrance of a transition can also be expressed by the 'delay' or increase of the lifetime relative to the
expected lifetime for a single particle transition, i.e. by a hindrance factor {\it{F$_{W}$}} defined as \cite{Loeb68}\\
\\
\\
\\
{\it{F$_{W}$\,=\,$\frac{T_{1/2\gamma}(experiment)}{T_{1/2\gamma}(Weisskopf estimate)}$}}
\\
\\
K - hindrance factors were compiled by K.E.G L\"obner \cite{Loeb68}, presented in dependancy of the multipolarity of the transition and the 
{\it{$\mid$$\Delta$K$\mid$}} values and compared with the 'empirical rule' of L.I, Rusinov \cite{Rusinov61}
{\it{lg F$_{W}$\,=\,2($\mid$$\Delta$K$\mid$-$\lambda$)}}.
Evidently the experimental values showed a large straggling for each $\mid$$\Delta$K$\mid$ value at each multipolarity. Agreement with the 
'Rusinov empirical rule' is more or less satisfactory for E2, E3, M2 and M4 transition, but disagreed significantly for E1, M1 and E4 transitions.
Recently an updated compilation of hindrance factors was presented by F.G. Kondev et al. \cite{Kondev15}. The results are shown similar to the style of
presentation of L\"obner in fig. 2.\\
Evidently two items are visible, already known from the presentation of L\"obner:\\
a) there is a general trend of increasing hindrance factors at increasing K difference\\
b) in general transitions of lower multipolarities exhibit higher hindrance factors.\\
To get some some quantitative measure for the hindrance we extracted mean hindrance factors $\mid$lgF$_{W}$$\mid$ for each transition multipolarity at each K difference.
The mean values were obtained by fitting a Gaussian to the data if sufficient amount of data was available or taking the 'mean value' in case of 
low statistics. In this case, however, values deviating significantly from the 'bulk' were not respected. The results are presented in table 1
and for E1, E2, (E3), M1 and M2 transitions are also shown in fig. 3.

\begin{table}
	\caption{Mean hindrance factors for individual transitions within the decay of K isomers. Values are obtained either from fitting Gaussians to the data
		or taking the arithmetic means. Results from the latter procedure are given in {\it{italic font}}. Data are taken from \cite{Kondev15}. }
	\label{tab:1}       
	\begin{tabular}{llllll}
		\hline\noalign{\smallskip}
		\noalign{\smallskip}\hline\noalign{\smallskip}
		$\mid$$\Delta$K$\mid$ & $\mid$lgF$_{W}$$\mid$(E1) & $\mid$lgF$_{W}$$\mid$(E2) & $\mid$lgF$_{W}$$\mid$(E3) & $\mid$lgF$_{W}$$\mid$(M1)  &
		 $\mid$lgF$_{W}$$\mid$(M2) \\
	    0 & - & - & - & {\it{2.71}} & - \\
	    1 & 4.96 & {\it{2.71}} & - & 3.56 & - \\
	    2 & 6.16 & 1.21 & - & 4.14 & 1.77 \\
	    3 & 7.00 & 2.18 & {\it{0.86}} & 5.41 & {\it{3.00}} \\
	    4 & 8.09 & 3.00 & {\it{1.37}} & 5.911 & 4.37 \\
	    5 & 9.54 & 3.36 & {\it{3.02}} & 6.12 & - \\
	    6 & 9.43 & 4.17 & {\it{3.70}} & 7.10 & {\it{6.81}} \\
	    7 & 9.32 & 5.23 & {\it{5.29}} & {\it{6.97}} & - \\
	    8 & 11.91 & 5.65 & {\it{7.44}} & {\it{6.36}} & {\it{10.6}} \\
	    9 & - & {\it{6.00}} & - & 5.32 & - \\
	    10 & {\it{10.23}} & 6.35 & {\it{4.28}} & {\it{6.83}} & {\it{8.11}} \\
	    11 & {\it{10.95}} & {\it{7.48}} & - & {\it{7.50}} & - \\
	    12 & - & {\it{8.12}} & - & {\it{8.19}} & - \\
	    13 & - & {\it{8.87}} & - & {\it{12.29}} & - \\
	    14 & - & {\it{9.75}} & - & - & - \\
	    15 & - & - & - & {\it{7.70}} & - \\
	    16 & - & - & - & - & - \\
	    17 & - & {\it{10.18}} & - & 7.20 & - \\
	   	\hline\noalign{\smallskip}
	    \hline\noalign{\smallskip}
	\end{tabular}
	\vspace*{0.cm}  
\end{table}

\begin{figure*}
	\centering
	\resizebox{0.75\textwidth}{!}{
		\includegraphics{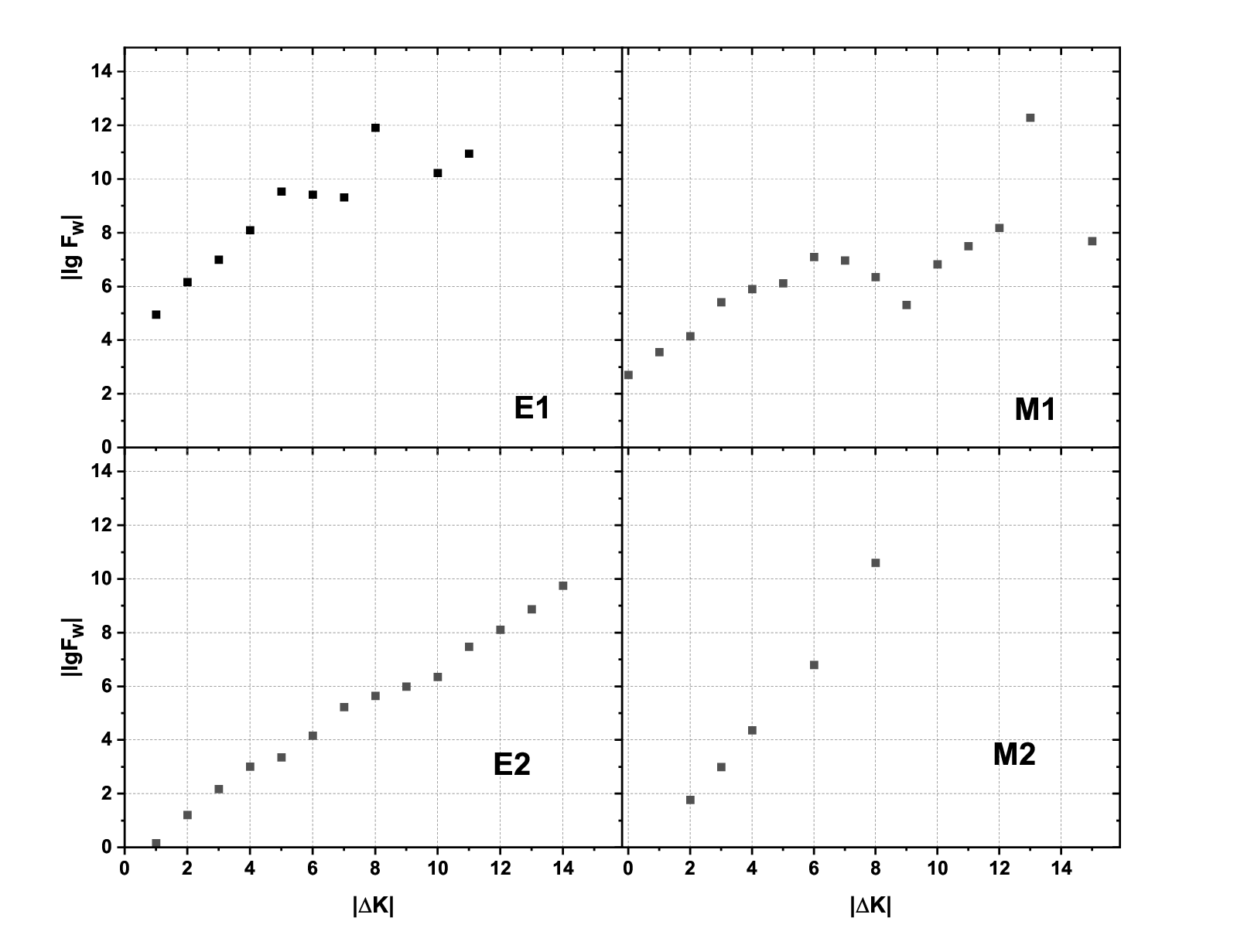}
	} 
	\caption{Mean hindrance factors $\mid$lgF$_{W}$$\mid$ for the decay of K isomers in dependence of the K difference {\it$\mid$$\Delta$K$\mid$} of initial and final
		states for transitions of multipolarities E1, E2, M1, M2. Data are taken from \cite{Kondev15}.
	}
	\label{fig:3}       
\end{figure*}

The trends already indicated in fig. 2 are seen more clearly in fig. 3. But definitely there is no unique trend that would suggest a simple scaling of
the mean hindrance factors {\it{$\mid$lgF$_{W}$$\mid$}} as a function of {\it{$\mid$$\Delta$K$\mid$}} as it had been originally suggested
by L.I. Rusinov \cite{Rusinov61}, who suggested a relation {\it{lg F$_{W}$ = 2($\mid$$\Delta$K$\mid$-$\lambda$)}}. But this item had already discussed by K.E.G. L\"obner \cite{Loeb68}.
For the E2 and M2 transitions we observe roughly a linear increase of {\it{$\mid$lgF$_{W}$$\mid$}} at increasing {\it{$\mid$$\Delta$K$\mid$}}, but slopes
are different. For E1 and M1 transitions the trend of increase seems to stop at {\it{$\mid$$\Delta$K$\mid$ $\approx$6}} and be followed by a plateau or even
a decrease up to {\it{$\mid$$\Delta$K$\mid$ $\approx$ 10}}, increasing again at {\it{$\mid$$\Delta$K$\mid$ $>$ 10}}. While for E1 transitions mean hindrance factors are higher than for E2 transitions 
in the whole range of {\it{$\mid$$\Delta$K$\mid$}} values, the steep increase of the mean hindrance factors for M2 transitions 
suggest higher values than for M1 transitions at {\it{$\mid$$\Delta$K$\mid$ $>$ 6}}. Thus in general one may state that hindrance factors certainly depend on details
of nuclear structure or structure of the K isomers and transition probabilities into lower lying levels and there is no simple relation between hindrance
factors, K differences between initial and final states and multipolarities of the transtions.\\

\vspace*{1.7cm}

\section{4. Experimental aspects}
To obtain detailed information on the decay path of K isomers measurement of the $\gamma$ rays emitted during the deexcitation process is indispensible.
The present technique for investigation of K isomers in the region Z\,$\ge$\,100 is producing the nuclei of interest by complete fusion reactions, 
separate the fusion products ('evaporation residues' (ER)) in-flight from the primary beam by magnetic (gas-filled separators) or electromagnetic 
field arrangements
(e.g. velocity filters) and implant the ER into an arrangement of silicon detectors ('stop detector') where their $\alpha$ decay or 
spontaneous fission (SF) - other radioactive decay modes do not play a role presently - is measured. Gamma rays are registered by an 
arrangement of Ge-detectors, placed close or around the stop detector. As production cross sections are low for these nuclei ($\sigma$ $\le$ 2 $\mu$b)
production rates are $\le$10/s at beam currents of 1 p$\mu$A (6x10$^{12}$ projectiles/s). On the other hand, depending on the set-up, one has a high background
(often $>>$100/s) of $\gamma$ radiation at the focal plane. Sources are e.g. $\gamma$s from the enviroment, $\gamma$s from nuclear reactions at the target
position or the 'beamstop', $\gamma$s from reactions with neutrons produced in nuclear reactios at the target position or the 'beamstop'. This simply means
that one has to discriminate $\gamma$ events from the decay of the isomeric state from background $\gamma$ events. A possibility to do so is given 
by the fact that the decay pattern of a K isomer is usually complicated, i.e. decay of the isomer into the ground state occurs via a couple of steps,
part of which are (highly) converted. So, following a technique applied at SHIP already at the end of the 1970ties \cite{HoM79}, G.D. Jones suggested to use the conversion electrons (CE) registered in the 'stop detector' as a kind of 'calorimeter' \cite{Jones02}. Indeed, using this 'technique' one gets rid of the 'background' $\gamma$ rays which are not in coincidence with CE. So, one is only concerned with two kinds of background: a) accidential coincidences between CE (or 'CE - like events)
and $\gamma$s and b) CE - $\gamma$ coincidence from 'other' nuclei. These could be 'neigbouring isotopes' also produced in the fusion-evaporation reactions
(e.g. from 2n - deexcitation channel of the reaction ($^{208}$Pb($^{48}$Ca,2n)$^{254}$No), if the product of the 
1n - deexcitation ($^{208}$Pb($^{48}$Ca,1n)$^{255}$No) is to be investigated), or isotopes produced in few nucleon transfer reactions and also passing the
separator. These can be discriminated by correlations with the $\alpha$ decay or SF of the considered isotope, i.e. (CE,$\gamma$) - $\alpha$/SF.
So data analysis searching for K isomers will thus be investigation of correlations between the implated nucleus (ER, implantation signal), decay of the isomer
((CE,$\gamma$) coincidences) and radioactive decay of the isotope by $\alpha$ decay or SF, hence correlations ER - (CE,$\gamma$) - $\alpha$/SF. 
Even more complicated decay paths, e.g. decay of isomer '1' into isomer '2' can be analyzed by correlations ER - (CE1,$\gamma$1) - (CE2,$\gamma$2) - $\alpha$,SF.
Clearly, proceeding to heavier nuclei production rates become smaller and thus the observation of $\gamma$ lines becomes less probable. Here it has been shown in recent years that observation of solely CE (ER - CE - $\alpha$/SF) is sufficient for identification of a K isomer and measuring its half-life. But estimation of the excition energy (here only limits can be given) and establishing decay pattern are hardly possible. \\
\\
\\

\section{5. Single Particle K Isomers}
As already mentioned above for nuclei with one unpaired nucleon the relation $\Omega$\,=\,K is valid. So, formally, single particle isomers in well
deformed nuclei may also be regarded as K - isomers. However, such a classification seems only meaningful, if the life-time or the decay are 
determined by the K - hindrance. To understand this issue, one should remind the above mentioned selection rule  
{\it{$\Delta$K}}\,=\,$\mid${\it{K$_{i}$-K$_{f}$}}$\mid$\,$\le${\it{$\Delta$L}}\,$\le${\it{$\lambda$}}, where {\it{K$_{i}$,K$_{f}$}} denote the K-values of the initial and the final state,
and {\it{$\Delta$L}} the angular difference  which finally defines the multipolarity of the transition. So in the case of single particle isomers one has to
distinguish the two cases {\it{K$_{i}$$<$K$_{f}$}} and {\it{K$_{i}$$>$K$_{f}$}}. As the Nilsson level {\it{K$_{f}$}} is the head of a rotational band with spins
{\it{$\Omega$(=\,K$_{f}$)}}, {\it{$\Omega$+1}}, {\it{$\Omega$+2}} ..., in the case of {\it{K$_{i}$$<$K$_{f}$}} 
the angular momentum difference {\it{$\Delta$L}}\,=\,{\it{$\mid$K$_{i}$-K$_{f}$$\mid$}} is the lowest within 
the rotational band, i.e. the transition {\it{K$_{i}$$\rightarrow$K$_{f}$}} has the lowest multipolarity, so according to the definitions given above, there is no
'K-hindrance' and the transition  {\it{K$_{i}$$\rightarrow$K$_{f}$}} can be regarded as a 'usual' single particle transition.\\
Contrary to this situation in the case of {\it{K$_{i}$$>$K$_{f}$}} angular momenta differences into excited members of the band are 
{\it{$\Delta$L}}\,=\,{\it{K$_{i}$-K$_{f}$-n}} (with n\,=\,1, 2 ...) and thus are lower than {\it{K$_{i}$-K$_{f}$}}. This means, that transitions of lower 
multipolarities from the level of
{\it{K$_{i}$}} into members of the rotational band built up on {\it{K$_{f}$}} are possible than for transtions into the bandhead having {\it{K$_{f}$}}. 
Consequently those transitions are 'K - hindered' according to the definitions given above. As illustrative examples (simplified) decay schemes for the
isomers $^{253m1}$No (5/2$^{+}$[622]) \cite{StH10}  and $^{251m}$Cf (11/2$^{-}$[725]) \cite{Ahm71} are shown in fig. 4.

 \begin{figure*}
 	\centering
	\resizebox{0.75\textwidth}{!}{
		\includegraphics{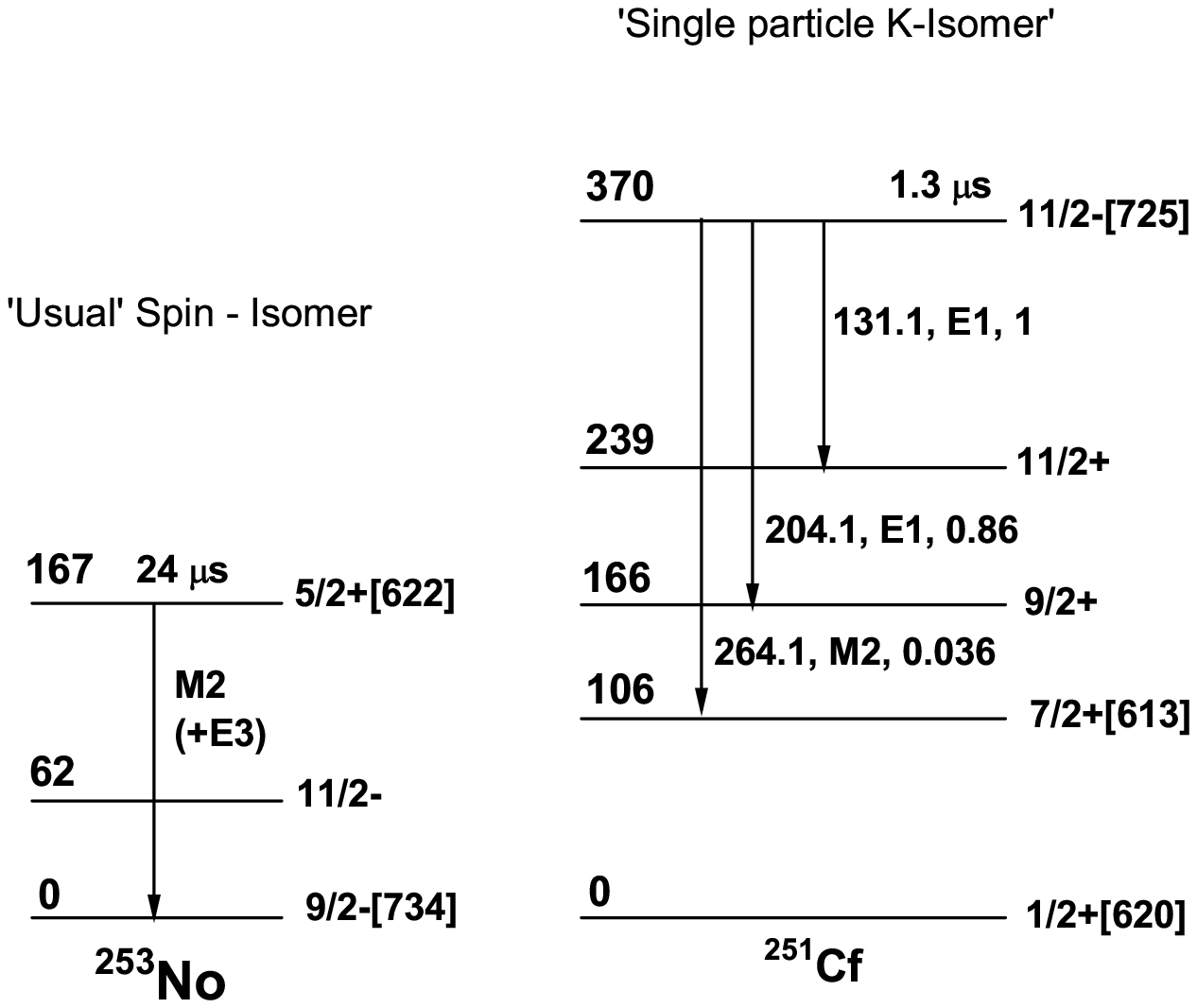}
	} 
	\caption{Simplified decay schemes of $^{253m1}$No \cite{StH10} and $^{251m}$Cf \cite{Ahm71}.
	}
	\label{fig:4}       
\end{figure*}

In the case of $^{253m1}$No the decay proceeds into the 9/2$^{-}$[734] ground-state via an M2 transition (with some E3 admixture). The half-life 
of T$_{1/2}$\,=\,24\,$\mu$s is quite close to the value of 5.9 $\mu$s for a single particle transition obtained from a Weisskopf estimation \cite{Fire96}. 
In the case of $^{251m}$Cf the decay of the isomeric state (T$_{1/2}$\,=\,1.3 $\mu$s) essentially populates the 9/2$^{+}$ and 11/2$^{+}$ members of the rotational band built up on the 7/2$^{+}$[613] Nilsson level by strongly hindered E1 - transitions for which half-lives of $<$10$^{-7}$ $\mu$s are expected on the basis of Weiskopf estimation. The not K - hindered M2
transition into the bandhead is only weak with a relative intensity of 0.036.

\section{6. 'Multi'-Quasiparticle K-Isomers}
Although, as discussed in the previous section, formally, the cases mentioned there, have to be regarded as 'K isomers', still, speaking about
K isomers usually high spin multi-quasiparticle states are meant. Thus in the following we will concentrate on such cases.
The occurence of those isomers is usually explained by breaking at least one nucleon pair and exciting at least one nucleon of each pair in a 
different level, while the different nucleons will couple their spins to high values of K.\\

 \begin{figure*}
 	\centering
	\resizebox{0.75\textwidth}{!}{
		\includegraphics{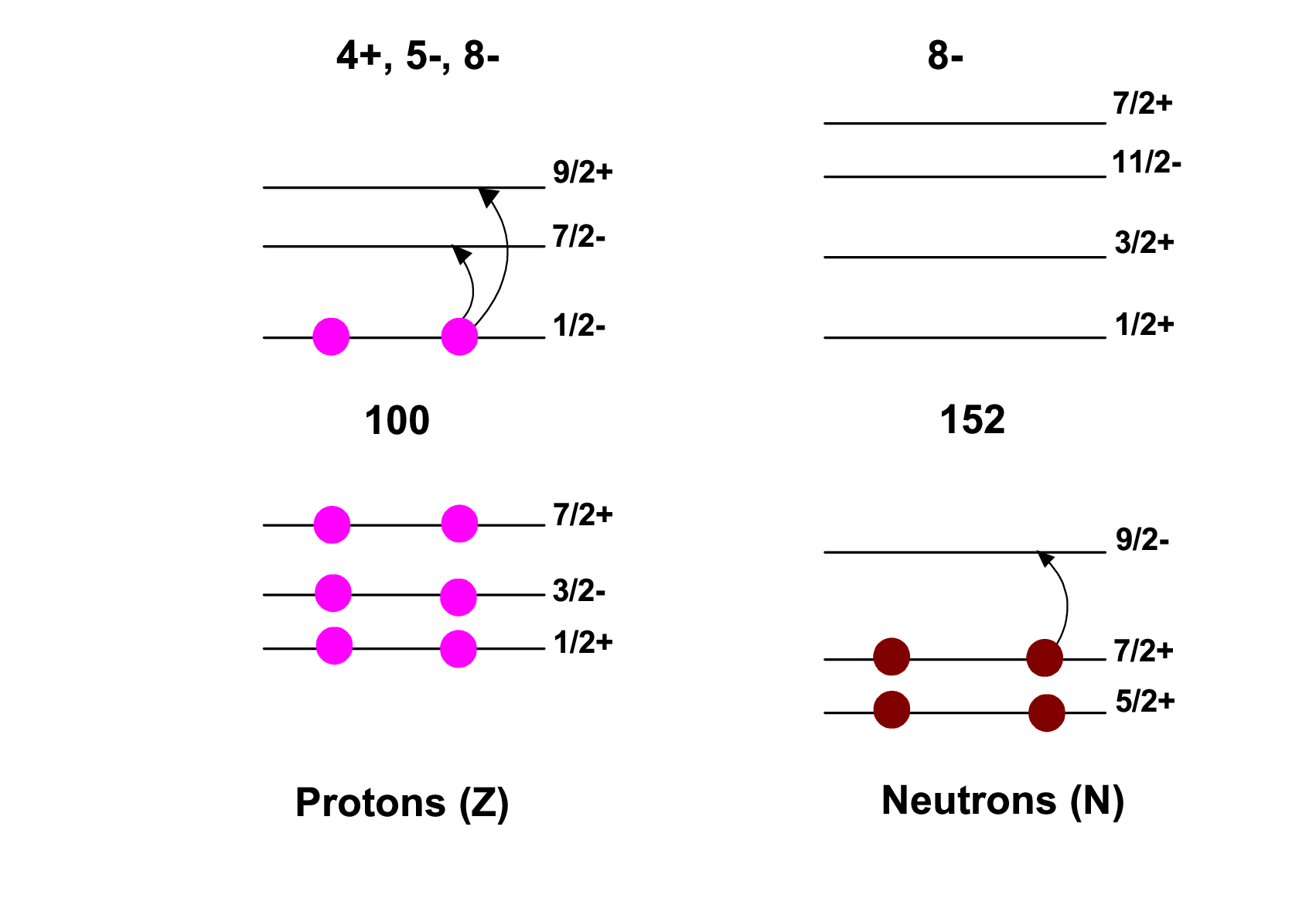}
	} 
	\caption{ Coupling schemes for protons and neutrons close to Z\,=\,100 and N\,=\,152 (schematically).
	}
	\label{fig:5}       
\end{figure*}

The possibilities of 'nucleon coupling' is schematically shown in fig. 5.
In left side of the figure proton levels are occupied up to the first level
(1/2$^{-}$) above the expected shell gap at Z\,=\,100. Due to the relatively small energy gaps, after breaking the pair, one proton may be excited into the
7/2$^{-}$ or into the 9/2$^{+}$ states while the other one remains in the 1/2$^{-}$
state. But it is also conceivable that both protons are removed from the 
1/2$^{-}$ state and one is excited into the 7/2$^{-}$ level, the other one
into the 9/2$^{+}$ state, resulting in possible configurations 
K$^{\pi}$ = 4$^{+}$ (1/2$^{-}$ $\otimes$ 7/2$^{-}$), 5$^{-}$  
(1/2$^{-}$ $\otimes$ 9/2$^{+}$), 8$^{-}$  
(7/2$^{-}$ $\otimes$ 9/2$^{+}$). The right side shows the situation 
for neutrons occupying levels up to the   
7/2$^{+}$ one still lying below the shell gap at N = 152. After breaking 
the neutron pair one of them can be excited into the 9/2$^{-}$ level, also
lying below the N = 152 shall gap, resulting in the configuration 
K$^{\pi}$ = 8$^{-}$. Due to the large energy difference excitation into 
a level above the N = 152 shell gap is less probable.

\subsection{{\bf 6.1 Early Investigations}}
First information on K isomers in the transfermium region came from A. Ghiorso et al. \cite{GhE73}, who observed $\alpha$ decay of $^{250}$Fm and
$^{254}$No correlated to mother activites of T$_{1/2}$\,=\,1.8$\pm$0.1 s ($^{250}$Fm) and T$_{1/2}$\,=\,0.28$\pm$0.04 s ($^{254}$No). They 
interpreted the 'mother activities' as isomeric states in the corresponding  nuclei, decaying into the ground-state by internal transitions,
but no information on excitation energies and decay pattern could be given. In \cite{GhE73} also possible configurations of the isomeric states
were discussed. Possible two-neutron configurations were $\nu$7/2$^{+}$[624]$\downarrow$ $\otimes$ $\nu$9/2$^{-}$[734]$\uparrow$ or
$\nu$9/2$^{-}$[734]$\uparrow$ $\otimes$ $\nu$7/2$^{+}$[613]$\uparrow$ both leading to K$^{\pi}$ = 8$^{-}$ states. But also two-proton states
were considered, $\pi$7/2$^{+}$[633]$\uparrow$ $\otimes$ $\pi$7/2$^{-}$[514]$\downarrow$ leading to a K$^{\pi}$ = 7$^{-}$ state for $^{250}$Fm
and $\pi$7/2$^{-}$[514]$\downarrow$ $\otimes$ $\pi$9/2$^{+}$[624]$\uparrow$ leading to a K$^{\pi}$ = 8$^{-}$ state for $^{254}$No.\\

Calculations performed almost twenty years later for $^{250}$Fm and $^{254}$No by V.G. Solov'ev et al. \cite{Sol91} 
resulted in each three  K$^{\pi}$ = 7$^{-}$ and K$^{\pi}$ = 8$^{-}$ 2-quasi-particle states in $^{250}$Fm, namely:\\
\\
a) ( K$^{\pi}$ = 7$^{-}$)  $\nu$5/2$^{+}$[622]$\uparrow$ $\otimes$ $\nu$9/2$^{-}$[734]$\uparrow$ (100$\%$) at  E$^{*}$ = 1.5 MeV \\
b) ( K$^{\pi}$ = 7$^{-}$) $\pi$7/2$^{-}$[514]$\downarrow$ $\otimes$ $\pi$7/2$^{+}$[633]$\uparrow$ (99$\%$) at  E$^{*}$ = 1.53 MeV \\
c) ( K$^{\pi}$ = 7$^{-}$)  $\nu$7/2$^{+}$[624]$\uparrow$ $\otimes$ $\nu$7/2$^{-}$[743]$\downarrow$ (100$\%$) at  E$^{*}$ = 1.9 MeV \\
d) ( K$^{\pi}$ = 8$^{-}$)  $\nu$7/2$^{+}$[624]$\downarrow$ $\otimes$ $\nu$9/2$^{-}$[734]$\uparrow$ (100$\%$) at  E$^{*}$ = 0.8 MeV \\
e) ( K$^{\pi}$ = 8$^{-}$)  $\nu$7/2$^{+}$[613]$\uparrow$ $\otimes$ $\nu$9/2$^{-}$[734]$\uparrow$ (100$\%$) at  E$^{*}$ = 1.6 MeV \\
f) ( K$^{\pi}$ = 8$^{-}$) $\pi$7/2$^{-}$[514]$\downarrow$ $\otimes$ $\pi$9/2$^{+}$[624]$\uparrow$ (99$\%$) at  E$^{*}$ = 1.7 MeV \\

Calculations for $^{254}$No resulted in three  K$^{\pi}$ = 8$^{-}$ 2-quasi-particle states
at E$^{*}$\,=\,(1-2) MeV, namely configurations:\\
\\
a)   $\nu$7/2$^{+}$[613]$\uparrow$ $\otimes$ $\nu$9/2$^{-}$[734]$\uparrow$ (99$\%$) with a 1$\%$ $\pi$7/2$^{-}$[514]$\downarrow$ $\otimes$
$\pi$9/2$^{+}$[624]$\uparrow$ admixture at E$^{*}$ = 1.4 MeV \\
b)   $\pi$7/2$^{-}$[514]$\downarrow$ $\otimes$ $\pi$9/2$^{+}$[624]$\uparrow$ (96$\%$) with a 3$\%$  $\nu$7/2$^{+}$[624]$\uparrow$ $\otimes$ 
$\nu$9/2$^{-}$[734]$\uparrow$ admixture at E$^{*}$ = 1.44 MeV \\
c)   $\nu$7/2$^{+}$[624]$\downarrow$ $\otimes$ $\nu$9/2$^{-}$[734]$\uparrow$ (97$\%$) with a 3$\%$ $\pi$7/2$^{-}$[514]$\downarrow$ $\otimes$
$\pi$9/2$^{+}$[624]$\uparrow$ admixture at E$^{*}$ = 1.7 MeV. \\

Further studies on both K isomers, $^{250m}$Fm and $^{254m}$No, were performed by Yu.A. Lazarev et al. \cite{LaL89}, who searched for fission branches,
but they could only give upper limits of b$_{SF}$ $\le$ 8.2x10$^{-7}$ for $^{250m}$Fm and b$_{SF}$ $\le$ 2.0x10$^{-3}$ for $^{254m}$No.\\
Another K isomer was discovered in the course of a $\beta^{-}$ decay study of $^{256}$Es by H.L. Hall et al. \cite{Hall89}. They measured a half-life 
T$_{1/2}$ = 70$\pm$5 ns for the previously known level at E$^{*}$\,=\,1425 keV in $^{256}$Fm, which they assigned to a K$^{\pi}$\,=\,7$^{-}$ state. They also observed two 
delayed fission events, which they attributed to the decay of the isomer.\\

In late 2000 the first K isomer in the transfermium region decaying by $\alpha$ emission, $^{270m}$Ds \cite{HoH01}, was observed at the velocity filter
SHIP at GSI, Darmstadt (Germany).\\

A breakthrough in K isomer spectroscopy came in the beginning of the 21st century, mainly thanks to higher available beam intensities for medium heavy projectiles like $^{48}$Ca, 
fast and efficient separation of the evaporation residues from the primary beam, and the availability of Ge - detector set-ups of high detection efficiency.
So $\gamma$ spectroscopic investigation of $^{250m}$Fm and $^{254m}$No became possible. As an additional filter to discriminate between $\gamma$ rays from the decay 
of the isomer and background radiation prompt coincidences between conversion electrons and $\gamma$ rays were searched for as suggested by 
G.D. Jones et al. \cite{Jones02} following a technique that already had been applied  by the end of the 1970ies at SHIP
\cite{HoM79} (see sect. 4). First decay schemes of $^{254m}$No were presented by R.-D. Herzberg et al. \cite{HeG06} at the RITU separator at the University of Jyv\"askyl\"a (Finland) and by S.K. Tandel et al. \cite{TaK06} at the FMA separator at ANL, Argonne (USA). In both experiments another isomeric state ($^{254m2}$No) with a half-life of
$\approx$170 $\mu$s was observed.\\

Short time later the first 'new' K isomer in the transfermiun region, $^{252m}$No \cite{Hes06} and the first even-Z - odd-A K isomer, $^{251m2}$No \cite{Hes06a}
were discovered at SHIP by F.P. Heßberger et al., and a further K isomer, $^{253m2}$No, was identified at SHIP  \cite{Hes06b}
and by A. Lopez-Martens et al. at the VASSILISSA separator at FLNR JINR, Dubna (Russia) \cite{LoH07}.
 \\
These positive results initiated an intense search for K isomers in the transfermium (or 'superheavy element') region at many facilities.\\
An overview of the presently known 2 - quasiparticle - K isomers in even - even nuclei is shown in table 2, 4 - quasiparticle are presented in table 3 
and an overview of odd-mass K isomers is given in table 4.\\

\begin{table}
\caption{'Safely' identified 2-quasiparticle - K isomers in even-even nuclei}
\label{tab:2}       
\begin{tabular}{lllllll}
\hline\noalign{\smallskip}
\noalign{\smallskip}\hline\noalign{\smallskip}
Isotope & E$^{*}$/MeV & half-life & decay mode & (assumed) & (assumed) configuration & Reference\\ 
&  & &  &spin/parity&  & \\ 
\hline\noalign{\smallskip}     
$^{270m}$Ds & $\approx$1.13 & 6.0$^{+8.2}_{-2.2}$ms  & $\alpha$  &$\nu$9$^{-}$& $\nu$11/2$^{-}$[725]$\uparrow$$\otimes$$\nu$7/2$^{+}$[613]$\uparrow$ &  \cite{HoH01}\\
&     &    &    &$\nu$10$^{-}$& $\nu$11/2$^{-}$[725]$\uparrow$$\otimes$$\nu$9/2$^{+}$[615]$\downarrow$ & \\
\hline\noalign{\smallskip} 
$^{266m}$Hs &    & 74$^{+354}_{-34}$ms   & $\alpha$ &   &  & \cite{Ack12} \\
\hline\noalign{\smallskip} 
$^{254m}$Rf &    & 4.7$\pm$1.1 $\mu$s &  IT, SF ? &$
\nu$8$^{-}$& $\nu$9/2$^{-}$[734]$\uparrow$$\otimes$$\nu$7/2$^{+}$[624]$\downarrow$  &  \cite{David15} \\
\hline\noalign{\smallskip} 
$^{256m1}$Rf & $\approx$1.120  & 25$\pm$2 $\mu$s & IT &$\pi$5$^{-}$& $\pi$1/2$^{-}$[521]$\downarrow$$\otimes$$\pi$9/2$^{+}$[624]$\uparrow$ & \cite{Jepp09,Berry10,Robi11}\\
 &  &  &  &  &  & \cite{Rub13,Khuy21a}\\
\hline\noalign{\smallskip} 
$^{256m2}$Rf & $\approx$1.40   & 17$\pm$2 $\mu$s & IT &$\pi$8$^{-}$& $\pi$7/2$^{-}$[514]$\downarrow$$\otimes$$\pi$9/2$^{+}$[624]$\uparrow$ & \cite{Jepp09,Berry10,Robi11}\\
 &  &  &  &  &  & \cite{Rub13,Khuy21a}\\
\hline\noalign{\smallskip} 
$^{258m1}$Rf &   & 2.4$^{+2.4}_{-0.8}$ ms   & IT  &   &  & \cite{Hess16}\\
\hline\noalign{\smallskip} 
$^{258m2}$Rf &   & 15$\pm$10 $\mu$s  & IT  &   &   & \cite{Hess16}\\
\hline\noalign{\smallskip} 
$^{250m}$No &    & 43$^{+22}_{-15}$ $\mu$s & IT  &$\nu$6$^{+}$& $\nu$5/2$^{+}$[622]$\uparrow$$\otimes$$\nu$7/2$^{+}$[624]$\downarrow$   & \cite{Peter06,Khuyag22}\\
\hline\noalign{\smallskip} 
$^{252m}$No & 1.254    &   110$\pm$10 ms & IT &$\nu$8$^{-}$& $\nu$9/2$^{-}$[734]$\uparrow$$\otimes$$\nu$7/2$^{+}$[624]$\downarrow$ & \cite{Sul07} \\
\hline\noalign{\smallskip} 
$^{254m1}$No & 1.295    &  275$\pm$7 ms & IT  &$\nu$8$^{-}$& $\nu$7/2$^{+}$[624]$\downarrow$$\otimes$$\nu$9/2$^{-}$[734]$\uparrow$  & \cite{Hes10,Clark10}\\
&  &   &  SF &  &  & \\
&  &   &\small{((2.0$\pm$1.2)$\times$10$^{-4}$}) &  &  & \\
&  &   & $\alpha$ &  &  & \\
&  &   &($\le$1$\times$10$^{-4}$) &  &  & \\
\hline\noalign{\smallskip} 
$^{256m}$No &     &   7.8$^{+8.3}_{-2.6}$ $\mu$s & IT & $\nu$5$^{-}$ & $\nu$11/2$^{-}$[725]$\uparrow$$\otimes$$\nu$1/2$^{+}$[620]$\uparrow$ & \cite{Kess21} \\
&     &   11.9$^{+21.7}_{-4.3}$ $\mu$s & IT &$\nu$7$^{-}$& $\nu$11/2$^{-}$[725]$\uparrow$$\otimes$$\nu$3/2$^{+}$[622]$\downarrow$ &  \\
\hline\noalign{\smallskip} 
$^{248m}$Fm &     & 10.1$\pm$0.6 ms   & IT &  &   & \cite{Ketel10}\\	 
\hline\noalign{\smallskip} 
$^{250m}$Fm & 1.199    & 1.92$\pm$0.05 s   & IT &$\nu$8$^{-}$& $\nu$9/2$^{-}$[734]$\uparrow$$\otimes$$\nu$7/2$^{+}$[624]$\downarrow$ & \cite{Green08}\\
\hline\noalign{\smallskip} 
$^{256m}$Fm & 1.425    & 70$\pm$5 ns  & IT, SF($\approx$2x10$^{-5}$) &$\nu$7$^{-}$ &  & \cite{Hall89} \\
\hline\noalign{\smallskip} 
$^{246m}$Cm & 1.179    & 1.12$\pm$0.24  & IT &$\nu$8$^{-}$& $\nu$9/2$^{-}$[734]$\uparrow$$\otimes$$\nu$7/2$^{+}$[624]$\downarrow$   & \cite{Fields68,Mult71,Yates75} \\
 &  &  &  &  &  & \cite{Rob08,Shir19}\\
\hline\noalign{\smallskip} 
$^{248m}$Cm & 1.461    & 146$\pm$18 $\mu$s  & IT & $\nu$8$^{-}$  & $\nu$7/2$^{+}$[624]$\downarrow$$\otimes$$\nu$9/2$^{-}$[734]$\uparrow$    & \cite{Shir19} \\
&   &  & & $\nu$8$^{-}$ & $\nu$7/2$^{+}$[613]$\downarrow$$\otimes$$\nu$9/2$^{-}$[734]$\uparrow$    &  \\
\hline\noalign{\smallskip} 
$^{244m}$Pu & 1.216 & 1.75$\pm$0.12 s & IT &$\nu$8$^{-}$&  $\nu$9/2$^{-}$[734]$\uparrow$$\otimes$$\nu$7/2$^{+}$[624]$\downarrow$  & \cite{Hota16} \\
\hline\noalign{\smallskip} 
\hline\noalign{\smallskip} 
* see table 4
\end{tabular}
\end{table}  

\begin{table}
\caption{'Safely' identified 4-quasiparticle - K isomers in even-even nuclei}
\label{tab:3}       
\begin{tabular}{llllll}
\hline\noalign{\smallskip}
\noalign{\smallskip}\hline\noalign{\smallskip}
Isotope & E$^{*}$/MeV & half-life / $\mu$s & K$^{\pi}$  & (assumed) configuration & Reference\\  
\hline\noalign{\smallskip}     
$^{250}$Fm & $\ge$1.530 & 8$\pm$2 & - &  -  &  \cite{HoH01}\\
\hline\noalign{\smallskip}     
$^{254}$No & 2.914 & 108$\pm$13 $\mu$s & 16$^{+}$ &  $\pi$7/2$^{-}$[514]$\downarrow$ $\otimes$ $\pi$9/2$^{+}$[624]$\uparrow$ 
$\otimes$ $\nu$7/2$^{+}$[624]$\uparrow$ $\otimes$ $\nu$9/2$^{-}$[734]$\uparrow$ &  \cite{Hes10,Liu14}\\
\hline\noalign{\smallskip}     
$^{254}$Rf & -  & 247$\pm$73 $\mu$s & 16$^{+}$ &  $\pi$7/2$^{-}$[514]$\downarrow$ $\otimes$ $\pi$9/2$^{+}$[624]$\uparrow$ 
$\otimes$ $\nu$7/2$^{+}$[624]$\downarrow$ $\otimes$ $\nu$9/2$^{-}$[734]$\uparrow$ &  \cite{David15,Liu14}\\
\hline\noalign{\smallskip} 
\hline\noalign{\smallskip} 
\end{tabular}
\end{table}  

\begin{table}
\caption{'Safely' identified multi - quasiparticle - K isomers in odd-mass nuclei. For K isomers in $^{253}$Rf see table 8.}
\label{tab:3}       
\begin{tabular}{llllll}
\hline\noalign{\smallskip}
\noalign{\smallskip}\hline\noalign{\smallskip}
Isotope & E$^{*}$/MeV & half-life & (assumed) K$^{\pi}$  & (assumed) configuration & Reference\\  
\hline\noalign{\smallskip}     
$^{249m}$Md & $\ge$ 0.910 & 2.4$\pm$0.3 ms & 19/2$^{-}$ & $\pi$7/2$^{-}$[514]$\downarrow$ $\otimes$ $\nu$5/2$^{+}$[622]$\uparrow$ 
$\otimes$ $\nu$7/2$^{+}$[624]$\downarrow$ & \cite{GoiT21}   \\
\hline\noalign{\smallskip} 
$^{251m}$Md & $\ge$ 0.844 & 1.37$\pm$0.6 s & 23/2$^{+}$ & $\pi$7/2$^{-}$[514]$\downarrow$ $\otimes$ $\nu$7/2$^{+}$[624]$\downarrow$ 
$\otimes$ $\nu$9/2$^{-}$[734]$\uparrow$ &  \cite{GoiT21}   \\
\hline\noalign{\smallskip}  
$^{251m2}$No & $\ge$ 1.7 & $\ge$2 $\mu$s &  &  &  \cite{Hes06a}  \\
\hline\noalign{\smallskip}
$^{253m2}$No & $\ge$ 1.36 & 715$\pm$3 $\mu$s & 25/2$^{+}$ & $\nu$9/2$^{-}$[734]$\uparrow$ $\otimes$ $\pi$7/2$^{-}$[514]$\downarrow$ $\otimes$  
$\nu$9/2$^{+}$[624]$\downarrow$ &  \cite{Hes06b,LoH07,LoM11,AntH11}   \\  
&  ($\ge$ 1.24) & & 19/2$^{+}$ & $\nu$9/2$^{-}$[734]$\uparrow$ $\otimes$ $\pi$1/2$^{-}$[514]$\downarrow$ $\otimes$  
$\nu$9/2$^{+}$[624]$\downarrow$ &  \cite{LoM11}   \\
\hline\noalign{\smallskip}
$^{255m2}$No & $\approx$1.3 & 2$\pm$1 $\mu$s & 21/2$^{+}$ & 
$\pi$9/2$^{+}$[624]$\uparrow$ $\otimes$ $\pi$1/2$^{-}$[521]$\downarrow$ $\otimes$ $\nu$11/2$^{-}$[725]$\uparrow$ &  
\cite{Hes10,Bronis22,Kessaci22}  \\
 &  & (1.2$^{+0.6}_{-0,4}$ $\mu$s) &  &  &   \\
\hline\noalign{\smallskip}
$^{255m3}$No & $\ge$1.5 & 92$\pm$13 $\mu$s & 27/2$^{+}$ &  
$\pi$9/2$^{+}$[624]$\uparrow$ $\otimes$ $\pi$7/2$^{-}$[514]$\downarrow$ $\otimes$ $\nu$11/2$^{-}$[725]$\uparrow$ & 
 \cite{Hes10,Bronis22,Kessaci22}    \\
& & 77$\pm$6 $\mu$s & &  &     \\
\hline\noalign{\smallskip}
$^{255m4}$No & $>$2.5 & 5$\pm$1 $\mu$s &  &  &  \cite{Kessaci22}    \\
\hline\noalign{\smallskip}
$^{255m2}$Lr & $>$ 1.6 (1.41)& 1.81$\pm$0.2 ms & 25/2$^{+}$ & $\pi$7/2$^{-}$[514]$\downarrow$ $\otimes$ $\nu$11/2$^{-}$[725]$\uparrow$ $\otimes$ 
$\nu$7/2$^{+}$[624]$\downarrow$ & \cite{Hau08,AntH08,Jepp08}  \\ 
\hline\noalign{\smallskip}
$^{255m3}$Lr & 0.74 & 10-100 ns & 15/2$^{+}$ &  $\pi$1/2$^{-}$[521]$\downarrow$ $\otimes$ $\pi$7/2$^{-}$[514]$\downarrow$ $\otimes$ $\pi$9/2$^{+}$[624]$\uparrow$ & 
\cite{Jepp08} \\
\hline\noalign{\smallskip}
$^{255m2}$Rf & 1.103  &  15$^{+6}_{-4}$ $\mu$s   &  19/2$^{+}$    &  $\nu$ 9/2$^{-}$[734]$\uparrow$ $\otimes$ $\pi$ 1/2$^{-}$[521]$\downarrow$ 
$\otimes$ $\pi$ 9/2$^{+}$[624]$\uparrow$     &  \cite{Mosat20,Chakma23}     \\  
&  &  29$^{+7}_{-5}$ $\mu$s  &   &    &  \cite{Chakma23}   \\  
\hline\noalign{\smallskip}
$^{255m3}$Rf &  1.303 &  38$^{+12}_{-7}$ $\mu$s  & 25/2$^{+}$  &  $\nu$ 9/2$^{-}$[734]$\uparrow$ $\otimes$ $\pi$ 7/2$^{-}$[514]$\downarrow$ 
$\otimes$ $\pi$ 9/2$^{+}$[624]$\uparrow$   &  \cite{Mosat20,Chakma23}   \\  
&  &  49$^{+13}_{-10}$ $\mu$s  &   &    &  \cite{Chakma23}   \\  
\hline\noalign{\smallskip}
$^{257m2}$Rf & 1.085   & 106$^{+}_{-}$ $\mu$s  &   21/2$^{+}$  &   & \cite{Jepp09,Qian09,Berry10,Riss13}    \\  
\hline\noalign{\smallskip} 
\hline\noalign{\smallskip} 
\end{tabular}
\end{table}

\subsection{{\bf 6.2 K isomers in $^{254}$No}}
Decay of the K isomers in $^{254}$No have been investigated at different facilities. After the pioneering experiments at RITU \cite{HeG06} and at the
FMA \cite{TaK06} detailed studies have been performed at SHIP \cite{Hes10} and at the BGS at LNBL, Berkeley (USA) \cite{Clark10}. Still no unambiguous 
configuration or decay scheme were obtained so far.\\
In \cite{HeG06} and  \cite{TaK06} the long-lived isomer $^{254m1}$No (275 ms \cite{Hes10}) was assigned as the 2-quasi-proton state of the configuration
K$^{\pi}$ = 8$^{-}$ ($\pi$7/2$^{-}$[514]$\downarrow$ $\otimes$ $\pi$9/2$^{+}$[624]$\uparrow$), while the short-lived isomer $^{254m2}$No (174 $\mu$s \cite{Hes10}), the decay of which was characterized
by two intense $\gamma$ lines of E\,=\,134 keV and E\,=\,605 keV, was attributed to a 4-quasi-particle state of spin and parity K$^{\pi}$ = 16$^{+}$  \cite{HeG06}
or K$^{\pi}$ = 14$^{+}$  \cite{TaK06}. The decay of the K$^{\pi}$ = 8$^{-}$ isomer was assumed to occur via a rotational band built up on a K$^{\pi}$ = 3$^{+}$ 
2-quasi-proton state of the configuration $\pi$1/2$^{-}$[521]$\downarrow$ $\otimes$ $\pi$7/2$^{-}$[514]$\downarrow$ into the ground-state rotational band. The decay of the K$^{\pi}$ = 16$^{+}$ 
(14$^{+}$) isomer was found to populate the K$^{\pi}$ = 8$^{-}$ isomer. While in \cite{HeG06} the decay path was left open, in \cite{TaK06} it was assumed, that the decay of  $^{254m2}$No would 
populate a rotational band built up on  $^{254m1}$No. The strong 605-keV line was not assigned to the decay of  $^{254m2}$No but 
from a level slightly below it.\\
The more detailed study at SHIP \cite{Hes10} confirmed the data of the privious studies. In addition a weak transition of  $^{254m1}$No into the ground-state
rotational band (8$^{+}$ - state) was observed. Also the assumption of populating a rotational band built up on $^{254m1}$No by decay of $^{254m2}$No \cite{TaK06}
was in-line with the data measured in \cite{Hes10}. Again the 605-keV lines was not assigned to the decay of $^{254m2}$No, for which in \cite{Hes10} also spin and parity K$^{\pi}$ = 16$^{+}$ was considered, but from an intermediate state populated
by the decay of the isomer. The reason for this interpretation was the observation of a 605-keV $\gamma$-transition in in-beam investigations at the RITU
separator at Jyv\"askyl\"a in prompt coincidence with $^{254}$No evaporation residues \cite{Herz09}. Thus it should be emitted from a level with a life-time of
$<$1$\mu$s. The decay scheme suggested in \cite{Hes10} is shown in fig. 6 (including some modifications discussed in the following). In addition decay of $^{254m1}$No by $\alpha$ emission and spontaneous fission was searched for in \cite{Hes10}. Small branches of 
b$_{\alpha}$ $\le$ 1x10$^{-4}$ and b$_{sf}$ = (2.0$\pm$1.2)\,x\,10$^{-4}$ were identified. \\
A closer inspection of the E = 605 keV line observed in the in-beam experiments, however showed, that it rather is a line doublet composed of the E = 604.21 keV and
E = 608.35 keV lines from transitions between excited states $^{74}$Ge populated by $^{74}$Ge(n,n'$\gamma$) reactions
with the detector material and $\gamma$ decays of the primarily populated states \cite{War16}.\\
A follow-up study at the BGS at LNBL confirmed the data obtained in the previous studies, but came to another conclusion on the configuration of
$^{254m1}$No and the decay path of $^{254m2}$No \cite{Clark10}. $^{254m1}$No was still assigned as a  K$^{\pi}$ = 8$^{-}$ but interpreted as a 
2-quasi-neutron state of the configuration $\nu$9/2$^{-}$[734]$\uparrow$ $\otimes$ $\nu$7/2$^{+}$[624]$\downarrow$ or  
$\nu$9/2$^{-}$[734]$\uparrow$ $\otimes$ $\nu$7/2$^{+}$[613]$\uparrow$.
The decay of the K = 16$^{+}$ isomer $^{254m2}$No was now interpreted not to populate directly the rotational band built up on  $^{254m1}$No, but via
a rotational band built up on a K = 10$^{+}$ state with the configuration $\nu$9/2$^{-}$[734]$\uparrow$ $\otimes$ $\nu$11/2$^{-}$[725]$\downarrow$. 
An experimental hint
for this interpretation was the observation of $\gamma$ - $\gamma$ coincidences for events within the E = 133 keV line, which had not been observed in the
previous studies \cite{HeG06,TaK06,Hes10} due to less 'statistics'. This finding proved the existance of two $\gamma$ of E $\approx$ 133 keV within
the decay path of $^{254m2}$No, which was not
in-line with the assumption of a 'direct' decay into the rotational band built up on $^{254m1}$No. The prompt decays from the K$^{\pi}$ = 10$^{+}$ 
2-neutron-quasiparticle state into members of the rotational band built up on the K$^{\pi}$ = 8$^{-}$ suggested that the latter would also be
2-neutron-quasiparticle state. The decay scheme proposed in \cite{Clark10} is shown in fig. 6b. \\
However, measurements of E2 / M1 - strengths by means of $\gamma$- und conversion electron (CE) spectroscopy performed at the RITU separator 
in Jyv\"askyl\"a \cite{War16} indicated that the decay path of $^{254m2}$No suggested in \cite{Clark10} is not compatible with the data, i.e. is unlikely, thus supporting 
the decay path suggested in \cite{Hes10}. On the other hand they showed that a 2-quasi-neutron configuration is more likely for 
the K$^{\pi}$ = 8$^{-}$ isomer \cite{War16}.
A modified decay scheme from \cite{Hes10} giving credit to the results in \cite{War16} is shown in fig. 6a.\\
Summarizing the discussion above, one has to state, that there is presently agreement in interpretation of the K$^{\pi}$ = 8$^{-}$ isomer $^{254m1}$No as
a 2-quasi-neutron configuration and its decay path seems clear. The shortlived isomer $^{254m2}$No obviously is a K$^{\pi}$ = 16$^{+}$ state, while its
decay path is still under discussion. But certainly some more precise measurements collecting higher 'statistics', which is not so hard to achieve will solve 
the problem.\\
In a recent review article by A. Lopez-Martens et al. \cite{LopM22} $\gamma$ spectra from the decay of both isomers, but no further information or details
on the structure or decay paths are given.\\

 \begin{figure*}
	\resizebox{0.99\textwidth}{!}{
		\includegraphics{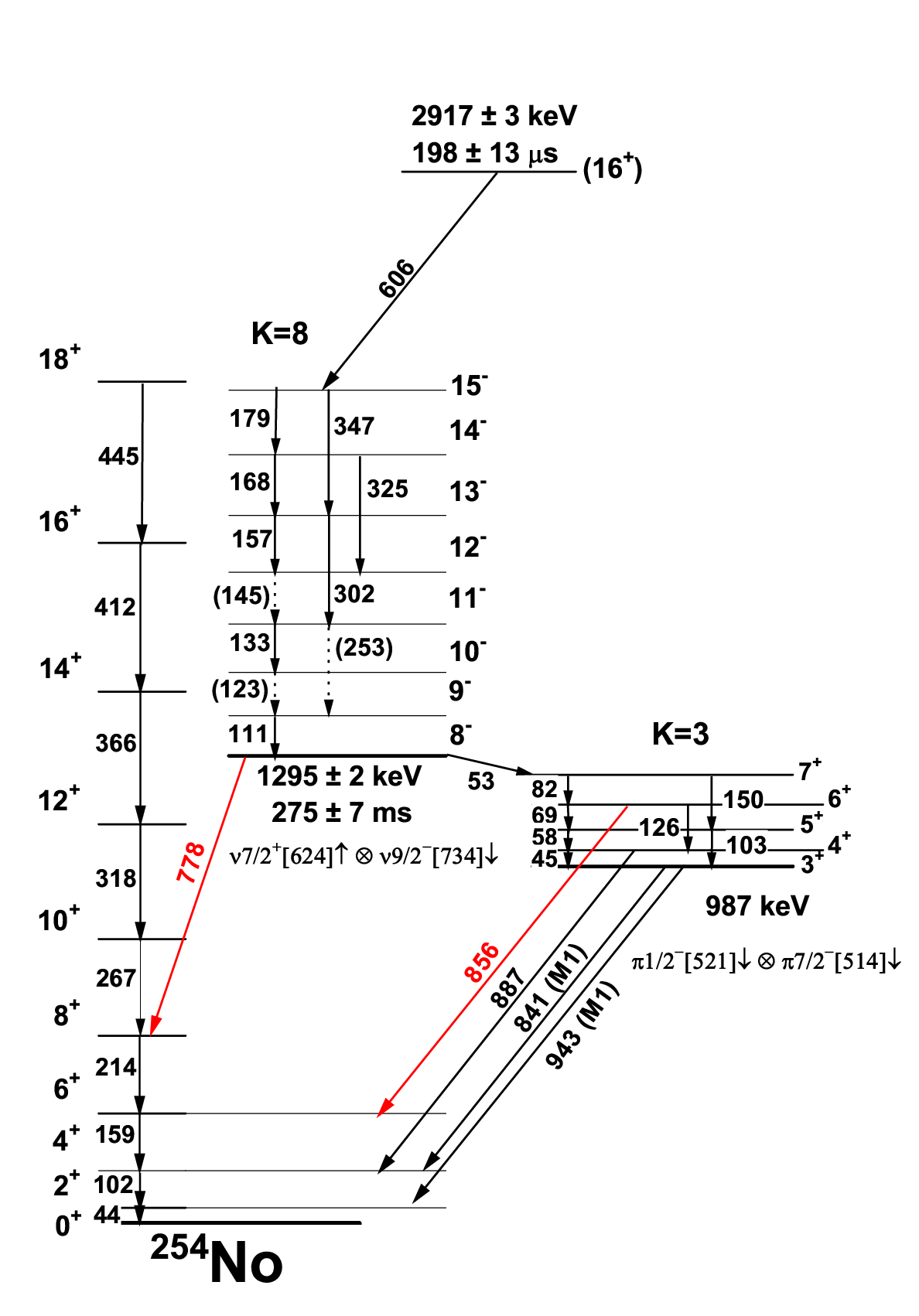}
	    \includegraphics{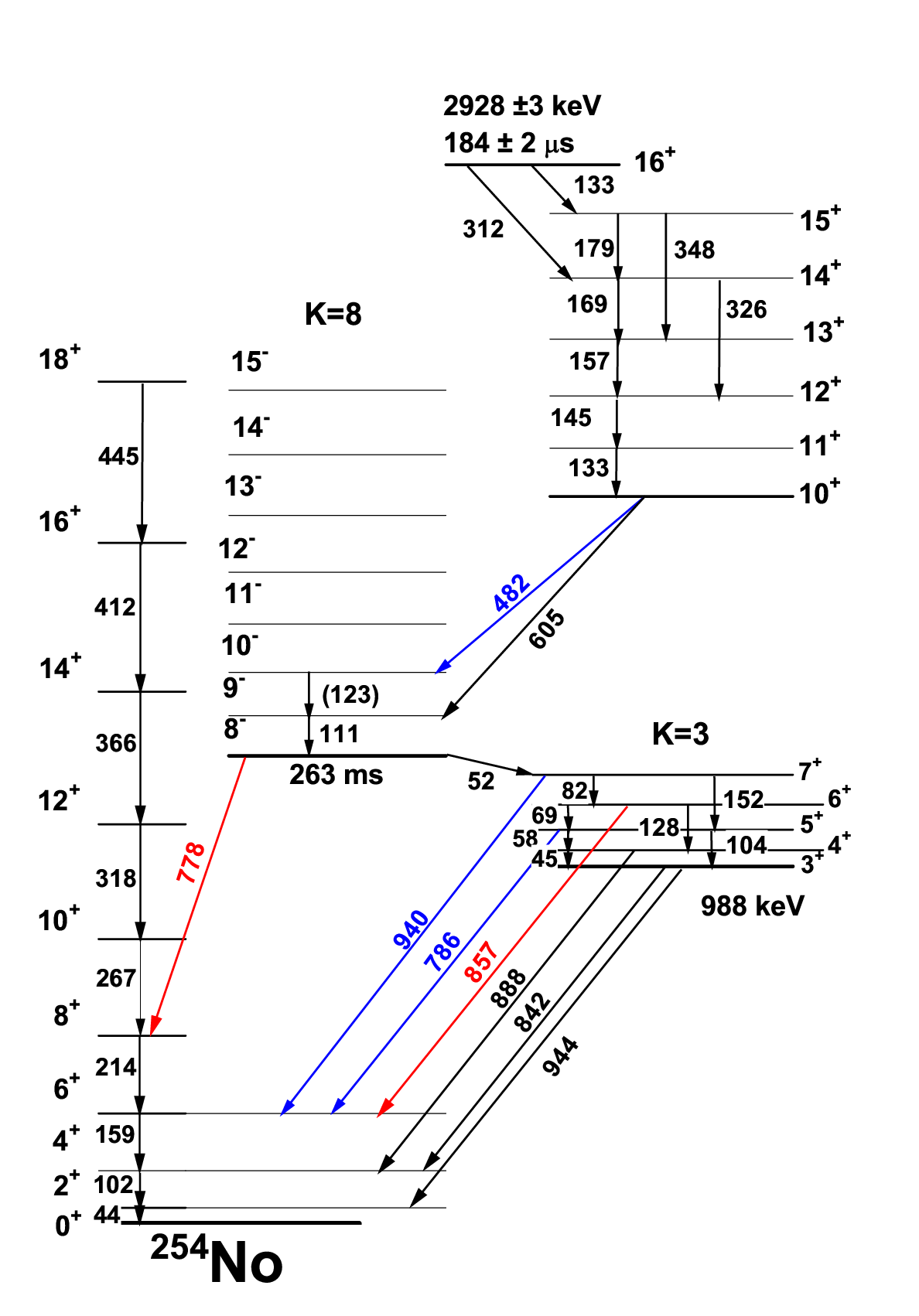}
	} 
	\caption{Left hand side: Decay scheme of $^{254m1,m2}$No as suggested by F.P. He\ss berger et al. (modified version from \cite{Hes10},
		see text for details); right hand side: Decay scheme of $^{254m1,m2}$No as suggested by R.M. Clark et al. \cite{Clark10}.
	}
	\label{fig:6}       
\end{figure*}

\begin{figure*}
	\resizebox{0.99\textwidth}{!}{
		\includegraphics{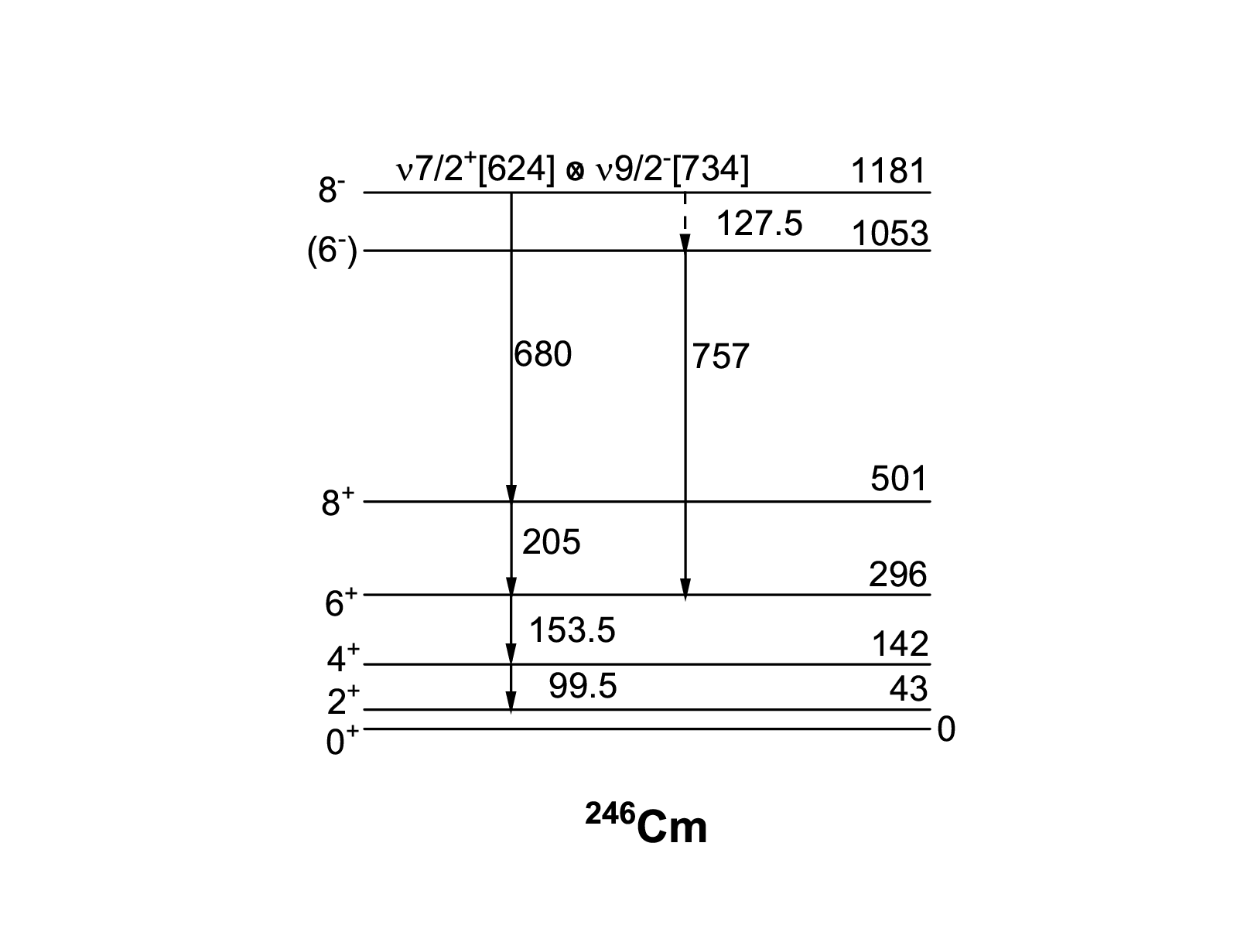}
    	\includegraphics{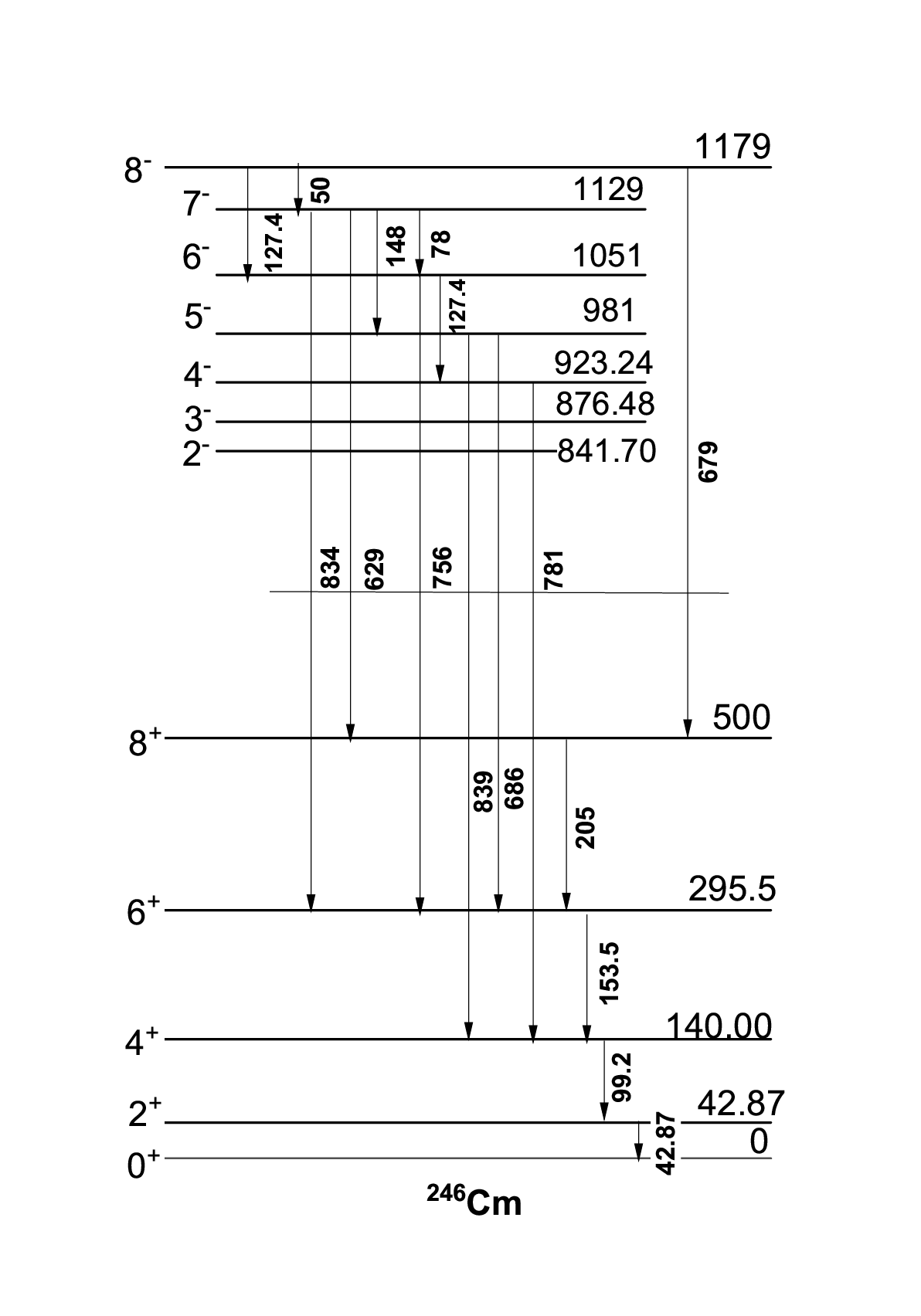}
	} 
	\caption{Left hand side: decay scheme of $^{246m}$Cm as suggested by P.R. Fields et al. \cite{Fields68};
		right hand side: decay scheme of $^{246m}$Cm as suggested by L.G. Multhauf et al. \cite{Mult71}.
	}
	\label{fig:7}       
\end{figure*}

\begin{figure*}
	\resizebox{0.99\textwidth}{!}{
		\includegraphics{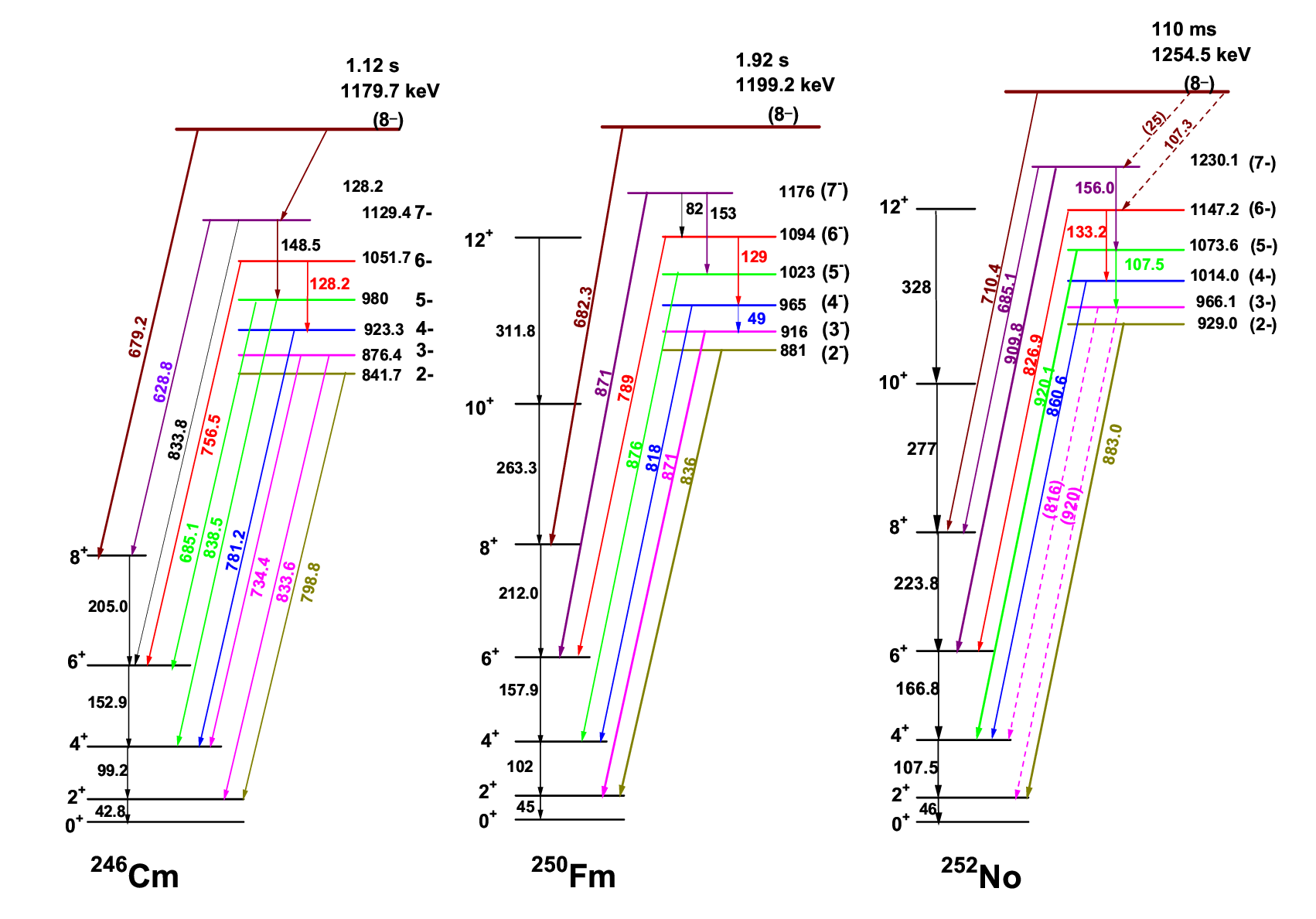}
	} 
	\caption{ Decay schemes of the N\,=\,150 isotones; a)  $^{246m}$Cm \cite{Rob08}; b) $^{250m}$Fm \cite{Green08}; c) $^{252m}$No \cite{Sul07}.
	}
	\label{fig:8}       
\end{figure*}

\begin{table}
	\caption{$\gamma$ transition in the decay of N\,=\,150 isotones}
	\label{tab:1}       
	\begin{tabular}{llll}
		\hline\noalign{\smallskip}
		\noalign{\smallskip}\hline\noalign{\smallskip}
		Transition & E$_{\gamma}$/keV ($^{246}$Cm) \cite{Rob08} & E$_{\gamma}$/keV ($^{250}$Fm \cite{Green08}) & E$_{\gamma}$/keV ($^{252}$No) \cite{Sul07}\\
		\hline\noalign{\smallskip}     
		8$^{-}$  $\rightarrow$ 8$^{+}$ &  679.2   & 682.3 &  710.4 \\
		8$^{-}$  $\rightarrow$ 7$^{-}$ &  128.2   &  (23)  &  (25) \\
		8$^{-}$  $\rightarrow$ 6$^{-}$ &   -   &  -  &  107.3 \\
		\hline\noalign{\smallskip}     
		7$^{-}$  $\rightarrow$ 8$^{+}$ &  628.8   &  -  &  685.1 \\
		7$^{-}$  $\rightarrow$ 6$^{+}$ &  833.8   &  871  &  909.8 \\
		6$^{-}$  $\rightarrow$ 6$^{+}$ &  756.5   &  789  &  826.9 \\
		5$^{-}$  $\rightarrow$ 6$^{+}$ &  685.1   &  -  &  - \\
		5$^{-}$  $\rightarrow$ 4$^{+}$ &  838.5   &  876  &  920.1 \\
		4$^{-}$  $\rightarrow$ 4$^{+}$ &  781.2   &  818 &  860.6 \\
		3$^{-}$  $\rightarrow$ 4$^{+}$ &  734.4   &  -  &  (816) \\
		3$^{-}$  $\rightarrow$ 2$^{+}$ &  833.6   &  871  &  (920) \\
		2$^{-}$  $\rightarrow$ 2$^{+}$ &  798.8   &  836  &  883.0 \\
		\hline\noalign{\smallskip} 
		7$^{-}$  $\rightarrow$ 5$^{-}$ &  148.5   &  153  &  156.0 \\
		7$^{-}$  $\rightarrow$ 6$^{-}$ &   -   &  82  &  - \\
		6$^{-}$  $\rightarrow$ 4$^{-}$ &  128.2   &  129  &  133.2 \\
		5$^{-}$  $\rightarrow$ 3$^{-}$ &   -   &  -   &  107.5 \\
		4$^{-}$  $\rightarrow$ 3$^{-}$ &   -   &  49  &  - \\
		\hline\noalign{\smallskip} 
		
	\end{tabular}
	
\end{table}

\subsection{{\bf 6.3 K isomers in  N\,=\,150 isotones $^{244}$Pu, $^{246}$Cm, $^{248}$Cf,  $^{250}$Fm, $^{252}$No, and $^{254}$Rf}}

{\bf K isomer in $^{244}$Pu}\\
The observation of a K isomer in $^{244}$Pu was reported by S.S. Hota et al. \cite{Hota16}. It was settled at E$^{*}$ = 1216 keV. It was interpreted
to decay with comparable intensities (see fig. 1 in \cite{Hota16}, but no values are given) either into the 8$^{+}$ state of the ground-state rotational band or
into the 7$^{-}$ state of an octupole band. In analogy to the heavier N\,=\,150 isotones ($^{246}$Cm \cite{Fields68}, $^{250}$Fm \cite{Green08},
$^{252}$No \cite{Sul07}) they assigned a  K$^{\pi}$ = 8$^{-}$ state
(configuration $\nu$7/2$^{+}$[624]$\downarrow$ $\otimes$ $\nu$9/2$^{-}$[734]$\uparrow$). A half-life of T$_{1/2}$ = 1.75$\pm$0.12 s was reported.\\

{\bf K isomer in $^{246}$Cm}\\
First identification of a K isomeric state in $^{246}$Cm was reported by P.R. Fields et al. \cite{Fields68}, which was populated
by $\beta^{-}$ - decay of $^{246}$Am (T$_{1/2}$\,=\,39 min). Based on calculations of C.J. Gallagher and V.G. Soloviev \cite{GalS62}
it was assigned as a 2-neutron-quasiparticle state with spin and parity K$^{\pi}$ = 8$^{-}$ and the configuration
$\nu$7/2$^{+}$[624]$\downarrow$ $\otimes$ $\nu$9/2$^{-}$[734]$\uparrow$. An excitation energy of E$^{*}$ = 1181 keV was given, but no half-life
was reported. The decay scheme published in \cite{Fields68} is shown in fig. 7 left side).
Evidently the decay of the K isomer is connected with two high energy $\gamma$ transitions; the stronger line of E\,=\,680 keV (I$_{rel}$\,=\,1)
was assigned as the
decay of the K$^{\pi}$\,=\,8$^{-}$ isomer into the 8$^{+}$ state of the ground-state rotational band, the weaker line 
E\,=\,757 keV (I$_{rel}$\,=\,0.25$\pm$0.02) was attributed to the decay of a tentative 6$^{-}$ level, populated by an assumed 127.5 keV transition 
8$^{-}$ $\rightarrow$ 6$^{-}$, into the 6$^{+}$ level of groud-state rotational band.\\
A more detailed study on the decay of the K$^{\pi}$\,=\,8$^{-}$ state was shortly after performed by L.G. Multhauf et al. \cite{Mult71}. They identified 
the 6$^{-}$ state decaying by the 756 keV transition as a member of an octupole band built up on a K$^{\pi}$\,=\,2$^{-}$ state at E$^{*}$\,=\,841.70 keV
on the basis of theoretical predictions and earlier suggestions \cite{Solov64,Neer70,Steph65}. The decay scheme presented by L.G. Multhauf et al. is shown in 
fig. 7 (right hand side). This interpretation was later supported 
by studies of S.W. Yates et al. \cite{Yates75}.\\
Further studies were performed by A.P. Robinson et al. \cite{Rob08} and U. Shirwadkar et al. \cite{Shir19} at the Argonne National Laboratory,
Argonne (USA). Robinson et al. also populated the isomeric state by $\beta^{-}$ - decay of $^{246}$Am (T$_{1/2}$\,=\,39 min), but obtained
a more detailed decay pattern, settled the isomer at an excitation energy of E$^{*}$\,=\,1179.7 keV,
but also did not report a half-life. U. Shirwadkar et al. produced the isomer by a 2n- transfer reaction in 
bombardments of $^{248}$Cm targets with a beam of $^{209}$Bi. They confirmed the decay pattern of \cite{Rob08} and in addition measured a half-life 
of T$_{1/1}$ = 1.12\,$\pm$\,0.24 s.\\

{\bf K isomer in $^{248}$Cf}\\
The situation is less clear in $^{248}$Cf. Using the $^{249}$Cf(d,t)$^{248}$Cf reaction K. Katori et al. \cite{Katori08} observed an 
excited level at E$^{*}$ = 1261 keV which they (tentatively) assigned as a K$^{\pi}$ = 8$^{-}$ state having a configuration 
$\nu$7/2$^{+}$[624]$\downarrow$ $\otimes$ $\nu$9/2$^{-}$[734]$\uparrow$. 
However, no half-life was measured and also no decay scheme was presented, as their experiment was not laid out for $\gamma$-ray measurements.\\
Recently the nuclear structure of $^{248}$Cf was investigated via in-beam spectroscopy at the Tokai Tandem Accelerator Laboratory of the Japanese Atomic Energy Agency (JAEA) at Tokai, Japan by R. Orlandi et al. \cite{Orlandi22}. Besides measuring the life-time of the already known I$^{\pi}$ = 2$^{-}$ collective state at
E$^{*}$ = 592 keV, two new isomeric states were identified. The higher lying on, at E$^{*}$ = 950 $\pm$ 300 keV has a half-life of T$_{1/2}$ = 11.6 $\pm$ 0.3 ns.
It feeds an isomeric state at E$^{*}$ = 900 $\pm$ keV with a half-life of T$_{1/2}$ $>$ 140 ns by an E\,=\,48 keV transition. As candidates for the
latter a 2-proton-quasiparticle configuration K$^{\pi}$ = 5$^{-}$ or a 2-neutron-quasiparticle configuration K$^{\pi}$ = 8$^{-}$,
forming a K isomeric state in $^{246}$Cm, $^{250}$Fm, and $^{252}$No, were considered. Since due to the applied experimental technique 
(in-beam spectroscopy) neither the half-life nor the decay could be measured, no further conclusions could be done.\\

{\bf K isomer in $^{250}$Fm}\\
A K isomer in $^{250}$Fm was first identified by A. Ghiorso et al. \cite{GhE73}. They reported a half-life of T$_{1/2}$\,=\,1.8$\pm$0.1 s, 
but due to the experimental technique used in \cite{GhE73} neither the excitation energy could be measured nor a decay scheme could be
established.\\
A short time after discovery of a K isomer in $^{252}$No results from a decay study of $^{250}$Fm based on $\gamma$ - ray spectroscopy were
reported by P. Greenlees et al. \cite{Green08}. The experiment was performed at the K130 cyclotron at the University of Jyv\"askyl\"a. 
The isotope was produced in the reaction $^{204}$Hg($^{48}$Ca,2n)$^{250}$Fm and separated from the primary beam by the gas-filled separator
RITU. The decay was measured using the GREAT spectrometer, installed in the focal plane of RITU. \\
They measured a half-life of T$_{1/2}$= 1.92$\pm$0.05 s, confirming the results of A. Ghiorso et al.  \cite{GhE73} and settled the excitation energy
at E$^{*}$ = 1199.2 keV. The decay scheme showed strong similarities with those of $^{246m}$Cm \cite{Mult71,Rob08,Shir19} and $^{252m}$No \cite{Sul07}.
It is shown in fig. 8, $\gamma$ energies and assigned transitins are given
in table 5. The isomer was settled as in the cases of $^{246m}$Cm and $^{252m}$No as 
a 2-neutron-quasiparticle configuration $\nu$7/2$^{+}$[624]$\downarrow$ $\otimes$ $\nu$9/2$^{-}$[734]$\uparrow$ 
resulting in a K$^{\pi}$ = 8$^{-}$ state. This assignment was supported by comparison of experimental and theoretical ratios of reduced transition
probabilities B(M$_{1}$)/ B(E$_{2}$) for transitions from states of initial spins and parities I$^{\pi}$ = 14$^{-}$, 15$^{-}$, 16$^{-}$ within
the band built up on the isomer. The experimental values (B(M$_{1}$)/ B(E$_{2}$))$_{exp}$ = 0.2\,-\,0.3 rather fit to those expected for a 
2-neutron-quasiparticle configuration, (B(M$_{1}$)/ B(E$_{2}$))$_{theo}$ = 0.32\,-\,0.38 than for a possible
2-proton-quasiparticle configuration $\pi$9/2$^{+}$[624]$\uparrow$ $\otimes$ $\pi$7/2$^{-}$[514]$\downarrow$,  
(B(M$_{1}$)/ B(E$_{2}$)$_{theo}$) = 0.67\,-\,0.77 \cite{Green08}.\\
Within the study of $^{250m}$Fm a second isomeric state with a half-life of T$_{1/2}$ = 8 $\pm$ 2 $\mu$s ($^{250m2}$Fm) was observed 
by ER - CE1 - CE2 - $\alpha$ correlations \cite{Ketel10}. Thus it can be assumed that decay of the short-lived isomer (at least partly) populates
the long-lived (1.92 s) isomer. A population probability of 6$\pm$3 $\%$ was estimated. From the energy distribution of the coincident CE 
its excitation energy can be assumed to be at least 350 keV above that of the 1.92 s isomer, i.e. E$^{*}$ $\ge$ 1530 keV. 
No information about the decay path or the configuration is given in \cite{Ketel10}.\\

{\bf K isomer in $^{252}$No}\\
A K isomeric state in $^{252}$No was first reported in \cite{Hes06}. Due to contaminations from decay of nuclei produced in a 
preceeding irradiation the $\gamma$ lines at E$^{*}$ $<$ 150 keV could not be unambiguously identified. So, a follow-up
experiment was performed at SHIP to obtain more precise data \cite{Sul07}. The excitation energy was estimated as E$^{*}$ = 1254 keV.
Based on the interpretation of the decay of $^{246m}$Cm by L.G. Multhauf et al. \cite{Mult71} the decay of the isomer was suggested to
occur essentially via a band built up on a K$^{\pi}$ = 2$^{-}$ state at E$^{*}$ = 929 keV, while the ground-state rotational band is only 
weakly populated by an 8$^{-}$ $\rightarrow$ 8$^{+}$ transition of E = 710.4 keV, whereas the corresponding $\gamma$ transitions in $^{246}$Cm (679.2 keV)
\cite{Mult71} and $^{250}$Fm (682.3 keV) \cite{Green08} appear much stronger (please note: only in \cite{Mult71} a relative intensity is given).
The decay scheme presented by B. Sulignano et al. \cite{Sul07} is shown in  fig. 8.\\
The results of \cite{Sul07} were short time later confirmed by A.P. Robinson et al. \cite{Rob08}.
Further information on the structure of the isomeric state was obtained in an in-beam study performed at the RITU - separator at the
University of Jyv\"askyl\"a, Finland, where the rotional band built up on the K$^{\pi}$ = 8$^{-}$ state up to I$^{\pi}$ = 22$^{-}$ was 
observed \cite{Sul12}. Based on the relationship between the gyromagnetic factor g$_{k}$ and the M1 / E2 intensity ratio
I(M1)/I(E2) it was shown that the latter (precisely: the range of the experimental values) was rather compatible
with the value g$_{k}$ = 0.01 expected for the already previously assumed 2-neutron-quasiparticle 
configuration $\nu$7/2$^{+}$[624]$\downarrow$ $\otimes$ $\nu$9/2$^{-}$[734]$\uparrow$
than with the value g$_{k}$ = 1.01 expected for the 2-proton-quasiparticle configuration 
$\pi$9/2$^{+}$[624]$\uparrow$ $\otimes$ $\pi$7/2$^{-}$[514]$\downarrow$ leading also to a  K$^{\pi}$ = 8$^{-}$ state \cite{Sul12}.\\
 
{\bf K isomer in $^{254}$Rf}\\
First indication of a K isomer in the next heavier N\,=\,150 isotone was found in an experiment performed at the FMA at ANL, Argonne, USA, where in four of 28 cases 
low energy signals were registered inbetween the implantation of the $^{254}$Rf residue produced in the reaction $^{206}$Pb($^{50}$Ti,2n)$^{254}$Rf into a double-sided silicon strip detector and the spontaneous fission of the ground-state \cite{David15} (T$_{1/2}$ = 23$\pm$3 $\mu$s \cite{Hess97}).
These signals were assigned to conversion electrons emitted during the decay of an isomeric state.
A more detailed study was performed in a follow-up experiment at the BGS at LBNL, Berkeley, USA.
Isomeric states were searched for by analyzing ER - CE - SF correlations. Two activities of T$_{1/2}$ = 4.7$\pm$1.1 $\mu$s and T$_{1/2}$ = 247$\pm$73 $\mu$s
were observed. In seven cases correlations ER - CE1 - CE2 - SF were registered and interpreted as events from feeding the short-lived isomer by decay of the long-lived
one \cite{David15}. Also a couple of $\gamma$ events were observed in prompt coincidence with CE assigned to the decay of the short-lived isomer:
E$_{\gamma}$ = 893 keV (5 events), E$_{\gamma}$ = 853 keV (3 events), and E$_{\gamma}$ = 829 keV (5 events). As E$_{\gamma}$ = 893 keV is similar to 
the energy of the 7$^{-}$ $\rightarrow$ 6$^{+}$ transition in the lighter N\,=\,150 isotones $^{250}$Fm (871 keV) and $^{252}$No (909 keV) one may speculate that
this line represents also the  7$^{-}$ $\rightarrow$ 6$^{+}$ transition in $^{254}$Rf. But certainly the data presented in \cite{David15} 
are too scarce to construct a decay scheme. Nevertheless the authors present possible configurations on the basis of calculations.
The short-lived isomer is attributed to 2-neutron-quasiparticle 
configuration $\nu$7/2$^{+}$[624]$\downarrow$ $\otimes$ $\nu$9/2$^{-}$[734]$\uparrow$ leading to a K$^{\pi}$ = 8$^{-}$ state as also in the
lighter isotones, the long-lived isomer is attributed to a 4-quasiparticle configuration 
$\nu^{2}$ (7/2$^{+}$[624],9/2$^{-}$[734]) $\otimes$ $\pi^{2}$ (7/2$^{-}$[514],9/2$^{+}$[624]) leading to a  K$^{\pi}$ = 16$^{+}$ state \cite{David15}.\\
The most striking feature, however, was the steep decrease of the half-lives. While those of $^{246}$Cm (1.12 s) and $^{250}$Fm (1.92 s) are comparable
a decrease by a factor od $\approx$17 is observed from $^{250}$Fm to $^{252}$No (111 ms) and even by a factor of $\approx$24000 from $^{252}$No to 
$^{254}$Rf (4.7 $\mu$s). As possible reasons for that behavior the authors give a decrease of the hindrance of the M1 decay branch 
from the K$^{\pi}$ = 8$^{-}$
isomer into the I$^{\pi}$ = 7$^{-}$ state of the octupole band and/or that the K$^{\pi}$ = 8$^{-}$ isomer and the I$^{\pi}$ = 8$^{-}$ state of the octupole band 
might be very close in energy leading to an accidential configuration mixing resultig in a shorter half-life.\\

\vspace*{1cm}
\subsection{{\bf 6.4 K isomers in  even-even nuclei with Z$\ge$98}}
Besides the cases discussed above search for K isomers was performed in most of the experimentally acessible nuclei in the region 
Z$\ge$100. In all cases the isomers essentially were identified via measuring the conversion electrons, in general only little numbers of $\gamma$ rays were 
observed. The collected data were in most cases not sufficient for establishing well based configurations and  decay schemes, however, in some cases (very 
speculative) partial decay schemes and configurations were presented.\\

{\bf K isomer in $^{248}$Cm}\\
The observation of a K isomer in $^{248}$Cm was reported by U. Shirwadkar et al. \cite{Shir19}. They gave a half-life of T$_{1/2}$ = 146$\pm$18 $\mu$s, the
excitation energy was settled at E$^{*}$ = 1461 keV. The isomer was assigned as a K$^{\pi}$ = 8$^{-}$ state, as possible configurations 
2-quasi-neutron states ($\nu$7/2$^{+}$[613]$\uparrow$$\otimes$$\nu$9/2$^{-}$[734]$\uparrow$ or 
$\nu$7/2$^{+}$[624]$\downarrow$$\otimes$$\pi$9/2$^{-}$[734]$\uparrow$) were considered.
The decay occured with similar intensities either directly into the 8$^{+}$ state of the ground-state rotational band (E$_{\gamma}$ = 954 keV) or by
an unobserved 7 keV-transition into the 8$^{+}$ - state of a $\gamma$ vibrational band (bandhead 2$^{+}$ at E$^{*}$ = 1048 keV). The further decay occured
from the 8$^{+}$ and 6$^{+}$ states of the $\gamma$ vibrational band into the corresponding level of the ground-state rotational band, i.e.
8$^{+}$ $\rightarrow$ 8$^{+}$ (E$_{\gamma}$ = 947 keV) or 6$^{+}$ $\rightarrow$ 6$^{+}$ (E$_{\gamma}$ = 985 keV). \\

{\bf Search for a K isomer in $^{246}$Fm}\\
A K isomer was searched for in $^{246}$Fm at SHIP, GSI using the production reaction $^{208}$Pb($^{40}$Ar,2n)$^{246}$Fm \cite{Venhart11}. Although a quite 
high number of $^{246}$Fm nuclei were produced, about 31000 $\alpha$ decays were observed, no signature for a K isomer in the half-life range between some
microseconds and about 100 milliseconds was found applying the method of ER - (CE,$\gamma$) - $\alpha$($^{246}$Fm) correlation.\\

{\bf K isomer in $^{248}$Fm}\\
An isomeric state in $^{248}$Fm was observed at the RITU - separator, University of Jyväskylä, using the production reaction $^{202}$Hg($^{48}$Ca,2n)$^{248}$Fm. It was identified via ER - CE - $\alpha$($^{248}$Fm) correlations \cite{Ketel10}. A half-life of T$_{1/2}$\,=10.1$\pm$0.6 ms was measured. Also a few $\gamma$ events, forming 
$\gamma$ lines of E = 808 keV and E = 904 keV were registered. These data, however, were not sufficient to establish a decay pattern.\\

{\bf K isomer in $^{250}$No}\\
A spontaneous fission activity of T$_{1/2}$ = 36$^{+11}_{-6}$ $\mu$s was observed by Yu.Ts. Oganessian et al. \cite{OgU01} in bombardments of $^{204,206}$Pb with
$^{48}$Ca and attributed to the decay of $^{250}$No. Further investigations of A.V. Belozerov et al. \cite{Beloz03} resulted in a splitting of the SF activity into 
two components with half-lives of T$_{1/2}$ = 5.6$^{+0.9}_{-0.7}$ $\mu$s and T$_{1/2}$ = 54$^{+14}_{-9}$ $\mu$s. Tentatively the shorter lived activity was assigned to
$^{250}$No, the longer lived one to $^{249}$No. However, it was not excluded, that both components could also be assigned to the same isotope, i.e. the decay of the ground-state 
and of an isomeric state. The latter assumption was proven in an experiment performed at the FMA at ANL, Argonne by D. Peterson et al. \cite{Peter06}. The authors
unambiguously could assign both activities to the same mass number A\,=\,250. They (tentatively) assigned the shorter lived component (T$_{1/2}$ = 3.7$^{+1.1}_{-0.8}$ $\mu$s) 
to the ground-state and the longer lived one (T$_{1/2}$ = 43$^{+22}_{-15}$ $\mu$s) to the isomer, for which they assumed a 2-neutron - quasiparticle 
configuration $\nu$5/2$^{+}$[622]$\uparrow$ $\otimes$ $\nu$7/2$^{+}$[624]$\downarrow$ leading to a K$^{\pi}$ = 6$^{+}$ state. It was,
however, not possible to decide if the isomeric state undergoes SF or decays by internal transitions into the ground state, and the measured half-life of the SF activity
is that of the ground-state delayed by the decay of the isomer.\\
To clarify the assigment and to search for a fission branch two more experiments were performed, one at the RITU - separator at the University of Jyv\"askyl\"a
\cite{Kallun20}, and one at the gasfilled-separator TASCA at GSI, Darmstadt \cite{Khuyag22}. In both experiments digital electronics was used for 
identifying CE, as a signature for internal transitions, in between implantation of the ER and the fission event. In both experiments it was clearly shown, 
that the longer lived isomeric state decays into the shorter lived as already assumed in \cite{Peter06}. Another feature of both experiments was search for direct fission of the isomeric state. In \cite{Kallun20} a population probability of 0.41$\pm$0.13, a relatively large upper limit of $\approx$0.5 for the fission branch, and an upper limit of 0.044 (including the data of \cite{Beloz03} 0.029) for the $\alpha$ branch were obtained.\\
In \cite{Khuyag22} also no unambiguous indication of SF of $^{250m}$No was observed and a considerably lower upper limit for SF of b$_{sf}$ $\le$ 0.035 was reported.
Also a lower population probability of 0.17$\pm$0.03 was obtained; in addition in that study indication of a so far unknown higher-lying isomeric state ($^{250m2}$No) with a half-life 
of T$_{1/2}$\,=\,0.7$^{+1.4}_{-0.3}$ $\mu$s on the basis of two registered ER - CE(1) - CE(2) - SF correlation was obtained, which decays via the known isomeric state
(now denoted as $^{250m1}$No) into the ground-state \cite{Khuyag22}.\\
In a recent study at the SHELS separator at FLNR JINR, Dubna (Russia) more than an order of magnitude higher number
of SF of $^{250}$No was registered \cite{KuzY20}. Also four $\gamma$ lines of E = 115, 176, 914, 1090 keV were observed.
Due to energy values similar to the 4$^{+}$ $\rightarrow$ 2$^{+}$ and the 6$^{+}$ $\rightarrow$ 4$^{+}$ transition 
energies within the ground-state rotational band of $^{252}$No (108, 167 keV) and $^{254}$No (101, 159 keV) \cite{HerG08},
the low energy $\gamma$ lines were (tentatively) attributed to transitions within the ground state rotational band of $^{250}$No.
No further discussion on the decay path, nor on SF of $^{250m1}$No are presented in \cite{KuzY20}.
It was, however, pointed to the result that the energy difference $\Delta$E\,=\,(1090-914) keV\,=\,176 keV \cite{KuzY20}, i.e. is the same as
for the candidate for the  6$^{+}$ $\rightarrow$ 4$^{+}$ transition. So it seems possible that the high energy $\gamma$ lines stem from the decay 
of the same level, populating the 6$^{+}$ and 4$^{+}$ of the ground state rotational band. Possible candidates for such an emitting level
could be an I$^{\pi}$\,=\,5$^{+}$ (M1 transitions 5$^{+}$ $\rightarrow$ 6$^{+}$, 5$^{+}$ $\rightarrow$ 4$^{+}$) or an
I$^{\pi}$\,=\,5$^{-}$ (E1 transitions 5$^{-}$ $\rightarrow$ 6$^{+}$, 5$^{-}$ $\rightarrow$ 4$^{+}$) state.\\

{\bf K isomer in $^{256}$No}\\
An isomeric state in $^{256}$No was identified in a study performed at SHELS separator at the FLNR-JINR Dubna, Russia \cite{Kess21}. The isotope was produced in the
reaction $^{238}$U($^{22}$Ne,4n)$^{256}$No. Fifteen correlations of the type ER - CE - $\alpha$($^{256}$No) were observed and assigned to the decay of an isomeric state. 
Half-lives of T$_{1/2}$\,=\,7.8$^{+8.3}_{-2.6}$\,$\mu$s or T$_{1/2}$\,=\,10.9$^{+21.7}_{-4.3}$\,$\mu$s, depending on the data selected for the half-life evaluation were
obtained. Thirteen photon events were observed in prompt coincidence with CE, five of them in the range of K or L X-rays of nobelium. On the basis of the energy sums 
of CE and photons a lower limit of 1089 keV for the excitation energy of the isomer was given. Only speculated could be about the struchture (spin, parity, configuration)
of the isomeric state, just a probable role of the 2-quasi-neutron states I$^{\pi}$\,=\,5$^{-}$ (configuration 
$\nu$11/2$^{-}$[725]$\uparrow$ $\otimes$ $\nu$1/2$^{+}$[620]$\uparrow$) and I$^{\pi}$\,=\,7$^{-}$ (configuration 
$\nu$11/2$^{-}$[725]$\uparrow$ $\otimes$ $\nu$3/2$^{+}$[622]$\downarrow$) were emphasized.\\

{\bf K isomers in $^{256}$Rf}\\
Several experiments to search for K isomeric states in $^{256}$Rf have been performed at different laboratories so far. The first report on observation of
K isomeric states came from an experiment performed at the BGS separator at LNBL Berkeley (USA) by H.B. Jeppesen et al. \cite{Jepp09}. The isotope was produced in the reaction
$^{208}$Pb($^{50}$Ti,2n)$^{256}$Rf. On the basis of observed ER - CE1 - CE2 - CE3 - SF($^{256}$Rf) correlations the existence of three isomeric states was 
concluded. Also a $\gamma$ line of E\,=\,900$\pm$1 keV was registered in correlations ER - (CE,$\gamma$) - SF($^{256}$Rf). It was attributed to the decay of the
head (K$^{\pi}$ = 2$^{-}$) of an octupole vibrational band into the I$^{\pi}$ = 2$^{+}$ level of the ground - state rotational band  \cite{Jepp09}. Estimated excitation energies,
half-lives and relative population intensities are given in table 6. The data were principally confirmed as by-products of studies primarily devoted to the 
investigation of $^{257}$Rf by J.S. Berryman et al. \cite{Berry10} and J. Rissanen et al. \cite{Riss13} also performed at LNBL Berkeley. The results
of \cite{Jepp09} were not reproduced in an experiment at the AMS separator at ANL, Argonne (USA), where only one isomeric state with a half-life 
of T$_{1/2}$ = 17$\pm$5 $\mu$s was registered \cite{Robi11} which tentatively was equated with ISO1 of \cite{Jepp09} (see table 6). The Berkeley data could be confirmed
in an experiment at the RITU separator at the University of Jyv\"askyl\"a (Finland) by J. Rubert \cite{Rub13}. As the 900-keV - $\gamma$-line was not
observed in the corresponding in-beam measured it was excluded that it may stem from the decay of a low - spin state, contrary to the tentative
assignment of \cite{Jepp09}. Recently results from two more studies were reported, an investigation at the gas-filled separator TASCA at GSI, Darmstadt (Germany)
by J. Khuyagbaatar et al. \cite{Khuy21a} and at the SHELS separator at FLNR-JINR Dubna (Russia) using the GABRIELA detector system \cite{LopM22}. J. Khuyagbaatar et al. oberved two isomeric 
states, in-line with the data for ISO1 and ISO2 (see table 6). The third isomeric state, however, was not observed, probably due its low population intensity
and the relatively low number of produced $^{256}$Rf nuclei. A considerably higher numbers of $^{256}$Rf decays than in the previous experiments was registered in
the experiment at SHELS \cite{LopM22}. The 900-keV $\gamma$ line is clearly observed, but no further details on obtained results are reported. 
Nevertheless one may expect enhanced information on the structure and decay properties and the structure of the K isomers in $^{256}$Rf when the data
from this study are fully analyzed.\\
The structure of the isomer is thus not unbambiguously established. Comparisons with predicted 2-quasiparticle states in $^{256}$Rf suggest that the 25\,\,$\mu$s
isomer might be a K$^{\pi}$ = 5$^{-}$ state  with the configuration $\pi$ 1/2$^{-}$[521]$\downarrow$ $\otimes$ $\pi$ 9/2$^{+}$[624]$\uparrow$ and the
17\,$\mu$s isomer a 2-quasiparticle state with a configuration $\pi$ 7/2$^{-}$[514]$\downarrow$ $\otimes$ $\pi$ 9/2$^{+}$[624]$\uparrow$ \cite{Robi11,Rub13}.
We will make use of these tentative assignments in the further discussion.

\begin{table}
	\caption{Excitation energies, half-lives and population intensities of K isomers in $^{256}$Rf}
	\label{tab:1}       
	\begin{tabular}{ccccccccccc}
		\hline\noalign{\smallskip}
		\noalign{\smallskip}\hline\noalign{\smallskip}
		& & \cite{Jepp09} & & \cite{Berry10} &  \cite{Robi11} &  &  \cite{Rub13} &  & \cite{Khuy21a} & \\
		\hline\noalign{\smallskip}     
		& E$^{*}$/keV & T$_{1/2}$/$\mu$ & i$_{rel}$ & T$_{1/2}$/$\mu$ &  T$_{1/2}$/$\mu$ & i$_{rel}$ &  T$_{1/2}$/$\mu$ & i$_{rel}$ &  T$_{1/2}$/$\mu$ & i$_{rel}$ \\
		\hline\noalign{\smallskip} 
		ISO1 & $\approx$1120 & 25$\pm$2 & 0.182 & 36.5$\pm$8.6 & 17$\pm$5 & 0.024 & 23$\pm$4 & 0.173 & 14$^{+6}_{-4}$ & 0.25 \\
		ISO2 & $\approx$1400 & 17$\pm$2 & 0.027 & 13.2$\pm$3.3 &  &  & 18$\pm$4 & 0.03 & 10$^{+5}_{-3}$ & 0.08 \\
		ISO3 & $>$2200 & 27$\pm$5 & 0.0013 &  &  &  & 27$\pm$6 & 0.013 &  &  \\
		
	\end{tabular}
	
\end{table}

\vspace{1.0cm}
{\bf K isomer in $^{258}$Rf}\\
Two isomeric states in $^{258}$Rf were identified in the course of an electron capture (EC) decay study of $^{258}$Db \cite{Hess16}, where ER - (CE,photon) - SF ($^{258}$Rf)
and ER - (CE) - SF ($^{258}$Rf) correlations were investigated to measure K - X - ray emission during the CE process to directly prove EC decay of $^{258}$Db and to identify EC - decay from 
both long-lived states in $^{258}$Db, that had been identified by $\alpha$ decay studies \cite{Hessb10,Hess16,Vosti19}. In addition, a couple of  ER\,-\,CE\,-\,CE\,-\,SF,
ER\,-\,CE\,-\,photon\,-\,SF correlations which could be assigned to the decay of two isomeric states in $^{258}$Rf with half-lives of T$_{1/2}$ = 15$\pm$10 $\mu$s and 
T$_{1/2}$ = 2.4$^{+2.4}_{-0,8}$ ms were registered. The data further indicated, that the short-lived isomer, at least to a notable fraction, decays into the long-lived one. The excitation 
enegies of both states remained uncertain.\\
Oberservation of isomeric states in $^{258}$Rf is not unexpected. Calculations of F.R. Xu et al. \cite{XuZ04} predict in $^{258}$Rf two 2-quasi-neutron states of 
I$^{\pi}$\,=\,10$^{-}$ (configuration $\nu$11/2$^{-}$[725]$\uparrow$ $\otimes$ $\nu$9/2$^{+}$[615]$\downarrow$) and 
I$^{\pi}$\,=\,9$^{-}$ (configuration $\nu$11/2$^{-}$[725]$\uparrow$ $\otimes$ $\nu$7/2$^{+}$[613]$\uparrow$) at E$^{*}$ $\approx$ 1.2 MeV and E$^{*}$ $\approx$ 1.1 MeV,
respectively, as well as a 2-quasi-proton state of 
I$^{\pi}$\,=\,7$^{-}$ (configuration $\pi$9/2$^{+}$[624]$\uparrow$ $\otimes$ $\pi$5/2$^{-}$[512]$\uparrow$ at E$^{*}$ $\approx$ 1.4 MeV.\\
As, tentatively, spin and parity of the long-lived state in $^{258}$Db (T$_{1/2}$ = 4.41 s) are assigned as I$^{\pi}$\,=\,5$^{+}$ or I$^{\pi}$\,=\,10$^{-}$
\cite{Vosti19}, population of high-spin states in $^{258}$Rf by EC decay can be expected. \\

{\bf K isomer in $^{266}$Hs}\\
The isotope $^{266}$Hs has been produced as $\alpha$ decay daughter of $^{270}$Ds in two experiments at SHIP
performed in autumn 2000 \cite{HoH01} and autumn 2010 \cite{Ack12,Ack15}. The isotope decays by $\alpha$ emission
with an energy E$_{\alpha}$ = 10.20$\pm$0.03 MeV and a half-life of T$_{1/2}$ = 2.6$^{+0.7}_{-0.5}$ ms
\cite{Hessb22}, evaluation of data from combined data of \cite{HoH01,Ack12}). It also has a significant
SF branch of $\approx$25 $\%$ \cite{Ack12}. In the 2010 - experiment a single $\alpha$ decay of 
E$_{\alpha}$ = 10.440 MeV with a correlation time of 
$\Delta$t($^{266}$Hs - $^{270}$Ds)\,=\,104.66 ms was registered \cite{Ack12}.
As these data are inconsistent with decay data of the ground-state of $^{266}$Hs, it was assigned to the decay of isomeric state
in $^{266}$Hs \cite{Ack12,Ack15}.\\

{\bf K isomer in $^{270}$Ds}\\
The isotope $^{270}$Ds was first time observed in an irradiation of $^{207}$Pb with $^{64}$Ni at SHIP. Energy and time distribution
of the $\alpha$ decays were found to be quite broad, and did not suggest to assign them to a single activity \cite{HoH01}: three decay events had energies of
E$_{\alpha}$ = 10.987 MeV, 11.075 MeV 1.925 MeV\footnote{In this case the $\alpha$ particle 'left' the detector depositing only part of its energy in it,
i.e. only an energy loss signal was registered.} and a life-time of $\tau$\,=\,0.15 ms, and three events had energies E$_{\alpha}$ = 12.147 MeV, 11.151 MeV, 10.954 MeV and a life-time $\tau$\,=\,8.6 ms. The theoretical $\alpha$ decay half-life \cite{PoI80,Rur84} estimated as T$_{1/2}$ = 0.62 ms for the short - lived activity
(for the mean energy 11.03 MeV from the two events registered with full energy), resulted in  a hindrance factor HF = T$_{\alpha}$(exp)/ T$_{\alpha}$(exp) = 1.6, i.e. it represents an unhindered transition \cite{Hessb22} while for the long-lived isomer values of
HF = 14824 (12.147 MeV), HF = 120 (11.151 MeV), HF = 43 (10.954 MeV) were obtained \cite{Hessb22}. This finding suggested to assign the short-lived activity to the ground-state decay,
the long-lived one to the decay of an isomeric state. On the basis of the highest observed $\alpha$-decay energy the excitation energy was settled at
E$^{*}$ $\approx$ 1.25 MeV.\\
Self-consistent Hartree - Fock - Bogoliubov calculations using Skyrme-SLy4 interaction resulted in 2-quasi - neutron states K$^{\pi}$ = 9$^{-}$ (configuration
$\nu$7/2$^{+}$[613]$\downarrow$ $\otimes$ $\nu$11/2$^{-}$[725]$\uparrow$) at E$^{*}$ = 1.31 MeV and 
K$^{\pi}$ = 10$^{-}$ (configuration
$\nu$9/2$^{+}$[615]$\downarrow$ $\otimes$ $\nu$11/2$^{-}$[725]$\uparrow$) at E$^{*}$ = 1.34 MeV. Assignment to the K$^{\pi}$\,=\,9$^{-}$ state was preferred using the argument 
of a lower hindrance for the 9$^{-}$ $\rightarrow$ 0$^{+}$ (ground state) - transition than for 10$^{-}$ $\rightarrow$ 0$^{+}$ (ground state) - transition.\\
A second experiment on investigation of the K isomer $^{270m}$Ds was performed at SHIP in autumn 2010 \cite{Ack12,Ack15}.
Although a factor of $\approx$3 more decays were registered, no new or enhanced informations on the
configuration, decay and decay path of the isomer were reported so far.
Improved half-lives using the data of both SHIP experiments, regarding 
$\alpha$ decays of E$_{\alpha}$ $\le$ 11.0 MeV and $\Delta$($\alpha$-ER) $\le$ 1 ms as decays of $^{270g}$Ds and
$\alpha$ decays of E$_{\alpha}$ $\ge$ 10.9 MeV and $\Delta$($\alpha$-ER) $>$ 1 ms as decays of $^{270m}$Ds
result in improved half-life values of T$_{1/2}$ = 0.17$^{+0.09}_{-0.04}$ ms for $^{270g}$Ds and
T$_{1/2}$ = 3.81$^{+1.39}_{-0.80}$ ms for $^{270m}$Ds \cite{Hessb22}.
\\

\begin{figure*}
	\centering
	\resizebox{0.71\textwidth}{!}{
		\includegraphics{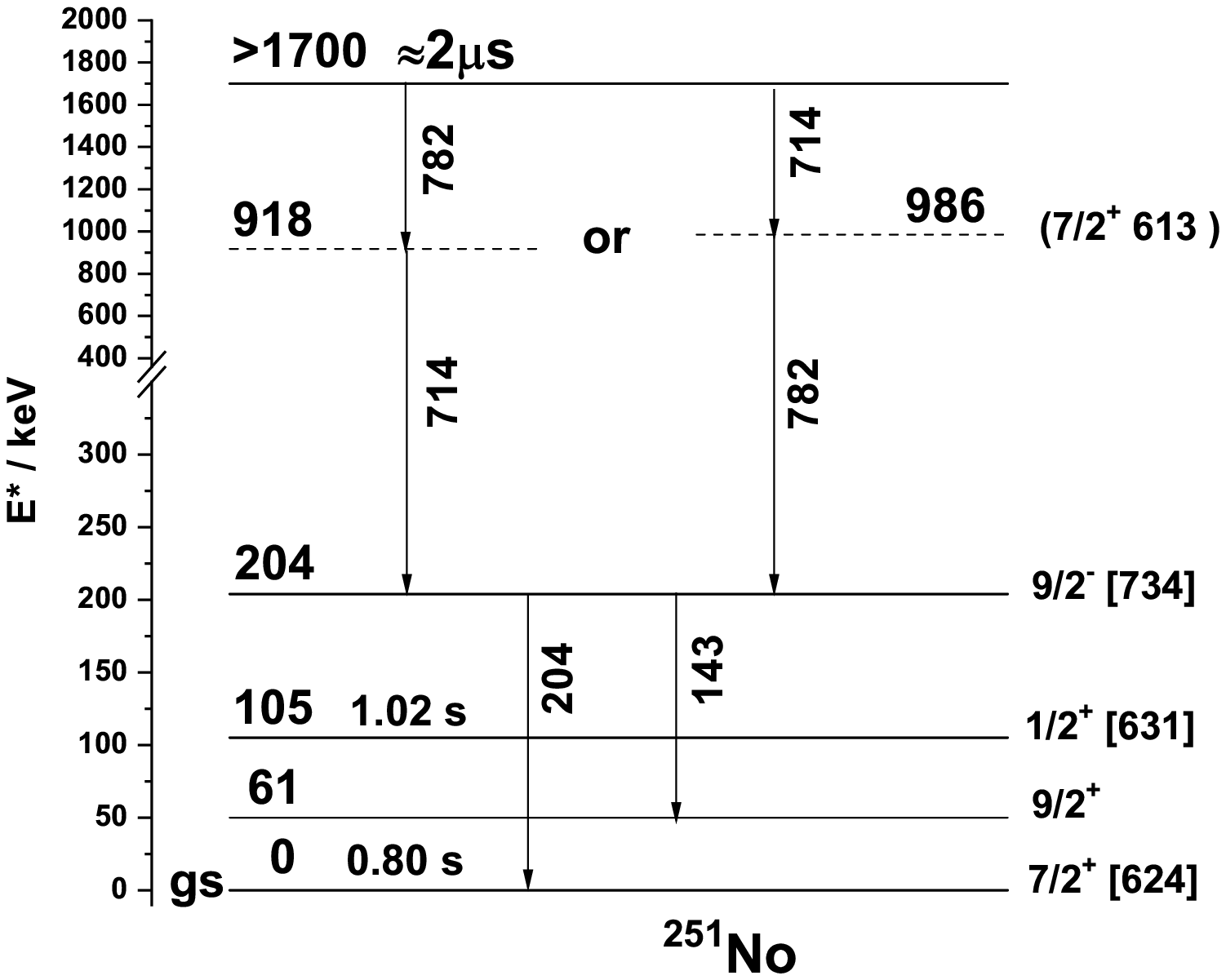}
	} 
	\caption{Partial decay pattern of $^{251m2}$No \cite{Hes06a}.}
	
	\label{fig:9}       
\end{figure*}

\begin{figure*}
	\resizebox{1.25\textwidth}{!}{
		\includegraphics{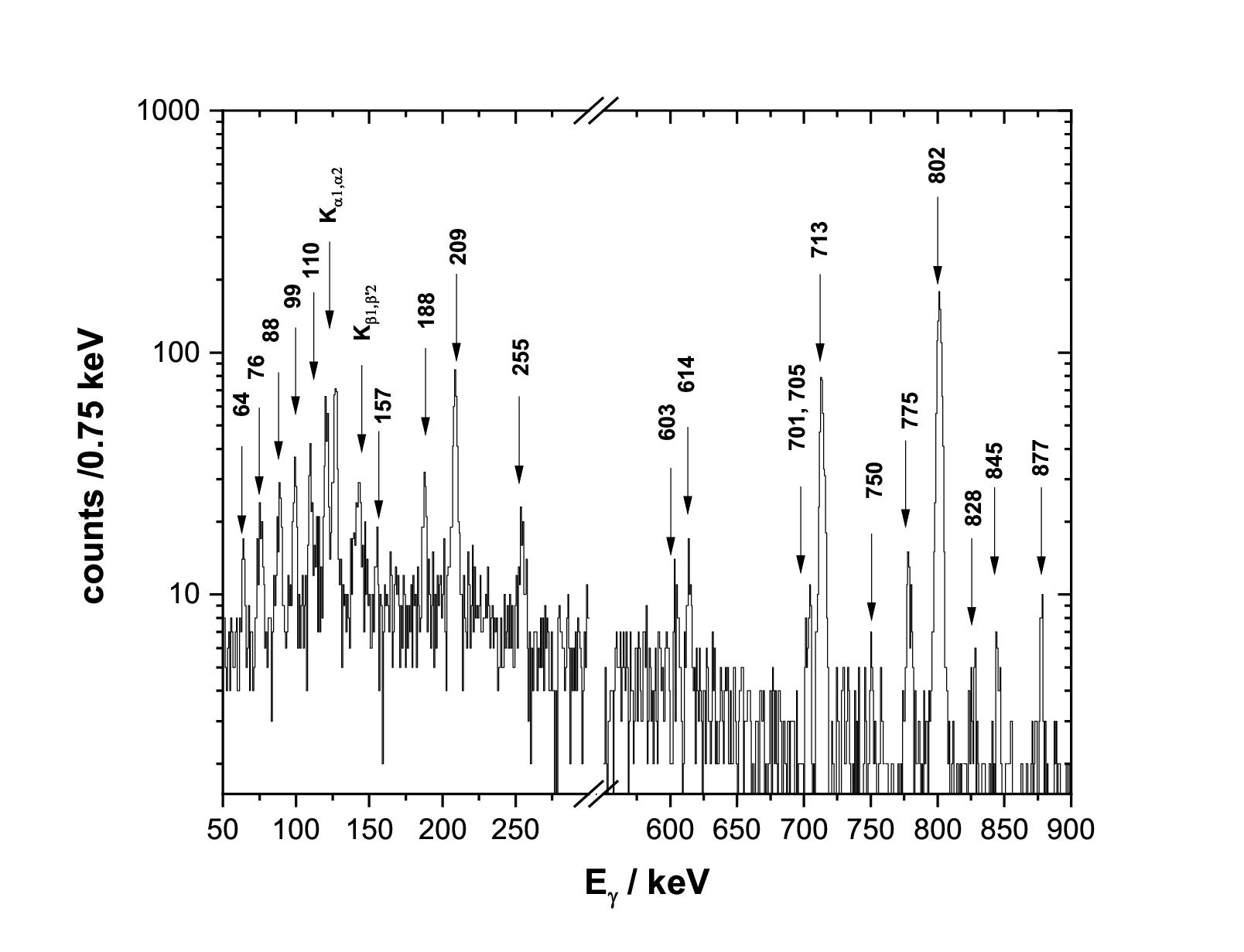}
			\includegraphics{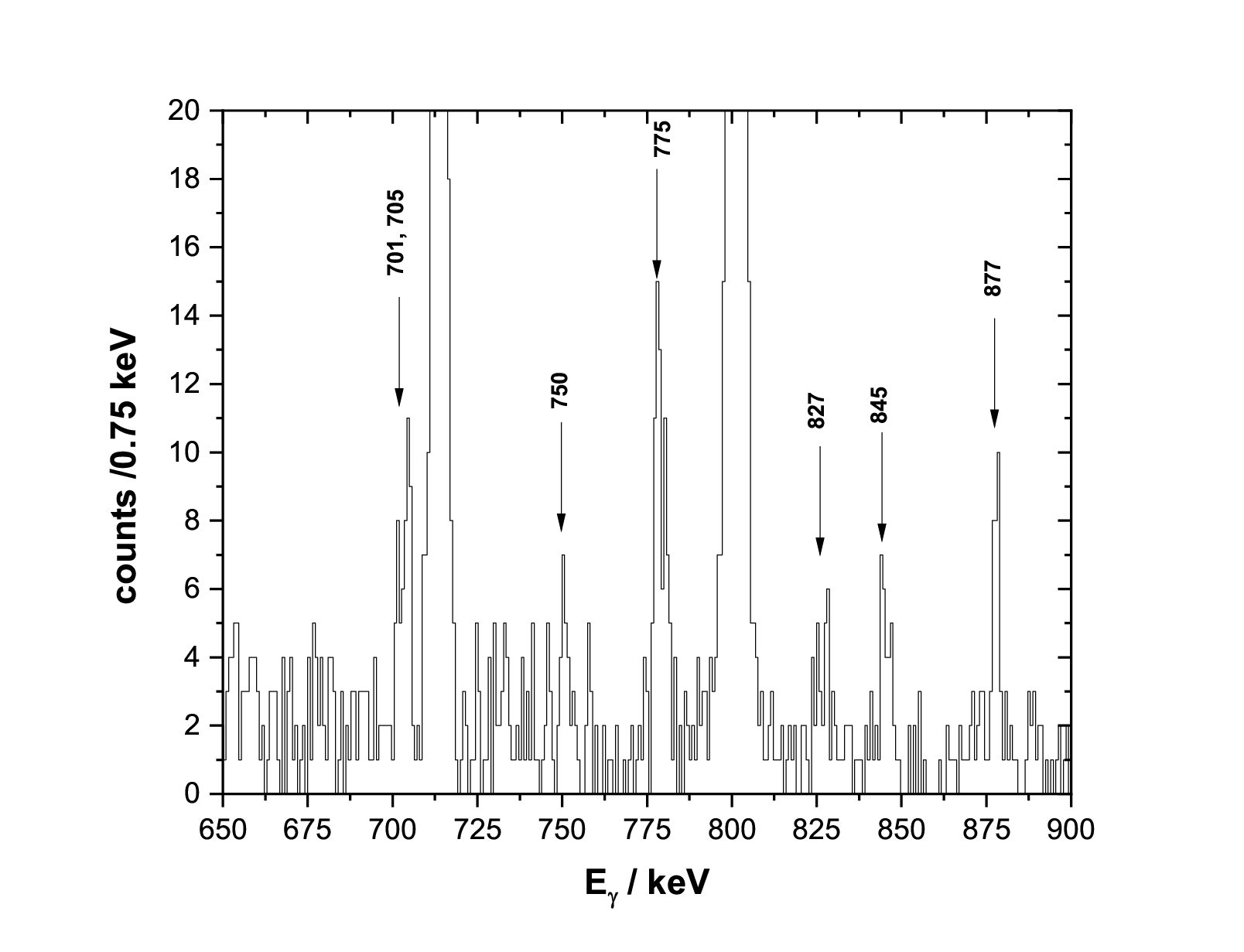}
	} 
	\caption{ $\gamma$ spectrum $^{253m2}$No measured at SHIP; left side: full range from 50-900 keV), right side: expanded, range 650-900 keV.}
	
	\label{fig:10}       
\end{figure*}

\begin{figure*}
	\centering
	\resizebox{0.75\textwidth}{!}{
		\includegraphics{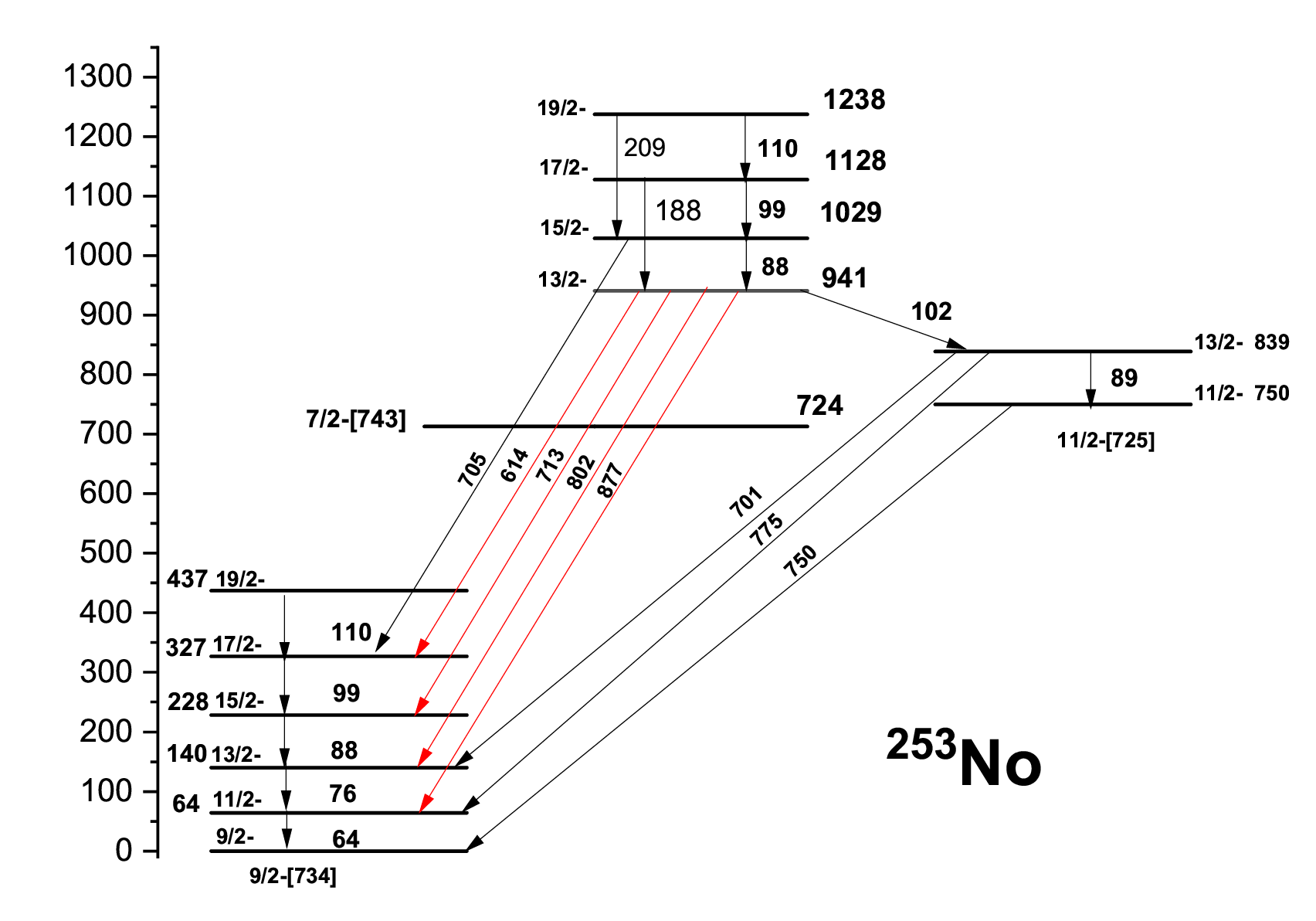}
	} 
	\caption{Decay scheme $^{253m2}$No based on the one proposed in \cite{LoM11} and from the SHIP Data \cite{AntH11}.}
	
	\label{fig:11}       
\end{figure*}

\subsection{{\bf 6.5 K isomers in  odd-mass nuclei with Z$\ge$100}}

{\bf K isomers in $^{249,251}$Md}\\
K  isomers in $^{249}$Md and $^{251}$Md were searched for in experiments performed at the University of Jyv\"askyl\"a, Finland using the RITU separator
and the GREAT detector system \cite{GoiT21}. The isotopes were produced in the reactions $^{203}$Tl($^{48}$Ca,2n)$^{249}$Md and $^{205}$Tl($^{48}$Ca,2n)$^{251}$Md.
The existence of K isomers was concluded from observed correlations ER - CE - $\alpha$($^{249,251}$Md. For $^{249m}$Md a 
half-life T$_{1/2}$ = 2.4 $\pm$ 0.3 ms was
obtained, for $^{251m}$Md a value of T$_{1/2}$ = 1.37 $\pm$ 0.6 s. For both isomers $\gamma$ decays were observed in coincidence with CE. 
In the case of $^{249m}$Md two $\gamma$ - lines of E = 175 $\pm$1 keV and E = 521.7$\pm$1.0 keV were indicated. Some more information was obtained for 
$^{251m}$Md. Here three $\gamma$ transitions of E = 216$\pm$1, 265$\pm$1, and 290$\pm$1 keV were observed, whose energies are similar to transitions within
the ground-state rotational band E = 214.8$\pm$0.5 keV (17/2$^{-}$ $\rightarrow$ 13/2$^{-}$),  E = 263.8$\pm$0.3 keV (21/2$^{-}$ $\rightarrow$ 17/2$^{-}$), 
and  E = 289$\pm$1 keV (23/2$^{-}$ $\rightarrow$ 19/2$^{-}$), as obtained from an in-beam study \cite{BriT20}.
Yet, it was stated, that also other transitions within the ground-state rotational band should have been observed, which was not the case, while other
observed $\gamma$ lines could not be placed in a level scheme (see \cite{GoiT21} for more details). Nevertheless the authors suggest, as possibly the
23/2$^{-}$ $\rightarrow$ 19/2$^{-}$ transition was observed, that decay of the isomeric state populates the 23/2$^{-}$ state of the ground-state rotational band.
A relatively strong $\gamma$ line of E = 389 keV was interpreted as the same as observed in the in-beam experiment and thus attributed to a 3/2$^{-}$
$\rightarrow$ 7/2$^{+}$ transition (see \cite{BriT20} for further details). 
Nevertheles, the quality of the data was not sufficient to draw conclusions on the decay path.
On the basis of the energy sum of CE and $\gamma$ rays lower limits of E$^{*}$ $\ge$ 910 keV ($^{249m}$Md) and E$^{*}$ $\ge$ 844 keV ($^{251m}$Md) were given
for the excitation energy of the isomers.\\
Based on (assumed) configurations of K isomers in neighbouring even - even nuclei, i.e. $^{248}$Fm and $^{250}$No in the case of $^{249}$Md and 
$^{250}$Fm and $^{252}$No in the case of $^{251}$Md and the ground state Nilsson levels of both isotopes configurations
$\pi$7/2$^{-}$[514]
$\downarrow$ 
$\otimes$ 
$\nu$5/2$^{+}$[622]$\uparrow$
$\otimes$ $\nu$7/2$^{+}$[624]$\downarrow$ resulting in I$^{\pi}$ = 19/2$^{-}$
for $^{249m}$Md and 
$\pi$7/2$^{-}$[514]$\downarrow$ $\otimes$ $\nu$7/2$^{+}$[624]$\downarrow$ $\otimes$ $\nu$9/2$^{-}$[734]$\uparrow$ resulting in I$^{\pi}$ = 23/2$^{+}$
for $^{251m}$Md were suggested.\\

{\bf K isomer in $^{251}$No}\\
In the course of a decay study of $^{251}$No a couple of $\gamma$ events with energies of E$_{\gamma}$ = 142.4, 203.1, 713.6 and 782.5 keV were observed in delayed
coincidence ($\Delta$t($\gamma$-ER)\,=\,(1-4)$\mu$s) with ER correlated to $\alpha$ decays of $^{251}$No \cite{Hes06a}. The two low energy lines were known to stem
from the decay of the 9/2$^{-}$[734] level in $^{251}$No, which is also populated by the $\alpha$ decay of $^{255}$Rf. That was seen as an indication that this level was
populated by the decay of an isomeric state with a half-life of T$_{1/2}$ $\ge$ 2 $\mu$s by the 'high' energy $\gamma$ transitions. As also $\gamma$ events above 700 keV
were observed it was concluded that the lower lying 1/2$^{+}$[631] isomeric state (E$^{*}$ $\approx$ 105 keV) and the ground state (7/2$^{+}$[624]) were not populated 
notably. From intensity considerations it was concluded that the high energy $\gamma$s do not stem from decay of the same nuclear level, but emitted in series via 
an intermediate level located at E$^{*}$ = 917 keV or E$^{*}$ = 985 keV. As a possible candidate the 7/2$^{+}$[613] - Nilsson level was suspected. In the neighbouring
N = 149 isotone $^{249}$Fm it is predicted at E$^{*}$ $\approx$ 900 keV \cite{ParS05}. No prediction is given for $^{251}$No in \cite{ParS05}. But as the 
predicted energy of this level increases from plutonium (Z=94) to fermium (Z=100) it may be located in $^{251}$No above 1 MeV, i.e. outside the range of predictions
in \cite{ParS05}. It should be noted, that in such a scenario the E1 transition 7/2$^{+}$[613] $\rightarrow$ 9/2$^{-}$[734] is expected to be faster 
than the M1 transition 7/2$^{+}$[613] $\rightarrow$ 7/2$^{+}$[624], which explains the decay into the 9/2$^{-}$[734] excited level and not into 7/2$^{+}$[624]
ground state. It should, however, be mentioned that this interpretation is still somewhat speculative. On the basis of the measured $\gamma$ energies a lower limit
of the excitation energy of 1.7 MeV was given. No speculation on spin, parity or configuration was presented in \cite{Hes06a}. The tentaive decay pattern presented there 
is shown in fig. 9.\\

{\bf K isomers in $^{253}$No}\\
Identification of a K isomer in $^{253}$No was first reported by F.P. He\ss berger \cite{Hes06b} and A. Lopez-Martens et al. \cite{LoH07}.
While A. Lopez-Martens et al. could only measure CE and gave a rough half-life T$_{1/2}$ = 700 $\pm$ 200 $\mu$s, 
F.P. He\ss berger identified the isomer by measuring CE - $\gamma$ - coincidences, suggested a decay scheme and presented 
more precise half-life values of T$_{1/2}$ = 715 $\pm$ 30 $\mu$s from the $\gamma$ lines and T$_{1/2}$ = 590 $\pm$ 40 $\mu$s 
from the K x-rays. Further, more detailed studies were performed as well at VASSILISSA, JINR, Dubna \cite{LoM11} as at SHIP, GSI,
Darmstadt \cite{AntH11}. The $\gamma$ spectra measured at SHIP are shown in fig. 10.
In \cite{LoM11} a partial decay scheme was presented. It is shown in fig. 11, supplemeted by the results from the SHIP
experiment \cite{AntH11}. The strong $\gamma$ - lines of E = 802 keV and E = 715 keV were interpreted as transitions from a
15/2$^{-}$ - state into the 13/2$^{-}$ - and 15/2$^{-}$ - states of the ground-state rotational band of $^{253}$No which is known
up to I$^{\pi}$ = 49/2$^{-}$ \cite{HerzM09}. On the basis of comparison of calculated and experimental gyromagnetic factors
$\mid$g$_{k}$-g$_{r}$$\mid$ performed in \cite{AntH11}, an I = 9/2 state was assumed as the possible bandhead. Another candidate for the
bandhead was the Nilsson level 11/2$^{-}$[725], which was not known at the time of publication of \cite{LoM11,AntH11}, but recently identified
at E$^{*}$ = 750 keV \cite{HaL22}. On the basis of the decay scheme presented in fig. 11 the 15/2$^{-}$ - level is located at E$^{*}$ = 942 keV, 
which would be $\approx$ 192 keV above the bandhead. This is quite similar to the very tentative difference betweem 11/2-
and 15/2- in $^{251}$Cf with $\Delta$E = 199 keV \cite{Fire96}. \\
On the basis of the decay scheme suggested by Lopez-Martens \cite{LoM11}, the results from the SHIP experiments \cite{AntH11},
recent nuclear structure investigations of $^{253}$No \cite{Hess16a,HaL22} and predicted Nilsson levels at E$^{*}$ $<$ 1 MeV
(see \cite{Asai15}), we here suggest the decay scheme  presented in fig. 11. It will be discussed in detail in the following.
We here shall just note, that besides the strong transitions of E$_{\gamma}$ = 713 and 802 keV, interpreted to populate the
15/2$^{-}$ and 13/2${-}$ members of the ground-state rotational band \cite{LoM11}, two more $\gamma$ - lines clearly observed in the SHIP
experiments, were assigned on the basis of energy balance to be emitted from the level at E$^{*}$ = 941 keV, namely E$_{\gamma}$ = 614 keV
populating the 17/2$^{-}$ state and  E$_{\gamma}$ = 877 keV
populating the 11/2$^{-}$ state.\\
As possible bandheads of the level at E$^{*}$ = 941 keV we will consider the Nilsson levels 11/2$^{-}$[725] and 7/2$^{-}$[743].
A bandhead with spin/parity 9/2$^{-}$ will not be considered as such a state (besides the ground-state of $^{253}$No) is not predicted
at an excitation energy below 1 MeV.\\
Let us first consider the 11/2$^{-}$[725] state. The bandhead was located at E$^{*}$ = 750 keV as mentioned above. The energy
difference of the 11/2$^{-}$[725] and the one at E$^{*}$ = 941 keV is $\Delta$E = 191 keV. Keeping I$^{\pi}$ = 15/2$^{-}$ as 
assigned by A. Lopez-Martens \cite{LoM11} the decay into the bandhead could occur by two M1 transitions 15/2$^{-}$ $\rightarrow$ 13/2$^{-}$ $\rightarrow$ 11/2$^{-}$.
Since $\Delta$E/2 = 95.5 keV (we do not assume that both transitions have the same energy) one of the transitions must have an 
energy E\,$>$\,95.5 keV, so its energy must be higher than E\,=\,88 keV the transition feeding the E$^{*}$ = 941 keV, which is in contradiction 
to the energy systematics in rotational bands. Here we have to remark,
that the feeding of the E$^{*}$ = 941 keV is established by $\gamma$ - $\gamma$ - coincidences. Therefore an 11/2$^{-}$[725] assignment 
for the bandhead is in contradiction to the experimental data and can be excluded.\\
The 7/2$^{-}$[743] Nilsson level was identified at E$^{*}$\,=\,724 keV \cite{Hess16a}, so the energy difference to the E$^{*}$ = 941 keV is 
$\Delta$E = 217 keV. Analyzing known cases (see \cite{Fire96}) one finds typical energy differences $\Delta$E(15/2\,-\,7/2)\,$>$\,230 keV for the
Nilsson levels 7/2$^{-}$[743], 7/2$^{+}$[624], and 7/2$^{+}$[613]. For $\Delta$E(13/2\,-\,7/2) values between 175 keV and 219 (236) keV 
are obtained with some tendency to increase at increasing atomic number. Specifically for $^{245}$Cm (the closest case of transition within the band
built up on the 7/2$^{-}$[743] level one obtains $\Delta$E(13/2$^{-}$\,-\,7/2$^{-}$ = 209 keV \cite{Ahmad15}. On this basis we prefer to assign
I$^{\pi}$\=\,13/2$^{-}$ to the level at E$^{*}$ = 941 keV.\\
In the SHIP - experiments also a weak line at E$_{\gamma}$\,=\,750 keV was observed that agrees with the excitation energy of the 11/2$^{-}$[725] 
Nilsson level. So it can be assigned to the decay of that level into the ground-state, vice versa requiring that the 11/2$^{-}$[725] is populated 
during the decay of the isomeric state.\\
In the SHIP experiment further two significant lines of E$_{\gamma}$\,=\,701 keV (a line doublet, consisting of E$_{\gamma}$\,=\,701 keV and
E$_{\gamma}$\,=\,705 keV) and E$_{\gamma}$\,=\,775 keV were observed. Their energy difference
$\Delta$E\,=\,74 keV is in-line with the energy difference $\Delta$E(13/2$^{-}$-11/2$^{-}$)\,=\,76 keV between the corresponding members of the 
ground-state rotational band. The energy sums E\,=\,140 keV + 701 keV = 841 keV and E\,=\,64 keV + 775 keV = 839 keV do not support an assigment
of the emitting state to a member of the band built up on the 7/2$^{-}$[743] state, specifically the 11/2$^{-}$ one as this state would lie $\approx$100 keV below the 
(assumed) 13/2$^{-}$ state, as discussed above in contradiction the to the energy systematics in rotational bands. So we tentatively assign it to the
13/2$^{-}$ state of the rotational band built up on the 11/2$^{-}$[725] Nilsson level. Thus the energy difference is $\Delta$E(13/2$^{-}$-11/2$^{-}$) = 89 (91) keV,
which is in-line with the energy difference $\Delta$E(13/2$^{-}$-11/2$^{-}$) = 88 keV for these members of the rotional band built up on the 
11/2$^{-}$[725] Nilsson level in $^{257}$Rf \cite{Riss13}\footnote{In fig. 11 the 13/2$^{-}$ energy obtained from the more precisely measured energy of the
E\,=\,775 keV line is given.}. Both Nilsson levels are assumed to be connected by a not observed M1 transition of E\,=\,102 keV (13/2$^{-}$ $\rightarrow$
13/2$^{-}$).\\
The energy of the line E$_{\gamma}$\,=\,705 keV fits quite well to the energy difference of the 15/2$^{-}$ member of the band built up on the
7/2$^{-}$[743] Nilsson level and the 17/2$^{-}$ member of the ground-state rotational band (702 keV) and is thus attributed to that transition.\\
The time distributions of the significant lines at  E\,=\,828 keV and E\,=\,845 keV do not fit to the half-life of $^{253m2}$No and are thus not attributed to decay
of the isomer.\\
Still some problems concerning the decay scheme should be mentioned:\\
a) strong K$_{\alpha_ {1},\alpha_{2},\beta_{1}}$ No - X-ray lines are observed. As the energies of the assumed and tentatively 
assigned low energy M1 transitions (see fig. 11) are below the K binding
energy of nobelium (E = 149.2 keV \cite{Fire96}) and as the high energy transitions (E $>$ 600 keV) are only little converted
($\epsilon_{K}$ $<$ 0.28 for E $>$ 600 keV \cite{Kib06}) it seems quite unlikely that the observed K X-rays stem from the 
decay of the K isomer discussed above. This assumption is supported by the fact, that prompt coincidences between the K X-rys were observed,
but no coincidences between K X-rays and $\gamma$ events. 
Also the K X-rays show a half-life of T$_{1/2}$ = (552$\pm$15) $\mu$s \cite{Hess23}, which is somewhat
lower than the value T$_{1/2}$ = (627$\pm$5) $\mu$s obtained from the $\gamma$ transitions \cite{AntH11}. 
This could be a hint for the existence of a further isomeric state. However, from the SHIP data due to the 
lack of delayed coincidences between K X-rays and gammas, we do not have an indication that such second K isomer
feeds the 'known' one \cite{Hess23}. Also from the lack of prompt coincidences between K X-rays and gammas
we do not have indication for a decay path bypassing the 'known' isomer. So the origin of the relatively strong
K X-ray lines remains uncertain and the assumption of the existence of a second isomer remains speculative.
The situation is similiar for the line at E\,=\,255 keV. For this transition we obtain half-life of T$_{1/2}$ = (540$\pm$30) $\mu$s \cite{Hess23},
in-line with that for the K - X-rays and thus related to the same source.\\
b) no transition that can be assigned to the 'direct' decay of the isomer was observed. \\
More detailed measurements are necessary to clarify the situation.\\  
c) lines at E\,=\,157 keV and E \,=\,603 keV cannot be placed un the decay scheme. \\

Possible configurations of the isomeric state were discussed in \cite{LoM11}. For the isomeric state a 3-quasiparticle - state
(1n $\otimes$ 2p - cofiguration) $\nu$9/2$^{-}$[734]$\uparrow$ $\otimes$ $\pi$7/2$^{-}$[514]$\downarrow$ $\otimes$ $\pi$9/2$^{+}$[624]$\uparrow$
leading to a K$^{\pi}$ = 25/2$^{+}$ - state or 
$\nu$9/2$^{-}$[734]$\uparrow$ $\otimes$ $\pi$1/2$^{-}$[521]$\downarrow$ $\otimes$ $\pi$9/2$^{+}$[624]$\uparrow$ leading to a K$^{\pi}$ = 19/2$^{+}$ - state
were considered, which were predicted to lie below an excitation energy of 1.5 MeV by calculations using a universal Woods-Saxon - potential
\cite{LoM11}. 
More detailed measurements are necessary to clarify the situation.\\  

\begin{table}
	\caption{Comparison of properties of isomers in $^{255}$No as reported by A. Bronis et al. \cite{Bronis22} and K. Kessaci \cite{Kessaci22}.
Note that $^{255m2}$No is denoted as $^{255m3}$No and $^{255m3}$No is denoted as $^{255m2}$No in \cite{Bronis22}. } 
	\label{tab:1}       
	\begin{tabular}{cccccccc}
		\hline\noalign{\smallskip}
		&   &  Bronis et al. \cite{Bronis22} &   &  &  & Kessaci \cite{Kessaci22}  &      \\
		\hline\noalign{\smallskip}     
		              &  E$^{*}$ / keV  &  T$_{1/2}$ / $\mu$s & I$^{\pi}$ & $|$ & E$^{*}$ / keV  &  T$_{1/2}$ / $\mu$s & I$^{\pi}$ \\
		\hline\noalign{\smallskip} 
		$^{255m1}$No  &   240-300 & 109$\pm$9 & 11/2$^{-}$[725] & $|$ & $\approx$200  & 86$\pm$8 & 11/2$^{-}$[725] \\
		$^{255m2}$No  &  $\ge$1500  & 1.2$^{+0.6}_{-0.4}$ & $\ge$19/2 & $|$ & $\approx$1.3& 2$\pm$1 & 21/2$^{+}$ \\
                $^{255m3}$No  &   1400-1600 & 77$\pm$6 & 19/2, 21/2, 23/2 & $|$ &$>$1500  & 92$\pm$13 & 27/2$^{+}$ \\
			$^{255m4}$No  &    & (not reported) &  & $|$ &   $>$2500  & 5$\pm$1 & - \\
		
	\end{tabular}
	\\
	
\end{table}

{\bf K isomer in $^{255}$No}\\
First indication for the existence of a K isomer in $^{255}$No was obtained from the observation of $\gamma$ transitions of
E$_{\gamma}$(1) = 742 keV, T$_{1/2}$(1) = 105$\pm$25 $\mu$s and E$_{\gamma}$(2) = 839 keV, T$_{1/2}$(1) = 130$\pm$25 $\mu$s
in a study at the velocity filter SHIP at GSI, Darmstadt (Germany) devoted to the investigation of $^{254m1,254m2}$No 
produced in the reaction $^{208}$Pb($^{48}$Ca,2n)$^{254}$No \cite{Hes10}. The transitions
could not be attributed to the K isomers in $^{254}$No as a) the half-lives were not in-line with those of the K isomers in $^{254}$No,
T$_{1/2}$($^{254m1}$No)\,=\,275$\pm$7 ms and T$_{1/2}$($^{254m2}$No)\,=\,198$\pm$13 $\mu$s \cite{Hes10}), and b) the intensity of both lines
increased relatively to those of the $\gamma$ lines attributed to the decay of $^{254m1,254m2}$No at decreasing excitation energy of the compound nuclei.
So it seemed straightforward to assign them to an isomeric state in $^{255}$No, the 1n - deexcitation channel of the reaction. 
A careful follow-up analysis of the data resulted in the identification of three K isomers in $^{255}$No \cite{Bronis22} (see table 4). \\
The lowest lying isomer $^{255m1}$No was attributed to the Nilsson level 11/2$^{-}$[725]. It represents the so far 'missing' link in the systematics
of that state, the energies of which are decreasing with increasing proton number, forming short-lived isomers (T$_{1/2}$ $\le$ 100 $\mu$s) up Z\,=\,102
\cite{AntH11,Bronis22}, a low-lying isomer at E$^{*}$\,=\,(70-74) keV \cite{Hess97,Hess16a,HaL22} in $^{257}$Rf and finally becoming the ground-state in 
$^{259}$Sg \cite{AntH15}. It belongs to the type of single particle K-isomers (see sect.5).\\
Another study of isomeric states in $^{255}$No was performed by K. Kessaci \cite{Kessaci22} (and coworkers) at 
the separator SHELS at the FLNR, JINR, Dubna (Russia). A significantly higher number of decays including also
a couple of prominent $\gamma$ lines were registered, allowing to establish a somewhat more detaild decay scheme.
Kessaci identified four isomeric states, the measured properties are summarized and compared with the data 
from Bronis \cite{Bronis22} in table 7. At first glance it may seem that data may be not in-line with each other,
but they have to be compared with respect to their limited quality.
Under this aspect one can state, that the half-lives agree sufficiently, indicating that the same isomers were observered. The same holds for
the excitation energies. Spin assignments are quite uncertain in \cite{Bronis22} for $^{255m2,255m3}$No and als no parities are given. Also
$^{255m4}$No was not reported in \cite{Bronis22}.\\
In \cite{Kessaci22} also configurations for $^{255m2,255m3}$No were suggested, namly
$\pi$9/2$^{+}$[624]$\uparrow$ $\otimes$ $\pi$1/2$^{-}$[521]$\downarrow$ $\otimes$ $\nu$11/2$^{-}$[725]$\uparrow$ resulting in I$^{\pi}$ = 21/2$^{+}$
for $^{255m2}$No and 
$\pi$9/2$^{+}$[624]$\uparrow$ $\otimes$ $\pi$7/2$^{-}$[514]$\downarrow$ $\otimes$ $\nu$11/2$^{-}$[725]$\uparrow$ resulting in I$^{\pi}$ = 27/2$^{+}$
for $^{255m3}$No.
\\

\begin{figure*}
\centering
\resizebox{0.99\textwidth}{!}{
\includegraphics{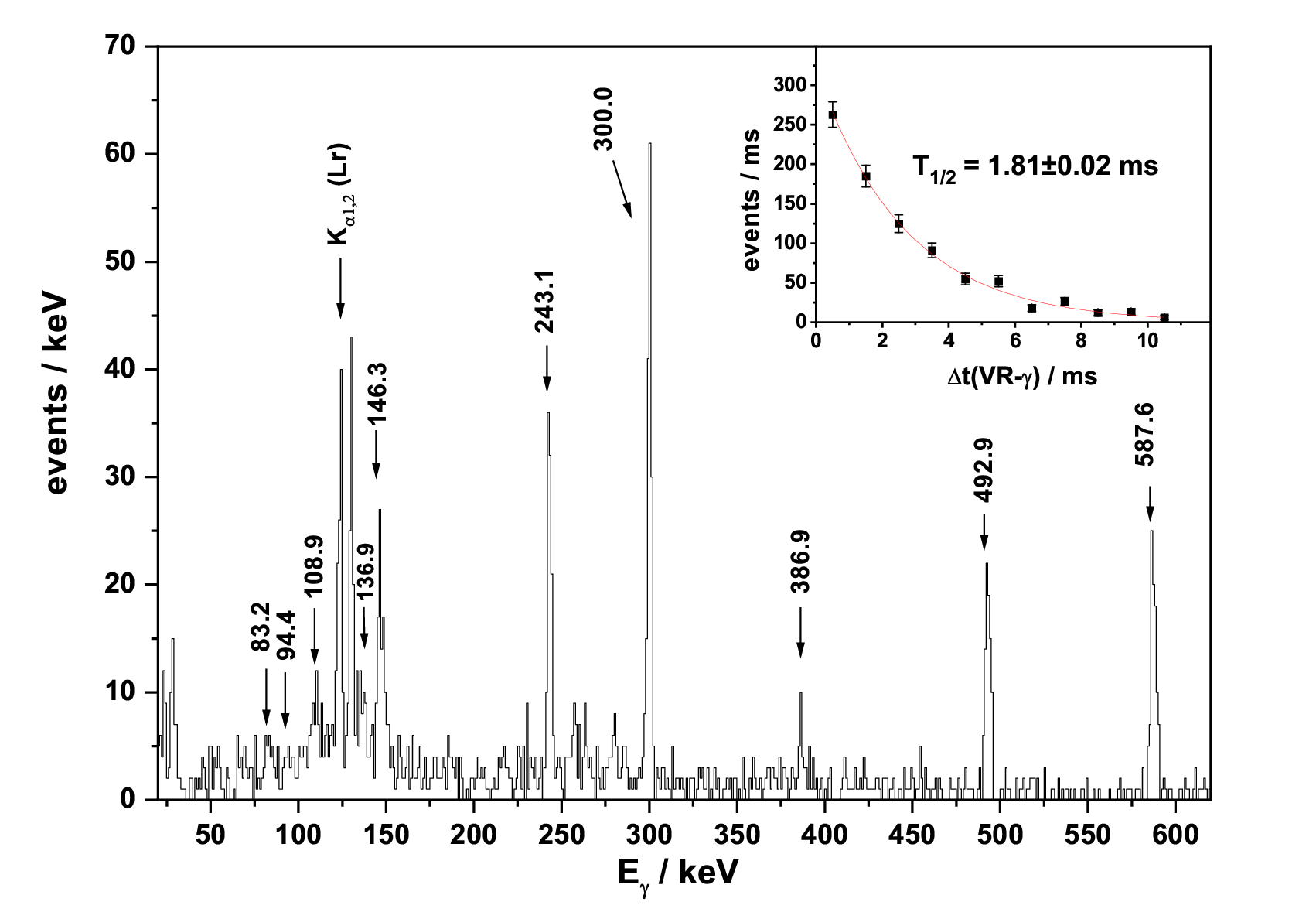}
} 
\caption{ Gamma - spectrum from the decay of the K isomeric state  $^{255m2}$Lr at SHIP \cite{Hessb22,AntH08}
Insert: time distribution $\Delta$t((CE,$\gamma$),ER) of events attributed to the decay of $^{255m2}$Lr.}

\label{fig:12}       
\end{figure*}

\begin{figure*}
\centering
\resizebox{0.99\textwidth}{!}{
\includegraphics{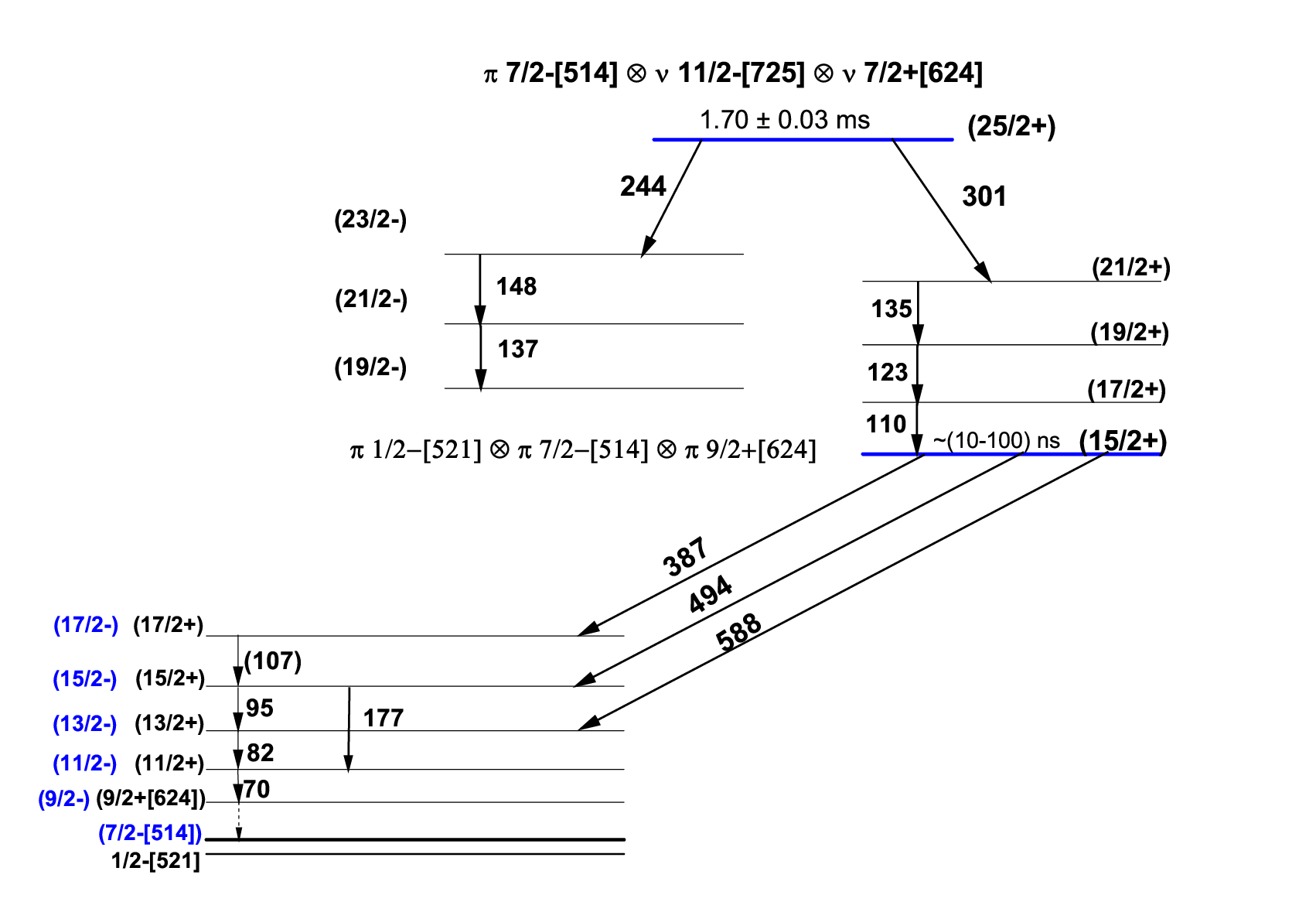}
} 
\caption{Decay scheme of $^{255m2}$Lr as suggested by Jeppesen et al. \cite{Jepp08}, and alternative level assignments (blue numbers).}

\label{fig:13}       
\end{figure*}

{\bf K isomer in $^{255}$Lr}\\
The first report on discovery of a multiparticle K isomer in $^{255}$Lr was presented by K. Hauschild et al. \cite{Hau08} from an experiment
performed at the VASSILISSA separator at JINR - FLNR, Dubna. Although the data were of limited
quality a couple of $\gamma$ transitions from the decay of the K isomer were reported. A half-life T$_{1/2}$ = 1.4$\pm$0.1 ms was measured,
the excitation energy was estimated as E$^{*}$ $>$ 720 keV.\\
Data of much higher quality were obtained in an experiment performed at SHIP at GSI, Darmstadt \cite{AntH08}.
The spectrum of $\gamma$ lines from the SHIP experiment is shown in fig. 12. Also $\gamma$ - $\gamma$ - coincidences could be established, 
specifically ($\gamma_{1}$(587.5 keV) - $\gamma_{2}$(109.1, 243.1, 300.0 keV) and  ($\gamma_{1}$(493.1 keV) - $\gamma_{2}$(243.1, 300.0 keV) - coincidences.
A half-life of T$_{1/2}$ = 1.81$\pm$0.02 ms was measured, the excitation energy was estimated as E$^{*}$ $>$ 1.6 MeV, the spin as I$>$21/2.
The decay of the isomeric state was predominantly followed by $\alpha$ decays of E$_{\alpha}$ = 8467 keV, attributed to the 7/2$^{-}$[514]$\downarrow$
isomeric state in $^{255}$Lr with a half-life of T$_{1/2}$\,=\,2.5 s \cite{Chat06}, while in the direct production $^{209}$Bi($^{48}$Ca,2n)$^{255}$Lr the decay
of the ground-state (1/2$^{-}$[521]$\downarrow$, E$_{\alpha}$\,=\,8373 keV, T$_{1/2}$\,=\,31 s \cite{Chat06,AntH08}) dominates. As it can be expected that decay of 
a high-spin K isomer preferrably populates a state with a high spin, this finding was seen as confirmation of the assignment of T$_{1/2}$\,=\,2.5 s - activity
to the high-spin Nilsson level (7/2$^{-}$[514]). The $\alpha$ decays of E$_{\alpha}$\,=\,8373 keV were interpreted as decays of the ground-state populated
by decay of the isomer via internal transitions, thus corroborating the assignement of the  low-spin Nilsson level (1/2$^{-}$[521]) to the ground-state
and the high-spin Nilsson state (7/2$^{-}$[514]) to the isomer as done in \cite{Chat06}.
The excitation energy of the 2.5 s - state is given in \cite{Chat06} as E$^{*}$ = 37 keV, which was confirmed in the direct measurement
at SHIPTRAP which resulted in E$^{*}$ =  32.2$\pm$2.5 keV \cite{Block22}. Energy differences between the 1/2$^{-}$[521] - bandhead and the 3/2$^{-}$ member of the rotational band 
are $\Delta$E = 19.9 keV in $^{251}$Bk and $\Delta$E = 18.45 keV in $^{249}$Bk \cite{Fire96}. As transition of the lowest multipolarity
between the isomer and the ground-state rotational band thus the E2 - transition 7/2$^{-}$ $\rightarrow$ 3/2$^{-}$ can be assumed for which even
for an energy difference of $\Delta$E = (10-15) keV a half-life $<<$1 s can be expected \cite{Fire96}. Thus the E2 - transition is strongly 
K hindered ($\Delta$K = 3) and the 7/2$^{-}$ isomer thus represent the type of a single particle K isomer discussed in sect. 5
and we here observe the case of decay of a multi particle K isomer into a single particle K isomer.\\
A further study of the K isomer in $^{255}$Lr was performed by H.B. Jeppesen et al. \cite{Jepp08}, who obtained data of similar quality as in \cite{AntH08}.
They obtained a half-life of T$_{1/2}$ = 1.70$\pm$0.02 ms. Also possible configurations were discussed and a decay scheme was proposed. It
is shown in fig. 13. The 1.7 ms - K isomer is attributed to an I$^{\pi}$ = 25/2$^{+}$ - state with a possible
configuration $\pi$7/2$^{-}$[514]$\downarrow$ $\otimes$ $\nu$11/2$^{-}$[725]$\uparrow$ $\otimes$ $\nu$7/2$^{+}$[624]$\downarrow$.
It decays via two paths, one feeding by the 244-keV $\gamma$ transition an I$^{\pi}$ =  23/2$^{-}$ level of a band with a not assigned band head.
The other one feeding by the 301-keV the I$^{\pi}$ =  21/2$^{+}$ level of a band with the band head  I$^{\pi}$ =  15/2$^{+}$ 
(possible configuration $\pi$1/2$^{-}$[521]$\downarrow$ $\otimes$ $\pi$7/2$^{-}$[514]$\downarrow$ $\otimes$ $\pi$9/2$^{+}$[624]$\uparrow$).
This state was assumed to be isomeric with a half-life (10-100) ns.
Both bands are connected by an (unobserved) 28-keV transition 19/2$^{-}$ $\rightarrow$ 17/2$^{+}$. 
The decay of the assumed I$^{\pi}$ =  15/2$^{+}$ isomeric state is interpreted  to populate a band of positive parity, 
built up on the 9/2$^{+}$[624]$\uparrow$ Nilsson level.
As shown in \cite{AntH08} the decay of the 1.8 ms K isomer populates the 7/2$^{-}$[514] isomeric state. Thus the decay scheme 
suggested in \cite{Jepp08} requires an E1 - transition 9/2$^{+}$[624] $\rightarrow$ 7/2$^{-}$[514]. As E1 transitions are only
weakly converted such a line must be quite strong compared to the M1 transition between the members of the rotational band built
up on the 9/2$^{+}$[624]. Such a line, however, is not observed at E $>$ 30 keV\footnote{at E$<$30 keV it might be hidden by the lines attibuted to 
L X-rays of lawrencium.}. Alternatively one could assume the decay into the rotational band built up on the 7/2$^{-}$[514] Nilsson level (blue numbers in fig. 13).
Leaving the spin assignment unchanged, in this scenario the 9/2$^{-}$ $\rightarrow$ 7/2$^{-}$ transition, the energy of which could be assumed
around 50 keV, would be unobserved. This, however, would not be surprising, as already the 11/2$^{-}$ $\rightarrow$ 9/2$^{-}$  transition is quite weak.
Since the conversion coefficients for M1 transitions rise from $\epsilon$ = 41 (at 70 keV) to $\epsilon$ = 110 (at 50 keV) \cite{Kib06}, in this scenario
the 9/2$^{-}$ $\rightarrow$ 7/2$^{-}$ would be simply too weak to be observed.\\

{\bf K isomer in $^{253}$Rf}\\
The first claim for discovery of $^{253}$Rf (at this time denoted as $^{253}$Ku) came from G.N. Flerov \cite{Flerov76} who assigned an 1.8 s - SF activity observed in 
irradiations of $^{206}$Pb with $^{50}$Ti to the 3n - deexcitation - channel of that reaction. More thorough investigations by F.P. He\ss berger et al. \cite{Hess97} at the velocity filter SHIP at GSI, Darmstadt, using a more suited experimental set-up disproved the the results of G.N. Flerov and identified a 
T$_{1/2}$ = 48 $^{+17}_{-10}$ $\mu$s SF activity as $^{253}$Rf\footnote{The 1.8 s SF activity erreanously attributed to $^{253}$Rf by Flerov probably was SF of 
$^{255}$Rf (T$_{1/2}$ = 1.68 $\pm$ 0.09 s \cite{Hes06a}).}. In this experiment also a second SF activity of T$_{1/2}$ = 11$^{+6}_{-3}$ms was observed. As 
only a few number of decays was observed and this half-life 
was quite similar to that of $^{256}$Rf (T$_{1/2}$ = 6.2$\pm$0.2 ms \cite{Hess97}) it was not ruled out that it could represent decays of $^{256}$Rf, produced
in reactions with $^{207}$Pb impurities in the target material. So it was not assigned to $^{253}$Rf.\\
The data of \cite{Hess97} were recently confirmed by J. Khuyagbaatar et al. \cite{Khuy21} and by A. Lopez-Martens et al. \cite{LopM22a}. Based on a higher number
of observed decays, the longer - lived SF activity was now definitely assigned to an isomeric state in $^{253}$Rf. The results are compared in table 8.

\begin{table}
	\caption{Half-lives of identified long-lived states in $^{253}$Rf}
	\label{tab:1}       
	\begin{tabular}{cccc}
		\hline\noalign{\smallskip}
		& T$_{1/2}$$^{253}$Rf(1)/$\mu$s &  T$_{1/2}$$^{253}$Rf(2)/ms &   T$_{1/2}$$^{253}$Rf(3)/ms \\
		\hline\noalign{\smallskip}     
		\cite{Hess97} & 48$^{+17}_{-10}$ & 11$^{+6}_{-3}$$^{a}$ &  \\
		\cite{Khuy21} & 44$^{+17}_{-10}$ & 12.8$^{+7.0}_{-3.4}$ &  \\
		\cite{LopM22a} & 52.8$\pm$4.4 & 9.9$\pm$1.2 &  0.66$^{+0.40}_{-0.16}$\\
		
	\end{tabular}
	\\
	$^{a}$not assigned to $^{253}$Rf
\end{table}

In addition A. Lopez-Martens et al. observed one more isomeric state with a half-life of T$_{1/2}$ = 0.66$^{+0.40}_{-0.18}$ ms, decaying into the
short-lived state in $^{253}$Rf. As the energy of the CE attributed to the decay of that state reached up to 1.02 MeV, that energy can be regarded as 
the lower limit of the excitation energy of the high lying isomer, very likely a K isomer.
The latter observation may have an interesting consequence.
As low lying states on the basis of systematics in N = 149 isotones and theoretical predictions \cite{ParS05} the Nilsson levels 
1/2$^{+}$[631]$\downarrow$ and 7/2$^{+}$[724]$\downarrow$ are expected. 
As decay of the
K isomer, probably having a high spin, is populating the state of 52.8 $\mu$s it was concluded that this is the high-spin (7/2$^{+}$) - state,
while that of 9.9 ms half-life is the low spin state (1/2$^{+}$), which means that contrary to cases known so far (see \cite{Hess17}) the high-spin state
has lower fission hindrance than the low-spin state. Tentatively the 1/2$^{+}$[631]$\downarrow$ Nilsson level was assigned to the ground-state
of $^{253}$Rf, but it was not excluded, that also the 7/2$^{+}$[724] level could be the gound state. 
If the first scenario could be verified, we here observe the same interesting
feature as in $^{255}$Lr (see above), namly the decay of a 2-quasiparticle K isomer into a single particle K isomer.\\

{\bf K isomer in $^{255}$Rf}\\
A K isomeric state was searched for by analyzing ER - (CE,CE-$\gamma$) - $\alpha$,SF ($^{255}$Rf) correlations. The isotope was produced 
in an experiment performed at the velocity filter SHIP at GSI, Darmatstadt using the reaction $^{207}$Pb($^{50}$Ti,2n)$^{255}$Rf. 
A low lying isomeric state with a half-life of T$_{1/2}$ = (50$\pm$17) $\mu$s had already been known in that isotope \cite{AntH15}.
It was attributed to the 5/2$^{+}$[622]$\uparrow$ Nilsson level. Such isomers are common in N = 151 isotones (see e.g. \cite{AntH15}).
They are known to decay via highly K converted M2 - transitions (with some E3 contribution) into the 9/2$^{-}$[734]$\uparrow$ ground state.
As no K X - rays were observed, it was concluded that its excitation energy is below the K binding - energy of rutherfordium 
(E$_{B}$ =  156.288 keV \cite{Fire96}) and on the basis of the measured CE energies it was settled at E$^{*}$ $\approx$ 135 keV \cite{AntH15}.
Therefore possible contributions of this isomer had to be respected in the analysis of the ER - (CE,CE-$\gamma$) - $\alpha$,SF correlations.
The analysis was complicated due to the fact, that using analogue electronics, as in this case, the signal of the CE is sitting on the
'tail' of the ER pulse, adulterating the energy of the CE\footnote{The 5/2$^{+}$[622] isomer was identified from $\alpha$($^{259}$Sg) - CE - $\alpha$,SF($^{255}$Rf)
correlations. Due to the much lower $\alpha$ particle pulses this problem was not evident in that case.}. The analysis resulted in the identification of 
two activities \cite{Mosat20}. One with a half-life of T$_{1/2}$ = 15$^{+6}_{-4}$ $\mu$s for CE with E$>$350 keV and one of T$_1/2$\,=\,(35$\pm$5)$\mu$s for
CE of E $<$ 350 keV. As discussed above, the latter activity might contain contributions of the 5/2$^{+}$ isomer. But in addition 18 events were found
in coincidence with photons, however without presence of a clear signature of a distinct $\gamma$ line. As no $\gamma$ events in coincidence with CE 
had been observed for the decay of the 5/2$^{+}$ isomer, these events were attributed to the decay of a different isomer. A half-life of 
T$_{1/2}$ = 38$^{+12}_{-7}$ $\mu$s was obtained \cite{Mosat20}. On the basis of the measured CE- and $\gamma$ energies the
T$_{1/2}$ = 15$^{+6}_{-4}$ $\mu$s isomer was settled as the lower lying one at E$^{*}$ = (0.9-1.2) MeV with K $>$ 17/2, the  
T$_{1/2}$ = 38$^{+12}_{-7}$ $\mu$s one was settled at E$^{*}$ = (1.15 - 1.45) MeV. Although no decay scheme could be established
a possible configuration of the lower lying isomer was discussed. A K$^{\pi}$\,=\,19/2$^{+}$ state with the configuration
$\pi$1/2$^{-}$[521]$\downarrow$ $\otimes$ $\pi$9/2$^{+}$[624]$\uparrow$ $\otimes$ $\nu$9/2$^{-}$[734]$\uparrow$ was considered.
The partial level scheme of $^{255}$Rf obtained on the basis of the study of P. Mosat et al. \cite{Mosat20} and the previous
decay study of $^{259}$Sg by S. Antalic et al. \cite{AntH15} is shown in fig. 14a.
The data were confirmed in an experiment performed at the FLNR - JINR Dubna by R. Chakma et al. \cite{Chakma23}, who in addition
due to registering a significantly higher amount of decays also clearly observed several $\gamma$ lines which allowed to establish 
the level and decay scheme shown in fig. 14b.

\begin{figure*}
	\centering
	\resizebox{0.75\textwidth}{!}{
		\includegraphics{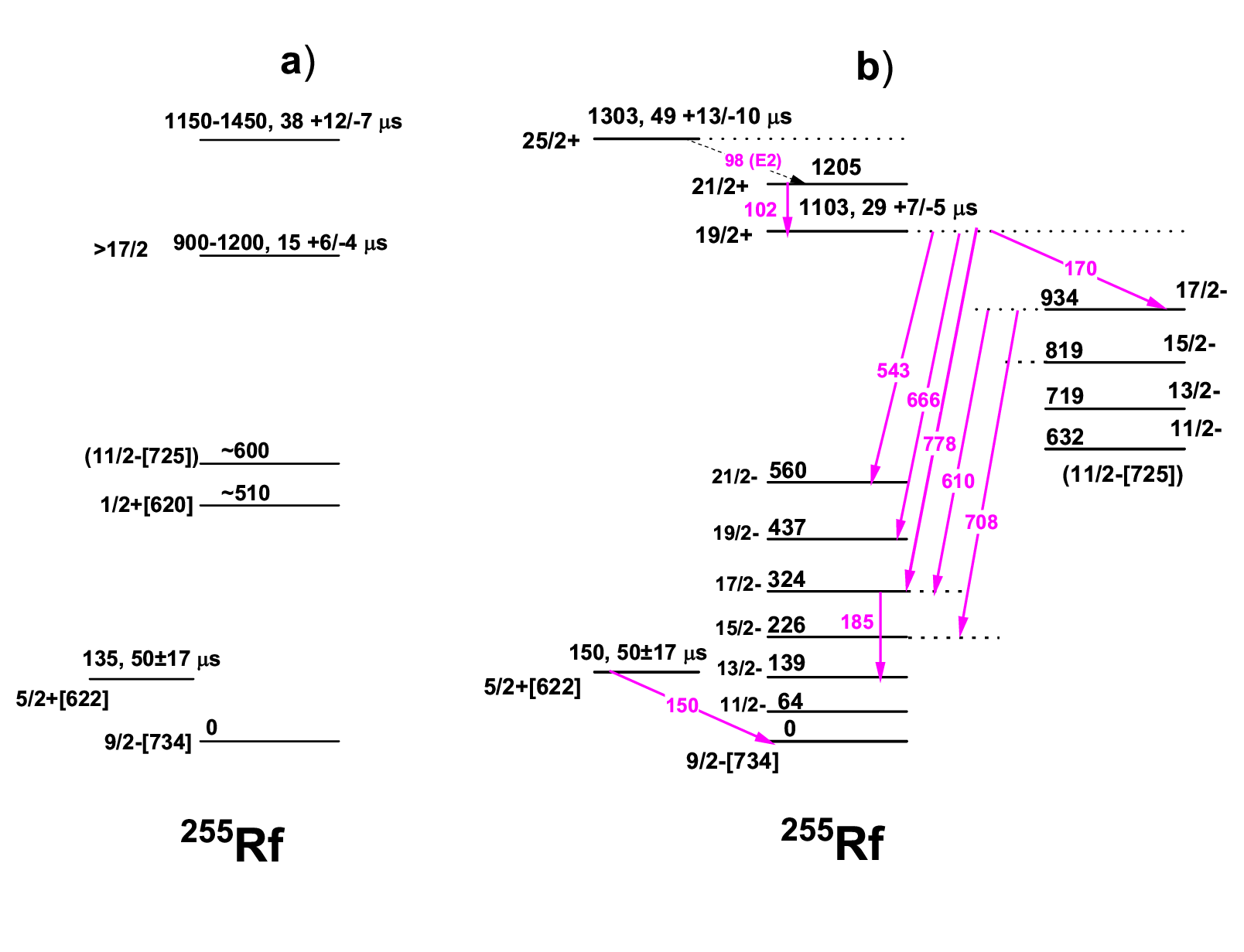}
	} 
	\caption{Partial level scheme of $^{255}$Rf and decay schemes of $^{255m1,255m2,255m3}$Rf as suggested by 
		S. Antalic et al. \cite{AntH15}, P. Mosat et al. \cite{Mosat20} (a) and R.Chakma et al.\cite{Chakma23} (simplyfied).}
	
	\label{fig:14}       
\end{figure*}

\begin{figure*}
	\centering
	\resizebox{0.75\textwidth}{!}{
		\includegraphics{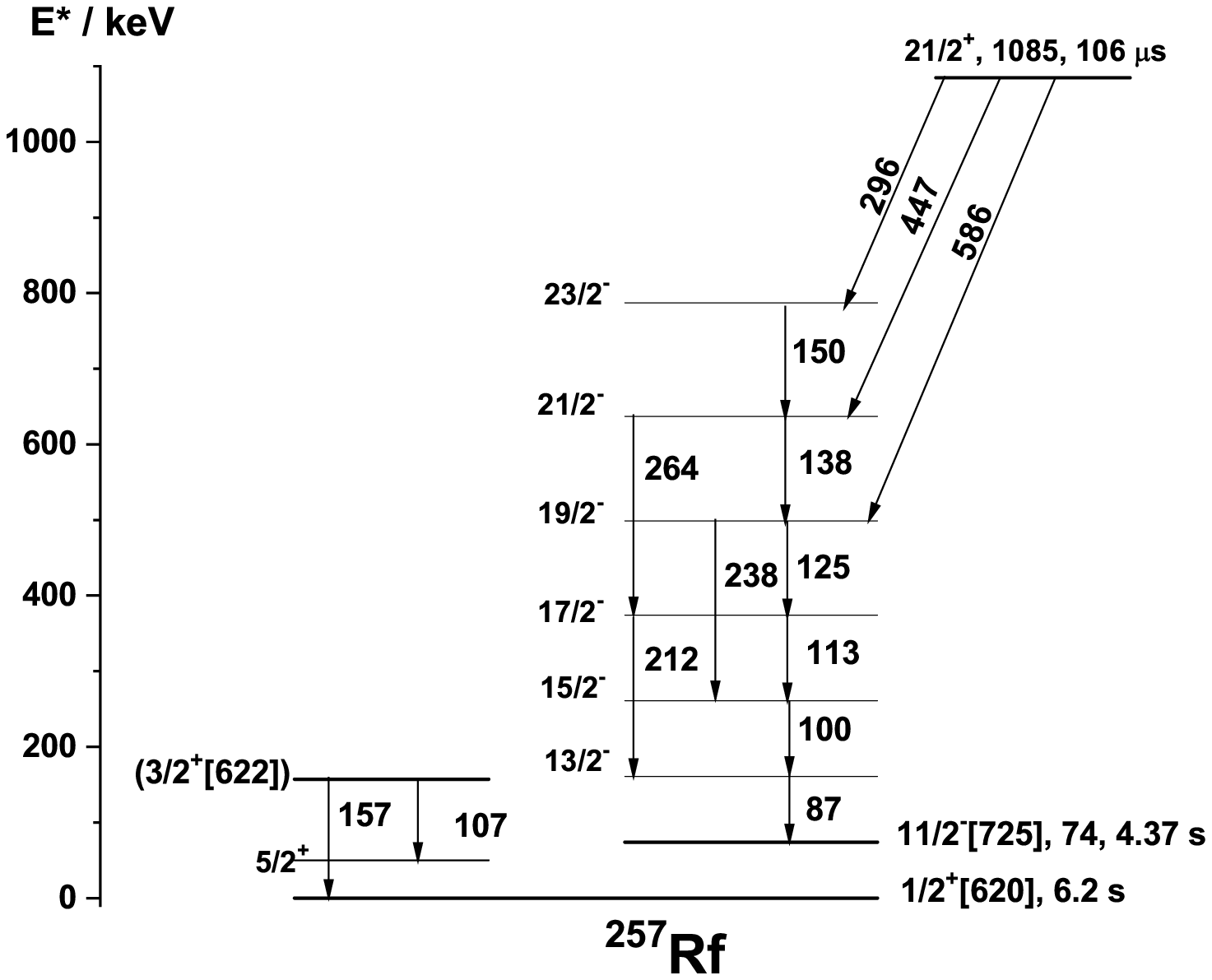}
	} 
	\caption{Partial level scheme of $^{257}$Rf and decay scheme of $^{257m2}$Rf based on the results of \cite{Riss13,HaL22,StH10}.
	Only clearly observed $\gamma$ transitions are given.}
	
	\label{fig:15}       
\end{figure*}

For the lower lying isomer $^{255m2}$Rf a half-life of T$_{1/2}$ = 29$^{+7}_{-4}$ $\mu$s was given and the excitation energy was 
settled at E$^{*}$ = 1103 keV. Spin and configuration of K$^{\pi}$\,=\,19/2$^{+}$ and
$\pi$1/2$^{-}$[521]$\downarrow$ $\otimes$ $\pi$9/2$^{+}$[624]$\uparrow$ $\otimes$ $\nu$9/2$^{-}$[734]$\uparrow$ were the same as in \cite{Mosat20}.
For the higher lying isomer a half-life and an excitation energy of  T$_{1/2}$ = 49$^{+13}_{-10}$ $\mu$s and  E$^{*}$ = 1238 keV were given.
Spin and configuration of I$^{\pi}$ = 25/2$^{+}$ and
$\nu$9/2$^{-}$[734]$\uparrow$ $\otimes$ $\pi$7/2$^{-}$[514]$\downarrow$ $\otimes$ $\pi$9/2$^{+}$[624]$\uparrow$ were suggested.\\
To place the observed $\gamma$ into the level scheme of $^{255}$Rf calculations of the bands built up on the 9/2$^{-}$[734] ground-state and the 
11/2$^{-}$[725] Nilsson level placed at E$^{*}$ = 632 keV were performed (see \cite{Chakma23} for details). The bands shown in fig. 14b represent
the results of these calculations.\\

{\bf K isomer in $^{257}$Rf}\\
The existence of a K isomer in $^{257}$Rf with a half-life of T$_{1/2}$ = 109 $\pm$13 $\mu$s was first mentioned by H.B. Jeppesen et al. \cite{Jepp09}
in a study mainly devoted to search for a K isomer in $^{256}$Rf (see sect. 6.4) performed at the BGS separator at LNBL, Berkeley.
But no further details were given.\\
Short time later results from an experiment performed at ANL Argonne were reoprted by J. Qian et al. \cite{Qian09}. They observed 
in bombardments of $^{208}$Pb with $^{50}$Ti 39 events of the type ER - CE - $\alpha$($^{257}$Rf) and measured a half-life of 
T$_{1/2}$ = 160$^{+42}_{-31}$ $\mu$s. Based on calculations possible spin and configuration were considered.
They favoured a 3-quasiparicle - state with K$^{\pi}$ = 27/2$^{+}$ 
($\pi$9/2$^{+}$[624]$\uparrow$ $\otimes$ $\pi$7/2$^{-}$[514]$\downarrow$ $\otimes$ $\nu$11/2$^{-}$[725]$\downarrow$)
but did not exclude a 3- quasiparticle - state with K$^{\pi}$ = 21/2$^{+}$ 
($\pi$9/2$^{+}$[624]$\uparrow$ $\otimes$ $\pi$1/2$^{-}$[521]$\downarrow$ $\otimes$ $\nu$11/2$^{-}$[725]$\downarrow$). \\
A further study, including also $\gamma$ ray measurements was performed by J.G. Berryman et al. \cite{Berry10} at the BGS at LNBL, Berkeley
who als observed some $\gamma$, specifically two 'high' energy ones of E\,=\,446, 585 keV.
They measured a half-life of T$_{1/2}$ = 134.9$\pm$7.7 $\mu$s. In \cite{Berry10} it  was shown that the decay of the isomeric state populates
the rotational band built up on the 11/2$^{-}$[725]$\uparrow$ Nilsson level in $^{257}$Rf, as already supposed in \cite{Qian09}.
Based on the L\"obner - systematics \cite{Loeb68} possible K difference of $\Delta$K = 4,5,6 were estimated. With this finding
as possible isomeric configurations 
K$^{\pi}$ = 21/2$^{+}$ 
($\pi$9/2$^{+}$[624]$\uparrow$ $\otimes$ $\pi$1/2$^{-}$[521]$\downarrow$ $\otimes$ $\nu$11/2$^{-}$[725]$\uparrow$)
or K$^{\pi}$ = 23/2$^{-}$ 
($\pi$7/2$^{-}$[514]$\downarrow$ $\otimes$ $\pi$5/2$^{-}$[512]$\downarrow$ $\otimes$ $\nu$11/2$^{-}$[725]$\uparrow$)
were considered\footnote{no parity assignemt was given in \cite{Berry10}.}.
The $\gamma$ lines of E = 446 keV and E = 586 keV were assigned as decays from the isomeric state into the I$^{\pi}$ = 23/2$^{-}$ and
I$^{\pi}$ = 21/2$^{-}$ states of the rotational band built up on the 11/2$^{-}$[725]$\uparrow$  Nilsson level.\\
Data of higher quality were obtained in a follow up experiment at the BGS at LNBL Berkeley by J. Rissanen et al. \cite{Riss13}.
The results of \cite{Berry10} were confirmed, but due to better statistics an enhanced decay scheme could be established.
It is shown in Fig. 15, energies are however modified with respect to the recently more precisely measured excitation energy
of the 11/2$^{-}$[725]$\uparrow$ level in $^{257}$Rf (E$^{*}$ = 74 keV) \cite{HaL22}. 
Spin and parity of the isomeric level were established as K$^{\pi}$ = 21/2$^{+}$, a half-life of T$_{1/2}$ = 106$\pm$6 $\mu$s was given.
The excitation energy is settled at E$^{*}$ = 1085 keV.

\vspace*{1cm}

\section{7. Discussion}

\begin{figure*}
	\resizebox{0.75\textwidth}{!}{
		\includegraphics{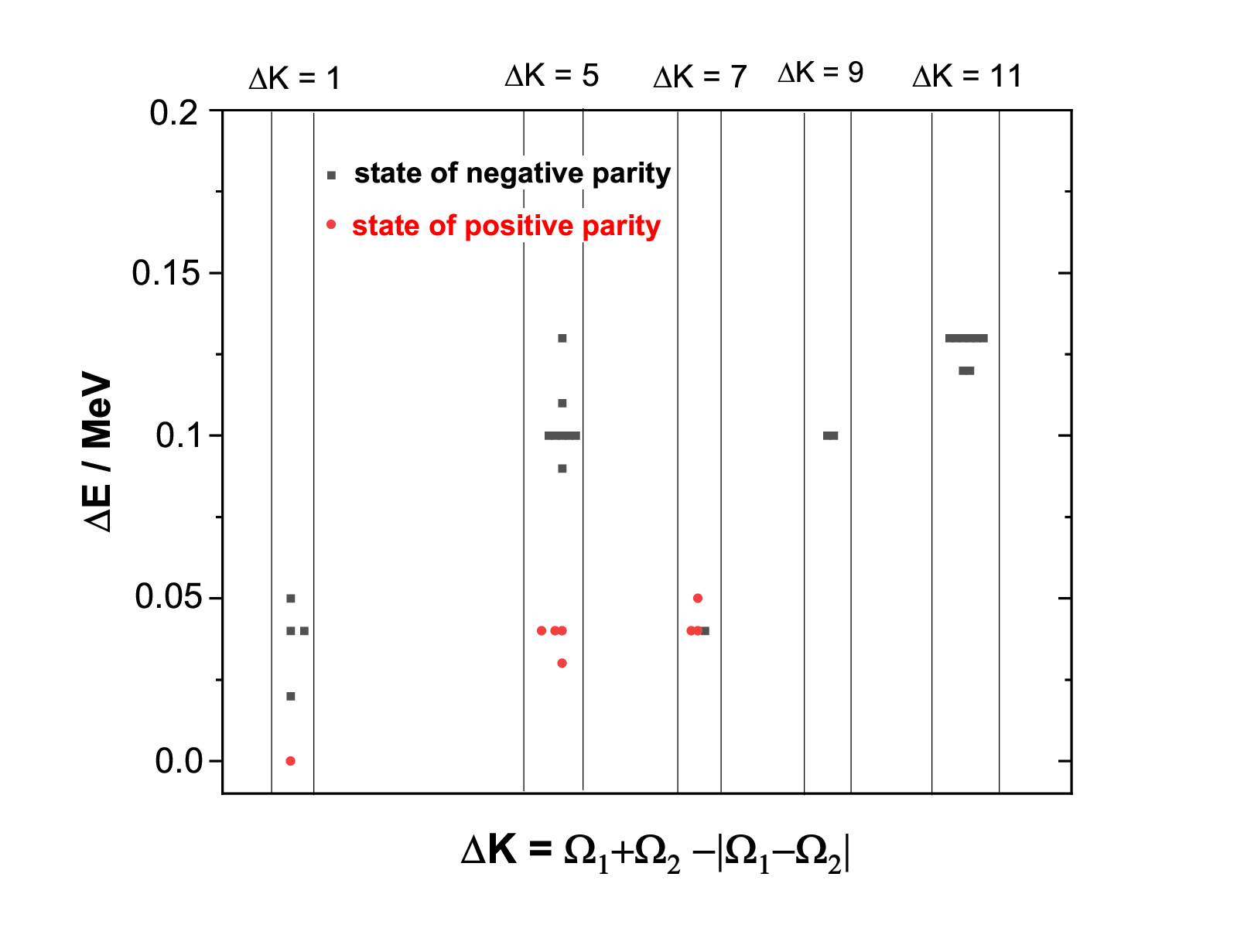}
	} 
	\caption{ Energy differences between states of K = $\Omega_{1} + \Omega_{2}$ and K = $\mid$$\Omega_{1} - \Omega_{2}$$\mid$ in 2- quasiparticle - configurations
		with anti-parallel spin projections according to the calculations in \cite{Prass15}; full squares (black): states of negative parity; full dots (red):
		states of positive parity.}
	
	\label{fig:16}       
\end{figure*}

\subsection{\bf 7.1 General considerations on occurence and the structure of K isomers in even - even nuclei}
Generally spoken, for the occurence of K isomers it is necessary
that below states of 'high' K values only states of sufficiently low
K values exist, so that decay of the former is delayed by strong
K hindrance. As according to the Gallagher rule (sect. 2.5) states
of different spin projection tend to maximize their K values, i.e.
states of K = $\Omega_{1}$ +$\Omega_{2}$ have the lowest excitation 
energies, such states can be assumed as ideal candidates for K isomers.
Indeed, as seen in table 2 all 2-quasiparticle - K isomers in even - even nuclei 
where the configuration is laid down with some certainty are 
configurations of spin-up ($\uparrow$) and spin-down ($\downarrow$)
states. 
However, things may be not so straightforward. In cases of coupling
states of low and high spins (e.g. $\nu$1/2$^{+}$[620]$\uparrow$ $\otimes$  
$\nu$11/2$^{-}$[725]$\uparrow$, with K = $\mid$$\Omega_{1}$ - $\Omega_{2}$$\mid$ = 5)
also states K = $\mid$$\Omega_{1}$ - $\Omega_{2}$$\mid$ may result in a K isomer,
if no states of sufficiently low K difference lie between them and the
K\,=\,0 ground state.\\
Calculations of excitation energies of K\,=\,$\Omega_{1}$\,+\,$\Omega_{2}$
and K\,=\,$\mid$$\Omega_{1}$\,-\,$\Omega_{2}$$\mid$ states (in the following denoted as
'K+' and 'K-' for better presentation) in the region of heaviest nuclei
(Z\,=\,104\,-\,110, N\,=\,160\,-\,168) have been investigated by 
V. Prassa et al. \cite{Prass15}. In all cases the K+ - states are located below 
the K- - states, irrespective of the spin - projections, $\uparrow$$\uparrow$,
$\downarrow$$\downarrow$ or $\uparrow$$\downarrow$. Differences in the excitation 
energy are, however, quite small. For cases with parallel spin - projections
($\uparrow$$\uparrow$, $\downarrow$$\downarrow$) one obtains values 
of $\Delta$E$^{*}$(K+,K-)\,$\le$\,0.020 MeV, with a mean value 
$\Delta$E$^{*}$(K+,K-)\,\,=\,0.018$\pm$0.005 MeV. In the case of 
anti-parallel spin - projections ($\uparrow$$\downarrow$)
the straggling of the energy differences ist larger, one obtains
values of $\Delta$E$^{*}$(K+,K-)\,\,=\,(0-0.13) MeV, resulting
in a mean value $\Delta$E$^{*}$(K+,K-)\,=\,0.083$\pm$0.041 MeV.
As seen in fig. 16, the energy differences are stronly dependent on 
the K difference $\Delta$K; a trend of increasing energy differences 
with increasing $\Delta$K values is evident. But fig. 16 shows also 
another trend. At the same $\Delta$K values the energy differences 
are larger for states with negative parity (full squares) than for states of positive
parity (full dots).

\subsection{\bf 7.2 Structure of K isomers in even - even nuclei - theoretical predictions}
The identification of about two dozens of new K isomers in the heavy actinide and transacitnide region (Z\,$\ge$\,100) within the last two decades 
asks for investigation of similarities and systematics in their structure and decay as well as comparison with theoretical calculations
as far as available. As the 2-quasiparticle states in even - even nuclei are due to breaking of a pair of nucleons and exciting them
into different levels, one may expect that they dependent on the single particle structure (Nilsson levels) and 
thus the K isomers 
exhibit similarities in their structure along the isotone lines in case of 2-quasi-neutron states, similar to even Z odd-mass nuclei which show similarities in their nuclear structure along the
isotone lines (N = const.). Indeed such a behavior is indicated
by similar decay pattern of the K isomers in N\,=\,150 isotones, as shown in fig. 8. But unfortunately detailed decay data a scarce for most of the cases. 
For odd-mass nuclei the situation is more difficult, as the K isomers are produced by coupling of a 2-quasiparticle state to a single particle 
state, so more possibilities are available. Due to this difficulties and the lack of theoretical calculations we will do with the discussion
of selected cases in even-even nuclei. Concerning theoretical results we will do with presenting and comparing them with experimental results
and omit a detailed discussion of the underlying theoretical framework, as this would go beyond the scope of this paper. \\
\\
\\
\begin{figure*}
	\centering
	\resizebox{0.85\textwidth}{!}{
		\includegraphics{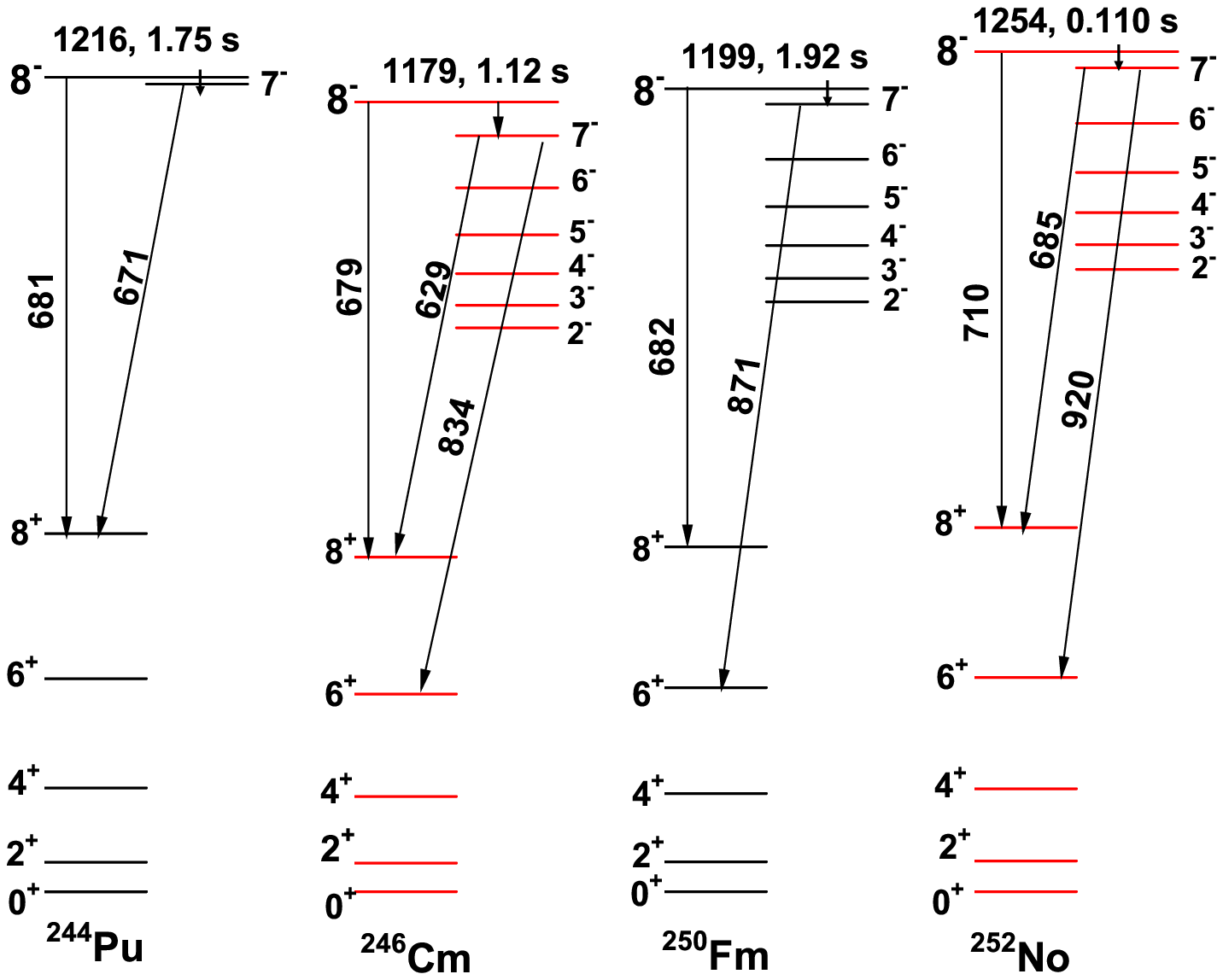}
	} 
	\caption{Simplified decay schemes of K isomers in N = 150 isotones $^{244}$Pu,  $^{246}$Cm,  $^{250}$Fm, and $^{252}$No. }
	\label{fig:17}       
\end{figure*}
\\
\\
\\
\begin{figure*}
	\centering
	\resizebox{0.75\textwidth}{!}{
		\includegraphics{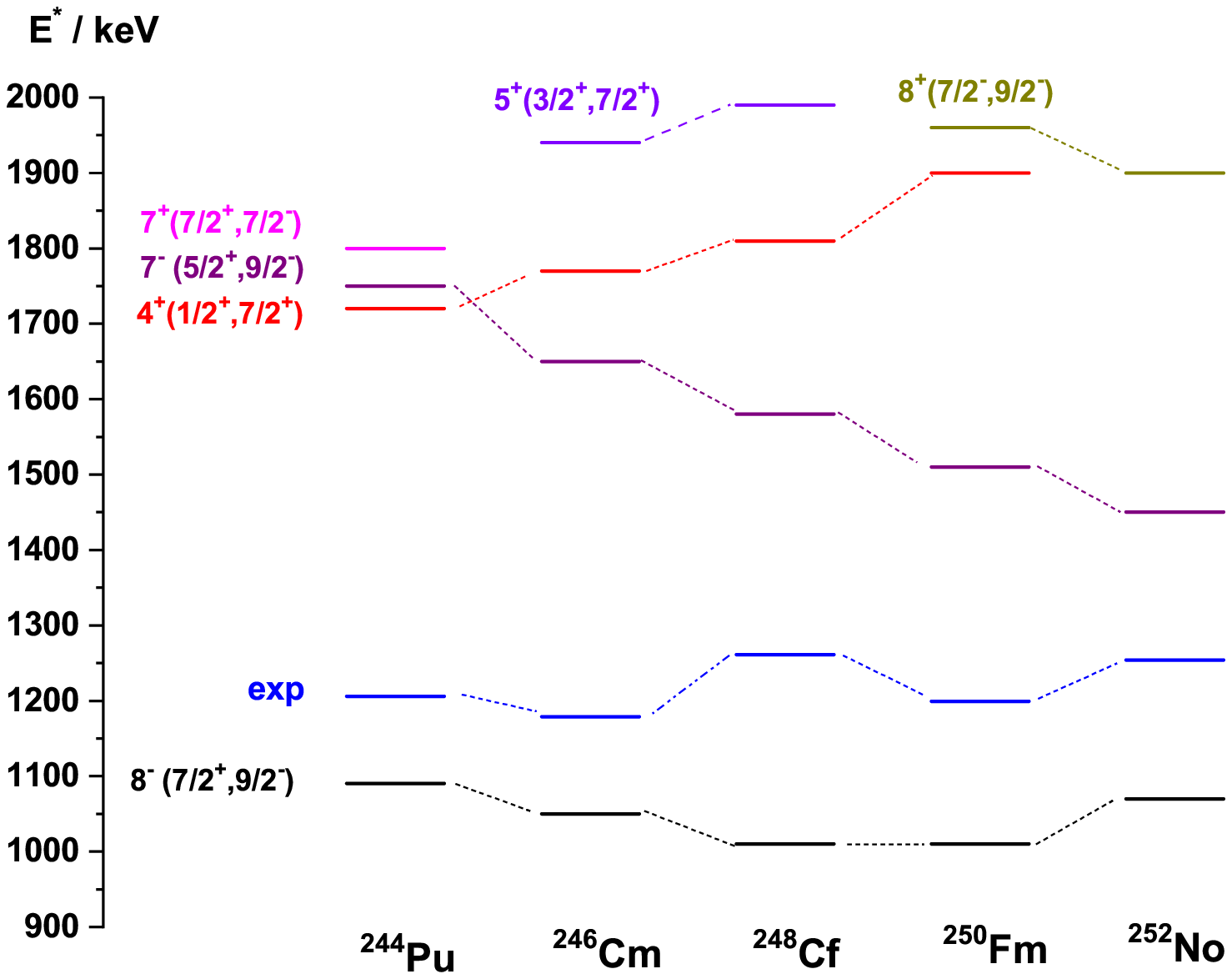}
	} 
	\caption{Comparison of predicted \cite{Dela06} excitation energies of 2-quasi-neutron - states in N = 150 isotones with 
		the experimental values for the K$^{\pi}$ = 8$^{-}$ isomers. }
	\label{fig:18}       
\end{figure*}
\\
\\
\begin{figure*}
\centering
\resizebox{0.75\textwidth}{!}{
\includegraphics{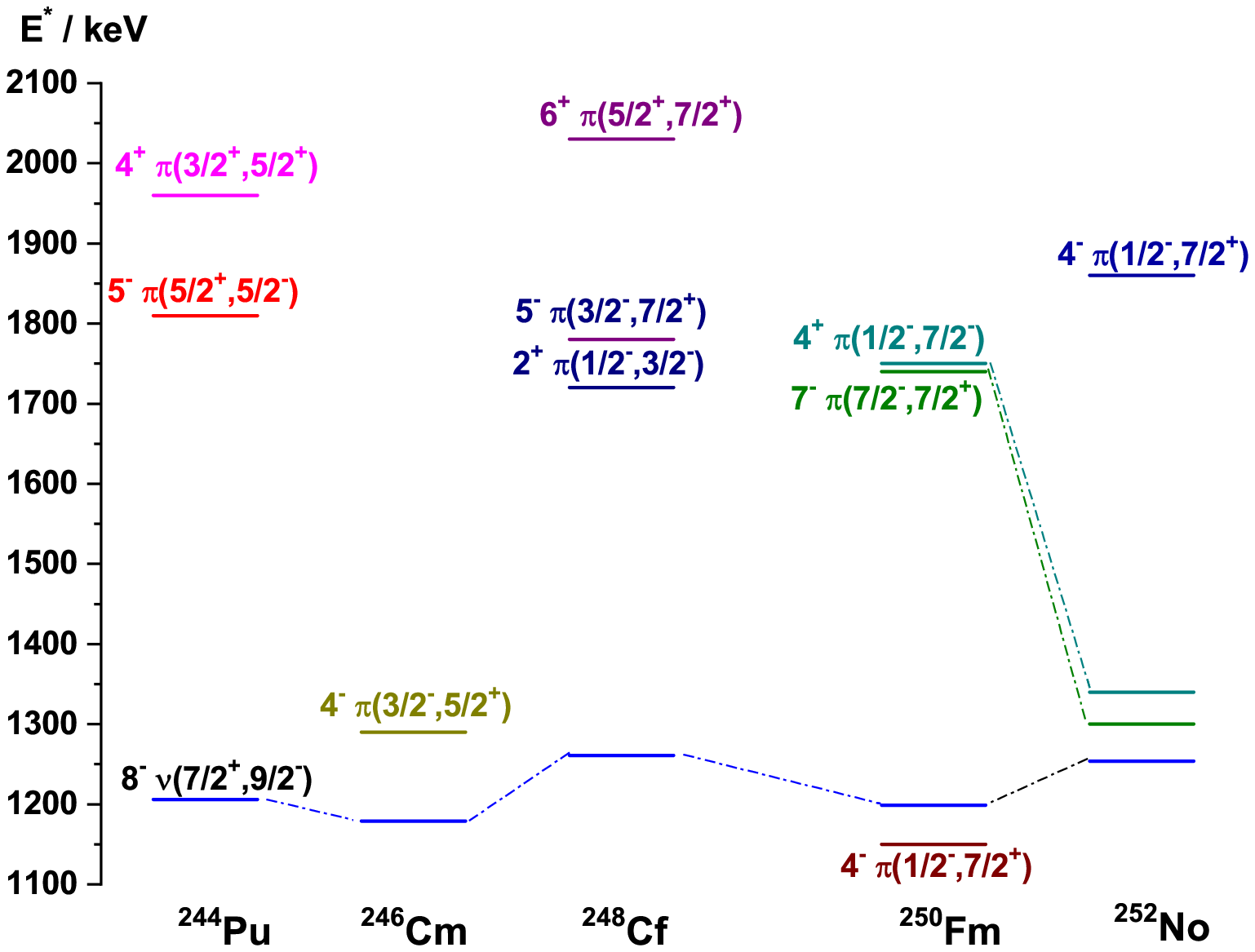}
} 
\caption{Comparison of predicted \cite{Dela06} excitation energies of 2-quasi-proton - states in N = 150 isotones with 
the experimental values for the K$^{\pi}$ = 8$^{-}$ isomers. }
\label{fig:19}       
\end{figure*}
\\
\\
\begin{figure*}
\centering
\resizebox{0.75\textwidth}{!}{
\includegraphics{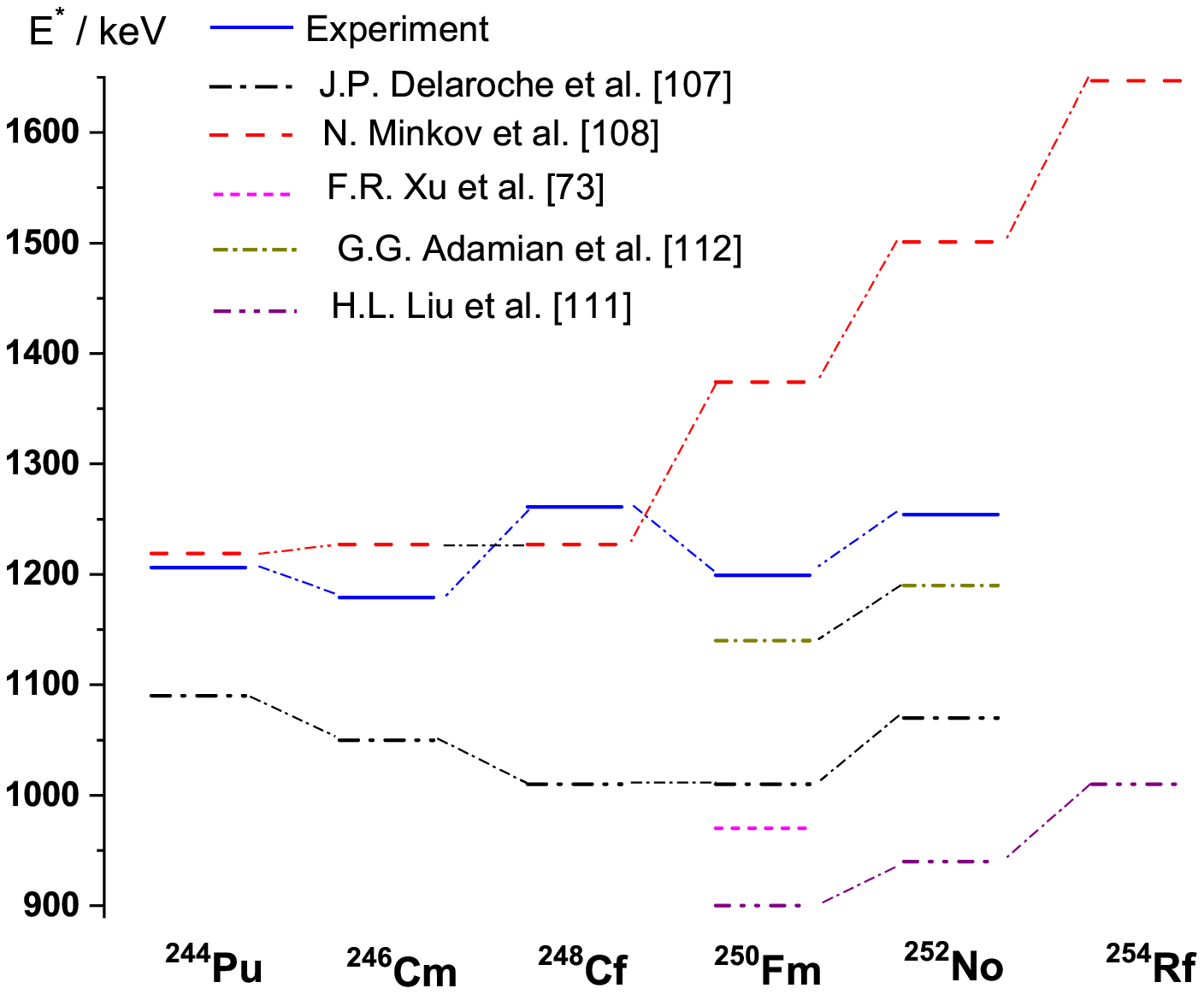}
} 
\caption{Comparison of predicted excitation energies of for the K$^{\pi}$ = 8$^{-}$ isomers in N = 150 isotones with 
the experimental values . }
\label{fig:20}       
\end{figure*}
\\

\vspace{-3.5cm}
{\bf 7.21 K isomers in N\,=\,150 isotones}\\
Detailed information on K isomers which allow for a systematic
examination of structure and decay of K isomers
are available for the even-Z N\,=\,150 isotones. The similar decay
schemes of $^{246m}$Cm, $^{250m}$Fm and $^{252m}$No are shown in fig. 8,
but also differences in the intensities of the relevant transitions
are pointed to. For sake of the further discussion we present simplified
decay schemes, showing only the 'most relevant' transitions for 
$^{244m}$Pu (for which a less detailed decay scheme was published),
$^{246m}$Cm, $^{250m}$Fm and $^{252m}$No in fig. 17. For all four cases the decay 
of the isomers occurs via two paths: a) decay of the isomer into the
I$^{\pi}$\,=\,8$^{+}$ level of the ground state rotational band. It
was, however, already remarked that in the case of $^{252m}$No this 
transition is quite weak, contrary to the situation for $^{244m}$Pu, 
$^{246m}$Cm and $^{250m}$Fm; b) decay of the isomer into the 
I$^{\pi}$\,=\,7$^{-}$ level of an octupole vibrational band with 
the bandhead I$^{\pi}$\,=\,2$^{-}$. Strong decay intensities are
observed into the I$^{\pi}$\,=\,8$^{+}$, 6$^{+}$ members of the 
ground-state rotational band, but also strong intraband transitions 
followed by decay from lower menbers of the octupole band into
lower members of the ground state rotational band are reported
such as transitions 5$^{-}$ $\rightarrow$ 4$^{+}$ and 
2$^{-}$ $\rightarrow$ 2$^{+}$ in $^{250m}$Fm \cite{Green08} and 
$^{252m}$No \cite{Sul07}. All these findings hint to the same
structure of the isomers and presently there is common agreement
on a 2-quasi-neutron- state $\nu$9/2$^{-}$[734]$\uparrow$ $\otimes$ 
$\nu$7/2$^{+}$[624]$\downarrow$ resulting in K$^{\pi}$\,=\,8$^{-}$. \\
Calculations of low lying 2 - quasiparticle - states in even-Z 
N\,=\,150 isotones have been performed by a couple of authors \cite{Dela06,AdaA10,Liu14,Minkov22,XuZ04}.
In fig. 18 the results from J.-P. Delaroche et al. \cite{Dela06} for 2-quasi-neutron - states 
are compared with the
experimental data for the N\,=\,150 isotones in the range Z\,=\,94\,-\,102.
In all cases the K$^{\pi}$\,=\,8$^{-}$ state is the lowest lying one. Other 
states are predicted at E$^{*}$ $>$ 1500 keV. The K$^{\pi}$\,=\,8$^{-}$ states,
however, are predicted at somewhat lower excitation energies with
$\Delta$E(exp,theo)\,=\,115-190 keV, except for $^{248}$Cf where the difference is 
$\Delta$E(exp,theo)\,=\,250 keV. But here one should keep in mind, that the
excitation energy of the isomer is not well established (see sect. 6.3).
In fig. 19 we compare the 2-quasi-proton - states predicted by J.-P. Delaroche et al.\cite{Dela06}
below E$^{*}$ = 2 MeV
with the experimental results. In general the predicted 2-quasi-proton - states have different
configurations, so no 'common trend' is observed as in the case of 2-quasi-neutron - states.
On the other side this behavior is not unexpected, as the isotones differ in the number of protons
leading also to different Nilsson-levels (proton single particle states) at low excitation 
energies, in contrast to nuclei within isotone lines, which are known to have similar
structure (similar Nilsson single particle levels) at low excitation 
energies. In this way the similar structure of the K isomers in the N\,=\,150 isotones
supports their interpretation as 2-quasineutron states.\\
Calculations for K isomers in N\,=\,150 isotones (but in general not for all nuclei)
were also performed by F.R. Xu et al. \cite{XuZ04}, G.G. Adamian et al. \cite{AdaA10}, H.L. Liu et al. \cite{Liu14},
and N. Minkov et al. \cite{Minkov22}. F.R. Xu et al. consider as a possible configuration of the 
K isomer in $^{250}$Fm a 2-quasi-proton state of I$^{\pi}$\,=\,7$^{-}$ (configuration
$\pi$ 7/2$^{+}$[633]$\uparrow$ $\otimes$ $\pi$ 7/2$^{-}$[514]$\downarrow$) at a calculated 
excitation energy of E$^{*}$\,=\,1.01 MeV(Note: the paper of F.R. Xu et al. was published
before $^{252m}$No was discovered and before detailed spectroscopic data for $^{250m}$Fm 
were published.) In $^{252}$No the I$^{\pi}$\,=\,7$^{-}$ was settled at a considerably 
higher excitation energy of E$^{*}$$\approx$1.5 MeV. F.R. Xu et al., however, remark that
their calculations show that I$^{\pi}$\,=\,8$^{-}$ 2-quasi-neutron states (configuration
$\nu$ 9/2$^{-}$[734]$\uparrow$ $\otimes$ $\nu$ 7/2$^{+}$[613]$\downarrow$) exist
systematically in N\,=\,150 isotones at excitation energies around 1 MeV; but only for 
$^{250}$Fm a definite value of E$^{*}$\,=\,0.97 MeV is given.\\
In fig. 20 the experimental excitation energies of the K isomers in the N\,=\,150 isotones
are compared with the results of the different calculations for 
K$^{\pi}$\,=\,8$^{-}$ 2-quasi-neutron states. The calculations of J.P. Delaroche et al.
reproduce the trend of quite stable excitation energies quite well, but deliver in general
about 115\,-\,190 keV too low values. The calculations of N. Minkov et al. 
deliver quite stable excitation energies for $^{244}$Pu, $^{246}$Cm and $^{248}$Cf 
and reproduce the experimantal values very well within $\Delta$E\,=\,$\mid$20$\mid$ keV, 
but for the Z$>$98 isotones the calculations result in steeply increasing excitation energies
and deliver excitation energies too high by $\approx$175 keV for $^{250m}$Fm and $\approx$250 keV 
for $^{252m}$No. Quite satisfying agreement between experimental and calculated values is obtained
by G.G. Adamian et al.. F.R. Xu et al., and H.L. Liu et al. obtain too low excitation energies 
for the cases they consider.\\
The calculations of H.L. Liu et al. may hint to a change of the structure of the K isomers going from
Z\,=\,102 to Z\,=\,104. The drastic change in the half-lives from $^{252m}$No to $^{254m}$Rf has been 
discussed in  sect. 6.3. In \cite{David15} as a possible reason the authors gave a decrease of the hindrance
of the M1 decay branch from the K$^{\pi}$\,=\,8$^{-}$ isomer into the I$^{\pi}$\,=\,7$^{-}$ state of the 
octupole band (indeed we observe in $^{252m}$No a strong increase of the transition
K$^{\pi}$\,=\,8$^{-}$ $\rightarrow$ I$^{\pi}$\,=\,7$^{-}$ (or I$^{\pi}$\,=\,6$^{-}$) compared to the
E1 transition K$^{\pi}$\,=\,8$^{-}$ $\rightarrow$ I$^{\pi}$\,=\,8$^{+}$ state of the ground state 
rotational band compared to the lighter isotones) and/or that the K$^{\pi}$\,=\,8$^{-}$ isomer and 
the I$^{\pi}$\,=\,8$^{-}$ might be very close in energy leading to an accidential configuration mixing 
resulting in a shorter half-life. We want to remind here, an indication for this already may be the drastic 
decrease of the isomeric half-life from $^{250m}$Fm to $^{252m}$No and also the low intensity 
of the E1 - transition 8$^{-}$ $\rightarrow$ 8$^{+}$ (ground-state rotational band) compared to the M1 - transition
8$^{-}$ $\rightarrow$ 7$^{-}$ (octupole vibration band). \\
The calculations of H.L. Liu et al. result in clear separation of the K$^{\pi}$\,=\,8$^{-}$ isomeric
state (there denoted as $\nu^{2}$8$^{-}_{1}$) from other states, $\pi^{2}$7$^{-}$, $\nu^{2}$6$^{+}$
in $^{250m}$Fm or $\nu^{2}$6$^{+}$, $\pi^{2}$5$^{-}$ in $^{252m}$No. Contrary, in $^{254m}$Rf the
K$^{\pi}$\,=\,8$^{-}$ state lies close in energy to a K$^{\pi}$\,=\,5$^{-}$ 2-quasi-proton state 
(configuration $\pi$ 1/2$^{-}$[521]$\downarrow$ $\otimes$ $\pi$ 9/2$^{+}$[624]$\uparrow$)
and an K$^{\pi}$\,=\,8$^{-}$ 2-quasi-proton state 
(configuration $\pi$ 7/2$^{-}$[514]$\downarrow$ $\otimes$ $\pi$ 9/2$^{+}$[624]$\uparrow$).
Thus a change of the configuration of the isomeric state from $^{252m}$No to $^{254m}$Rf
could also be the reason for the drastic change in the half-lives.\\

\begin{figure*}
\centering
\resizebox{0.75\textwidth}{!}{
\includegraphics{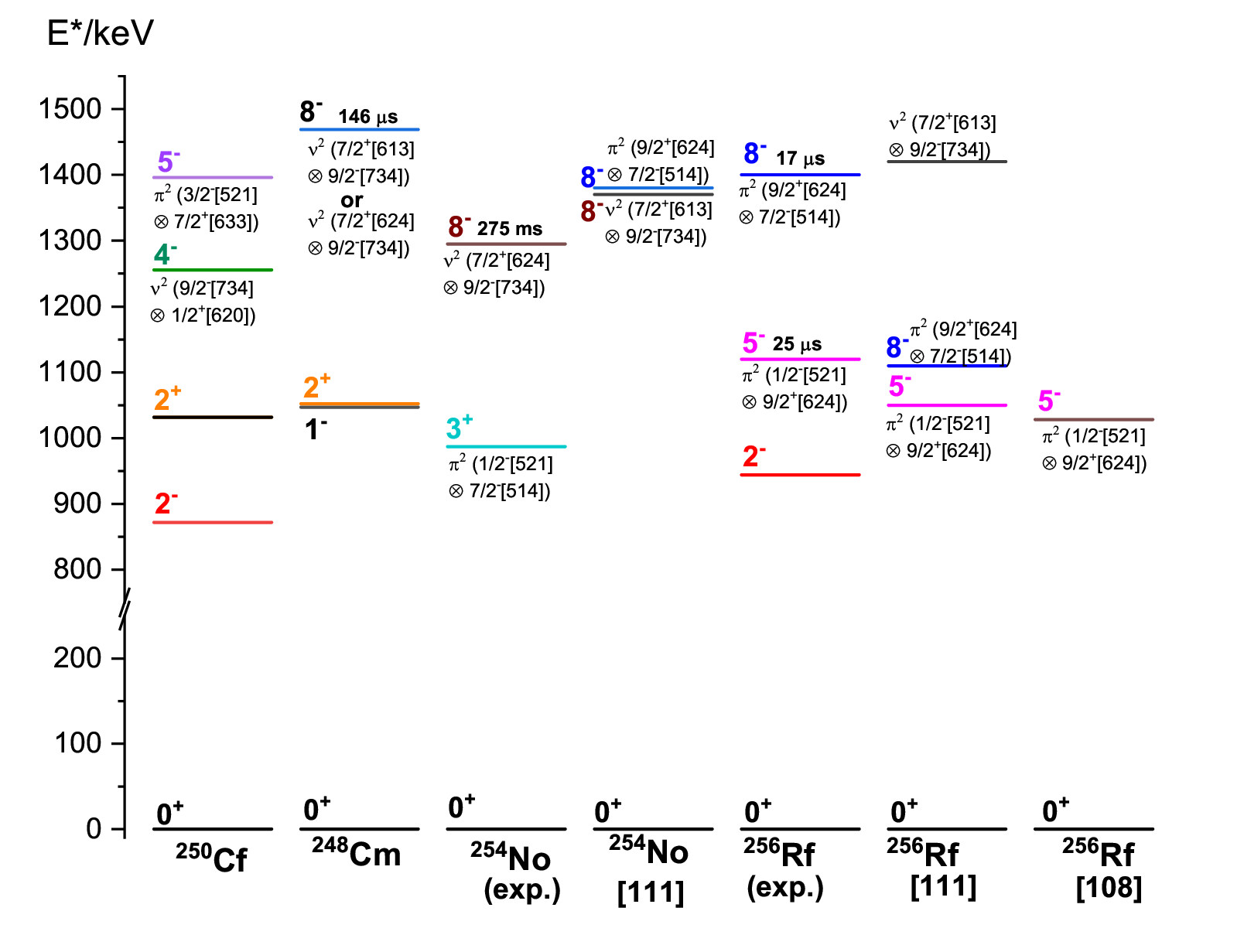}
}
\caption{Experimental and caluclated levels in $^{250}$Cf, $^{248}$Cm, $^{254}$No, $^{256}$Rf.}

\label{fig:21}       
\end{figure*}

{\bf 7.22 K isomers in N\,=\,152 isotones}\\
While for the 2-quasiparticle isomers in the N\,=\,150 isotones $^{244}$Pu, $^{246}$Cm, $^{250}$Fm, $^{252}$Fm, and $^{254}$Rf the isomeric state was attributed to the same configuration
the situation is completely different for the 2-quasiparticle isomers in the N\,=\,152 isotones $^{248}$Cm $^{254}$No and $^{256}$Rf.
The situation is shown in fig. 21.  For $^{248}$Cm and 
$^{254}$No each one 2-quasiparticle isomer was observed with spin and parity assigned as K$^{\pi}$ = 8$^{-}$; while in $^{254m1}$No the 
2-quasi-neutron configuration $\nu$7/2$^{+}$[624]$\downarrow$$\otimes$ $\nu$9/2$^{-}$[734]$\uparrow$ is preferred for $^{248}$Cm 
also the 2-quasi-neutron state  $\nu$7/2$^{+}$[613]$\uparrow$$\otimes$ $\nu$9/2$^{-}$[734]$\uparrow$ is considered \cite{Shir19}. Also decay paths
are different. While $^{248m}$Cm is interpreted to decay predominantly into the 8$^{+}$ and 6$^{+}$ members of a $\gamma$ vibrational band with an I$^{\pi}$ =  2$^{+}$
bandhead at E$^{*}$ = 1049 keV, $^{254m1}$No decays predominantly into the 7$^{+}$ member of the rotational band built up on a 2-quasi-proton K$^{\pi}$ = 3$^{+}$ state 
with a configuration $\pi$1/2$^{-}$[521]$\downarrow$ $\otimes$ $\pi$7/2$^{-}$[514]$\downarrow$. Also half-lives differ by a factor of roughly 2000. 
In $^{256}$Rf two isomeric states at E$^{*}$ $<$ 1.5 MeV are reported. A lower lying tentative K$^{\pi}$ = 5$^{-}$ state 
($\pi$1/2$^{-}$[521]$\downarrow$$\otimes$$\pi$9/2$^{+}$[624]$\uparrow$ configuration)
at E$^{*}$ $\approx$ 1.120 MeV
and higher lying tentative K$^{\pi}$ = 8$^{-}$ state ($\pi$7/2$^{-}$[514]$\downarrow$$\otimes$$\pi$9/2$^{+}$[624]$\uparrow$) at 
E$^{*}$ $\approx$ 1.4 MeV. While the K$^{\pi}$ = 8$^{-}$ isomer decays into the K$^{\pi}$ = 5$^{-}$ one (or members of the rotational
band built up on it), the K$^{\pi}$ = 5$^{-}$ isomer is interpreted to decay into a band built up on a
I$^{\pi}$ = 2$^{-}$ octupole vibrational state located at E$^{*}$ $\approx$ 944 keV. Indeed such a state has been identified
in the N\,=\,152 isotone $^{250}$Cf \cite{Fire96}.\\
H.L. Liu et al. \cite{Liu14}  predict for both isotopes $^{254}$No and $^{256}$Rf 2-quasi-proton- and 2-quasi-neutron states 
E$^{*}$ $<$ 1.5 MeV, namely 
K$^{\pi}$ = 8$^{-}$($\nu$) ($\nu$7/2$^{+}$[613]$\downarrow$$\otimes$ $\nu$9/2$^{-}$[734]$\uparrow$),
K$^{\pi}$ = 8$^{-}$($\pi$) ($\pi$9/2$^{+}$[624]$\uparrow$$\otimes$ $\pi$7/2$^{-}$[734]$\downarrow$), and
K$^{\pi}$ = 5$^{-}$($\pi$) ($\pi$1/2$^{-}$[521]$\downarrow$$\otimes$$\pi$9/2$^{+}$[624]$\uparrow$)(not shown in fig. 21. for better presentation)
In $^{254}$No all three states are predicted very close in energy E$^{*}$ $\approx$ 1.38 MeV and the authors
prefer the K$^{\pi}$ = 8$^{-}$($\pi$) state as the isomeric one.\\
In $^{256}$Rf the K$^{\pi}$ = 5$^{-}$($\pi$) (E$^{*}$\,=\,1.05 MeV) and  K$^{\pi}$ = 8$^{-}$($\pi$) (E$^{*}$\,=\,1.11 MeV)
are quite close in energy, the  K$^{\pi}$ = 8$^{-}$($\nu$) is predicted at a significantly higher energy E$^{*}$\,$\approx$\,1.5 MeV),
leading to preferation of the K$^{\pi}$ = 5$^{-}$($\pi$) state as the lower isomeric one.\\

Calculations of N. Minkov et al. \cite{Minkov22} predict for $^{254}$No the 2-quasi-proton -state 
K$^{\pi}$ = 8$^{-}$ ($\pi$) ($\pi$9/2$^{+}$[624]$\uparrow$$\otimes$ $\pi$7/2$^{-}$[514]$\downarrow$)
at E$^{*}$ = 1.914 MeV, i.e. 639 keV above the experimental value. For $^{256}$Rf they predict the 
K$^{\pi}$ = 5$^{-}$ state at E$^{*}$\,=\,1.028 MeV and the K$^{\pi}$ = 8$^{-}$ state at E$^{*}$\,=\,1.748 MeV
about 350 keV above the assumed value.\\
J.-P. Delaroche et al. \cite{Dela06} who predict the excitation energies of the K$^{\pi}$ = 8$^{-}$ 
2-quasi-neutron isomers in the N\,=\,150 isotones sufficiently well, do not predict either a 2-quasi-neutron - state 
or a 2-quasi-proton - state of  K$^{\pi}$ = 8$^{-}$ 
in $^{254}$No at
an excitation energy  E$^{*}$\,$<$\,2 MeV.
The drastic decrease in the half-lives from $^{254m1}$No to $^{256m1,256m2}$Rf can thus be understood
due to the lower K - difference of the initial and final states involved in the decay. 
But again we here want to point to the quite low half-life of $^{248m}$Cm with respect to the high difference 
of $\Delta$K\,=\,6 between the assumed isomeric state and the $\gamma$ vibrational state.
This could be a hint, that probably the isomer might be a state of a lower K value.
\\
\\
{\bf 7.23 K isomer in $^{270}$Ds}\\
As mentioned in sect. 6.4, the K isomer $^{270m}$Ds was identified by its $\alpha$ decay \cite{HoH01}. The excitation energy (E$^{*}$ = 1.13 MeV) was estimated 
on the basis of the highest observed $\alpha$ decay energy attributed to the isomer (12.147 MeV) and the mean $\alpha$ decay energy of the ground state (11.03$\pm$0.05 MeV). On the basis of the hindrance factor for the $\alpha$ decay a spin difference between ground state (0$^{+}$) and isomeric state of $\Delta$I = 10$\pm$2 $\hbar$ was estimated. Calculations resulted in possible K isomeric states of K$^{\pi}$ = 9$^{-}$, 10$^{-}$,
while K$^{\pi}$ = 9$^{-}$ was slightly preferred in \cite{HoH01}. Such a procedure, however, is not unambiguous as the preferred K assignment was 
mainly based on the hindrance of the $\alpha$ decay due to the angular momentum difference between the initial state and the final state,
assumed as the ground state. But even for the $\alpha$ decays of highest energy one cannot assume a priori decay into the ground state. Even
only considering hindrance of $\alpha$ decay due to angular momentum difference (leaving structural hindrance and hindrance due to parity change aside) one has to respect decreasing hindrance due to decreasing angular momentum difference and increase of $\alpha$ decay half-lives due to
lower Q - values for decays into excited members of the ground-state rotational band. The situation is shown in table 9.

\begin{figure*}
	\centering
	\resizebox{0.50\textwidth}{!}{
		\includegraphics{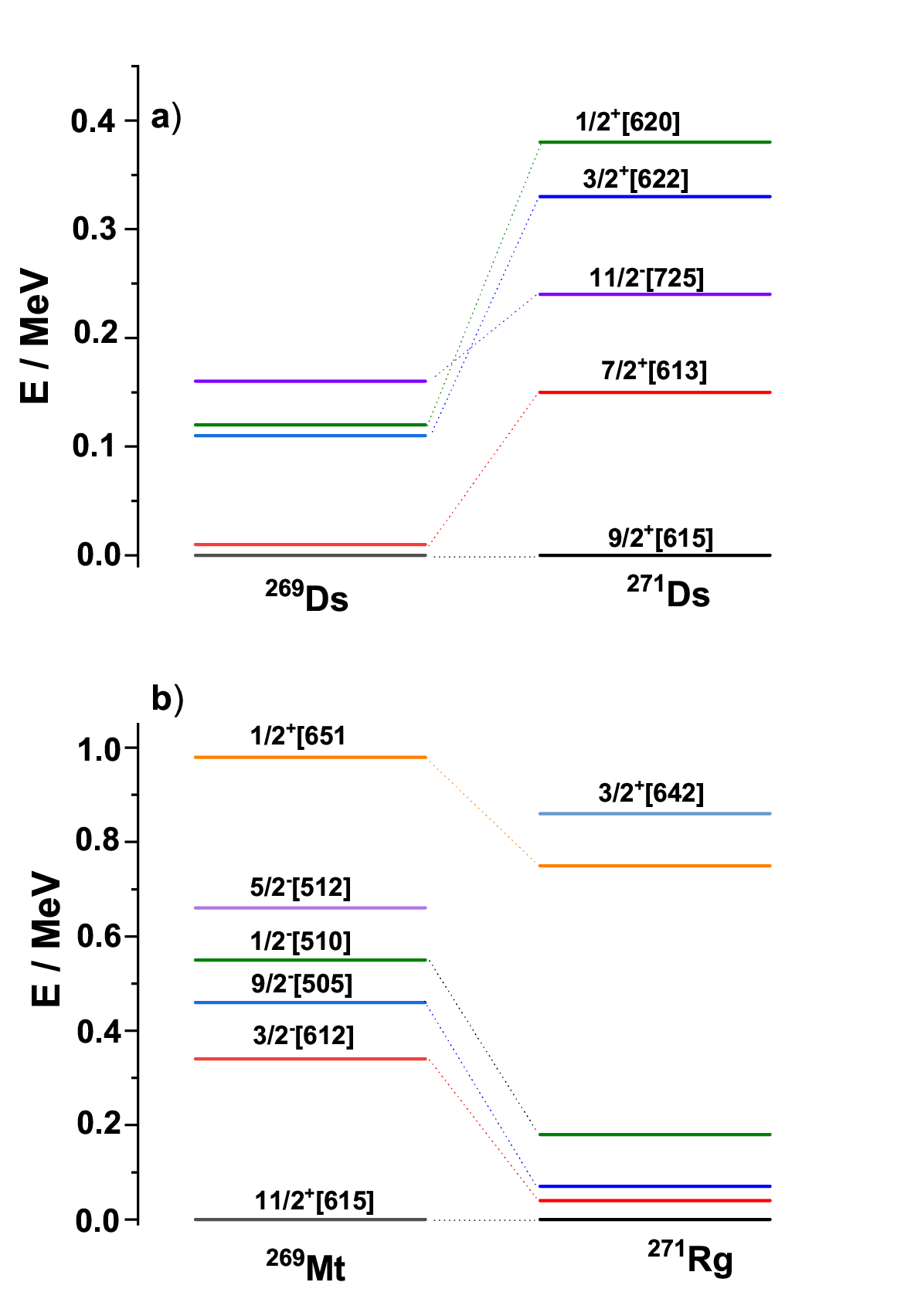}
	} 
	\caption{Predicted low lying Nilsson levels, a) neutron single particle levels in $^{269,271}$Ds \cite{ParS05}; b) proton single particles levels in $^{269}$Mt and $^{271}$Rg \cite{Park04}. }
	
	\label{fig:22}       
\end{figure*}

\begin{figure*}
	\resizebox{0.75\textwidth}{!}{
		\includegraphics{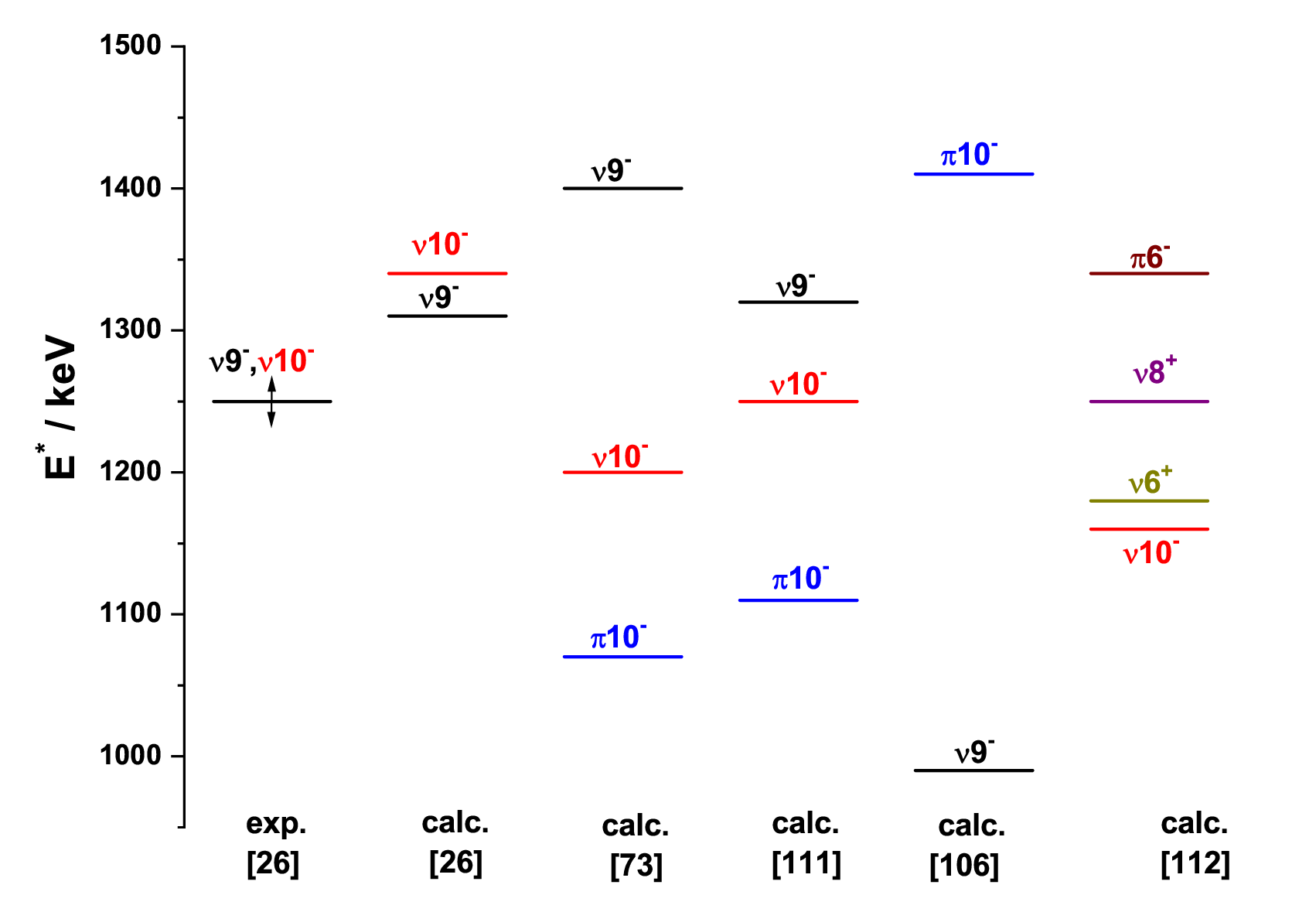}
	} 
	\caption{Comparison of predicted low lying state of high K numbers in $^{270}$Ds 
		\cite{HoH01,XuZ04,Liu14,Prass15,AdaA10} with the tentative experimental assignement \cite{HoH01}. In the case of \cite{Liu14} levels of 
		K$\le$8 in the range of E$^{*}$\,=\,(1.0-1.5) MeV are omitted for better presentation.}
	
	\label{fig:23}       
\end{figure*}

\begin{table}
	\caption{Estimated hindrance factors of $\alpha$ decays of $^{270m}$Ds into the ground-state rotational band of $^{266}$Hs}
	\label{tab:1}       
	\begin{tabular}{cccccccc}
		\hline\noalign{\smallskip}
		Level & E$_{\alpha}$/MeV & T$_{\alpha}$/$\mu$s & $\Delta$L & HF($\Delta$L) & T$_{\alpha}$/T$_{\alpha}$(gs) & HF(m) & T$_{\alpha}$(m)/ms\\
		\hline\noalign{\smallskip}     
		0$^{+}$ & 12.15 & 37 & 9 & 936 & 1 & 936 & 35 \\
		2$^{+}$ & 12.105 & 44 & 7 & 71 & 1.19 & 84.4 & 3.1 \\
		4$^{+}$ & 12.005 & 71 & 5 & 9.8 & 1.91 & 18.7 & 0.69 \\
		6$^{+}$ & 11.845 & 148 & 3 & 2.5 & 4.0 & 10 & 0.37 \\
		8$^{+}$ & 11.627 & 419 & 1 & 1.2 & 11.32 & 13.6 & 0.50 \\
		
	\end{tabular}
\end{table}

Here E$_{\alpha}$ denotes the $\alpha$ decay energy into the considered level, taking excitation energies of the levels typically for
heavy nuclei (see e.g. fig. 6). T$_{\alpha}$ is the theoretical $\alpha$ decay half-life according to \cite{PoI80,Rur84}. HF($\Delta$L) is the 
hindrance due to angular momentum change $\Delta$L as suggested by J.O. Rasmussen \cite{Ras59}. T$_{\alpha}$/T$_{\alpha}$(gs) is the ratio
of the partial $\alpha$ decay half-lives for the decay into the considered level and into the ground state. HF(m) denotes the 'mean'
hindrance factor, HF(m) = HF($\Delta$L) x (T$_{\alpha}$/T$_{\alpha}$(gs)), and T$_{\alpha}$(m) the resulting partial $\alpha$ half-life.
Although this treatment is quite crude, 
evidently $\alpha$ decay into the ground state is strongly hindered compared to $\alpha$ decay into low lying members of the 
ground-state rotational band, which makes the estimation
of the excitation energy of the isomeric state by using the $\alpha$ decay energy uncertain and the given value of E$^{*}$ = 1.13 MeV is rather 
a lower limit. It should here just be noted, that a small $\alpha$ decay branch for $^{254m1}$No is reported in \cite{Hes10} with an energy
somewhat smaller than expected from the excitation energy of $^{254m1}$No and its ground-state $\alpha$-decay energy.\\
The discussion on the structure of the isomeric state is based on the coupling of low lying single particle levels. For the sake of the further
discussion in fig. 22 calculated single particle levels in neighbouring odd-mass nuclei are shown. Data for the even-Z nuclei $^{269,271}$Ds are taken from \cite{ParS05}, for the odd-Z nuclei $^{269}$Mt and $^{271}$Rg  are taken from \cite{Park04}.\\
Evidently states of high K values can be formed by couplings 
$\nu$7/2$^{+}$[613]$\uparrow$ $\otimes$ $\nu$11/2$^{-}$[725]$\uparrow$ resulting in K$^{\pi}$ = 9$^{-}$,
$\nu$9/2$^{+}$[615]$\downarrow$ $\otimes$ $\nu$11/2$^{-}$[725]$\uparrow$ resulting in K$^{\pi}$ = 10$^{-}$, 
$\pi$9/2$^{-}$[505]$\downarrow$ $\otimes$ $\pi$11/2$^{+}$[615]$\uparrow$ resulting in K$^{\pi}$ = 10$^{-}$.\\
Calculations on the possible structure of the isomer, besides of those performed in \cite{HoH01} were later performed by F.R. Xu et al. \cite{XuZ04},
H.L. Liu et al. \cite{Liu14}, V. Prassa et al. \cite{Prass15} and G.G. Adamian et al. \cite{AdaA10}. The results are compared with those of \cite{HoH01} in Fig. 23.
Besides S. Hofmann et al. \cite{HoH01} also F.R. Xu et al. \cite{XuZ04} and H.L. Liu et al. \cite{Liu14} predict 2-neutron\,-\,quasi-particle\,-\, states of K$^{\pi}$\,=\,9$^{-}$, 10$^{-}$ at exciation energies E$^{*}$\,=\,(1.0-1.5) MeV, but contrary to \cite{HoH01} the  K$^{\pi}$\,=\,10$^{-}$ state is predicted below the K$^{\pi}$\,=\,9$^{-}$ state in \cite{XuZ04,Liu14}.
V. Prassa et al. \cite{Prass15} predict
only the K$^{\pi}$\,=\,9$^{-}$ state, G.G. Adamian only the K$^{\pi}$\,=\,10$^{-}$ state below E$^{*}$\,=\,1.5 MeV. 
In addition F.R. Xu et al. \cite{XuZ04}, H.L. Liu et al. \cite{Liu14} and V. Prassa et al. \cite{Prass15} predict a
K$^{\pi}$\,=\,10$^{-}$ 2-quasi-proton\,-\,state in the range E$^{*}$\,=\,(1.0-1.5) MeV,
In \cite{Liu14} the K$^{\pi}$\,=\,10$^{-}$ 2-quasi-neutron\,-\,state is favoured as the isomeric one due to the, according to the Gallagher rule favored coupling in energy ($\uparrow\downarrow$ configuration) in contrast to the unfavoured coupling of the K$^{\pi}$\,=\,9$^{-}$ state 
($\uparrow\uparrow$ configuration), while the  K$^{\pi}$\,=\,10$^{-}$ 2-quasi-proton\,-\,state is not considered as the isomeric one. 
Besides the K$^{\pi}$\,=\,10$^{-}$ 2-quasi-neutron\,-\,state G.G. Adamian et al. also consider a K$^{\pi}$\,=\,6$^{+}$ 2-quasi-neutron\,-\,state
(configuration $\nu$1/2$^{-}$[761]$\downarrow$ $\otimes$ $\nu$11/2$^{-}$[725]$\uparrow$) as a possible isomeric one.
In addition they obtain at E$^{*}$ $<$ 1.5 MeV a K$^{\pi}$\,=\,8$^{+}$ 2-quasi-neutron\,-\,state (confguration not given, but probably $\nu$7/2$^{+}$[613]$\uparrow$ $\otimes$ $\nu$9/2$^{+}$[604]$\downarrow$ as expected in $^{268}$Ds) and a
K$^{\pi}$\,=\,6$^{-}$ 2-quasi-proton\,-\,state; here also no configuration is given, but with respect to the calculated Nilsson levels shown in fig. 22 a possible configuration could be
$\pi$11/2$^{+}$[615]$\uparrow$ $\otimes$ $\nu$1/2$^{-}$[510]$\uparrow$.\\

{\bf 7.3 $\alpha$ decay of K isomers}\\
$\alpha$ decay of K isomers has been reported so far only for three
cases $^{254m1}$No \cite{Hes10}, $^{270m}$Ds \cite{HoH01,Ack12,Ack15} and
$^{266m}$Hs \cite{Ack12,Ack15}.\\
For $^{254m1}$No two $\alpha$ events with individual energies of
E = 9369 keV and E = 9336 keV were observed, while for the transition
$^{254m}$No $^{\alpha}_{\rightarrow}$ $^{250g}$Fm a value
E$_{\alpha}$ = 9370$\pm$10 keV was expected. The lower energy of the
second decay may be regarded as a hint that not the ground state of $^{250}$Fm
was populated, but a low lying member of the ground-state rotational band.
A branching ratio b$_{\alpha} \le$ 1x10$^{-4}$ and a partial $\alpha$ decay half-life
of $\ge$2750 s were obtained. The theoretical half-life \cite{PoI80,Rur84} for a 
9370-keV transition is T$_{\alpha}$ = 3.63 ms, resulting in a hindrance
factor HF $\ge$ 7.6x10$^{5}$, which is a factor $\ge$2435 or $\ge$26600 higher
than hindrance factors expected for $\Delta$L\,=\,8 or $\Delta$L\,=\,6 transitions 
(P$_{L}$/P$_{0}$ = exp[-2.027L(L+1)Z$^{-1/2}$A$^{-1/6}$])\cite{Ras59}. 
But here we also have to consider that besides structural hindrance according to the
selection rules for $\alpha$ decay for transitions between states of 
different parities (e.g. 8$^{-}$ $\rightarrow$ 0$^{+}$, 2$^{+}$) only odd values of angular
momentum change are allowed, which introduces strong hindrance between states of
even values of angular momentum difference \cite{SeaL90}.\\
$^{270m}$Ds is insofar a specific case of a K isomer as it decays 
to a large extent by $\alpha$ emission. As the half-lives of the
ground state (0.17 ms) and the isomeric state (3.81 ms) \cite{Hessb22} differ by 
roughly a factor of 22 a rough estimate of the $\alpha$ branching can be obtained
from the ratio of $\alpha$ - particles in the energy range of
the ground state $\alpha$ decays having correlation times a factor of 
five longer ($>$1 ms) than the half-life of the ground state, which
results in a branching b$_{\alpha}$\,$\approx$\,0.93 \cite{Hessb22}, resulting in a partial
$\alpha$ decay half-life of T$_{\alpha}$\,=\,4.1 ms. 
Assuming events with energies E$_{\alpha}$ $>$ 12 MeV as decays into the ground state 
(or a 'low' member of the ground-state rotational band) one obtains a branching ( 6 out of 15 
decays) of 0.4 and hence a partial $\alpha$ decay half-life of T$_{\alpha}$ $\approx$  4.1/0.4 ms $\approx$ 10.25 ms.
For the $\alpha$ events of 
the highest energies with a mean value of 12.136 MeV one obtains a theoretical $\alpha$ decay 
half-life of T$_{\alpha}$ = 3.9x10$^{-7}$ s resulting in a hindrance factor
HF $\approx$ 26300. Considering transitions 10$^{-}$, 9$^{-}$ $\rightarrow$ 0$^{+}$ one expects
hindrance factors of 4800 and 947 for $\Delta$L = 10 and $\Delta$L = 9 transitions \cite{Ras59}.
Evidently the differences between the 'total' hindrance factors (from the ratio of the
experimental and theoretical $\alpha$ half-lives) and those due to the angular momentum 
change are much smaller than in the case of $^{254m1}$No. If one now simply writes the total
hindrance factor HF(tot) as the product of the angular momentum hindrance factor HF($\Delta$L) 
and the 'nuclear structure based hindrance' HF(struct) as HF(tot) = HF($\Delta$L) x HF(struct)
one obtains HF(struct)($^{270m}$Ds) $<$$<$ HF(struct)($^{254m1}$No) (see table 10).\\

\begin{table}
	\centering
	\caption{Hindrance of $\alpha$ decay for transition from the K isomer into the ground-state or low lying members
		of the ground-state rotational band of the daughter nucleus for $^{254m1}$No and $^{270m}$Ds. }
	\label{tab:1}       
	\begin{tabular}{ccc}
		\hline\noalign{\smallskip}
		& $^{254m1}$No & $^{270m}$Ds\\
		\hline\noalign{\smallskip}     
		T$_{\alpha}$(calc) & 3.63 ms  & 0.39 $\mu$s \\
		T$_{\alpha}$(exp) & $\ge$2750 & 10.25 ms \\
		HF & $\ge$7.58 x 10$^{5}$ & 26148 \\
		HF($\Delta$L) & 28.5 ($\Delta$L = 6) & 932 ($\Delta$L = 9) \\
		& 311.2 ($\Delta$L = 8) & 4260 ($\Delta$L = 10) \\
		HF(struct) & 26581 ($\Delta$L = 6) & 28 ($\Delta$L = 9) \\
		& 2435 ($\Delta$L = 8) & 6 ($\Delta$L = 10) \\
	\end{tabular}
\end{table}

$\alpha$ decay of $^{270m}$Ds was investigated theoretically by R.M Clark and D. Rudolph \cite{Clark18}
using the 'superfluid tunneling model'(STM). They were able to reproduce the ground-state $\alpha$ decay
half-lives of all known even-even nuclei in the range Z\,=\,100\,-\,118.
Applying their model to the $\alpha$ decay of $^{270m}$Ds they obtain a partial $\alpha$ decay half-life 
of 36 ms for a $\Delta$L = 9 transition, when reducing the pairing gap parameter to sixty percent of the
value used to reproduce the half-lives for the ground-state $\alpha$ - decay of the even - even nuclei,
which caused an increase of the half-life by a factor of $\approx$136. 
This value is not in severe disagreement with the experimental partial half-life of 11 ms. 
For the decay of $^{270m}$Ds into the assumed isomer in $^{266}$Hs they obtain a partial $\alpha$ decay 
half-life of 31 ms, which can be compared with the experimental value of 63 ms, assuming a relative
intensity of 0.07 (one out of fifteen decays) for that transition. Thus, the agreement of the results
of R.M. Clark and D. Rudolph with the experimental data is not so bad.
It should be noted that reduction of 
the pairing gap parameter does not mean simply a reduction of the pairing, but includes also 
reduction of the decay probability due to nuclear structure 'effects' \cite{Clark18}. Consquently the authors conclude
'we find that the effects of nuclear structure of the multi-quasiparticle isomer, which is accounted for in
the STM model by a reduction of the pairing gap parameter must be included' \cite{Clark18}.\\
In conclusion we have to state that the quite low values of the 'structural' hindrance factors are in disagreement
with 'very high' hindrance factors expected for transitions violating by the selection rules, in this
case for $\alpha$ decay with an even value for angular momentum change and a parity change, i.e. for
the transition 10$^{-}$ $\rightarrow$ 0$^{+}$, 2$^{+}$. In other words, the $\alpha$ decay hindrance factors do not
support a K$^{\pi}$\,=\,10$^{-}$ configuration of the isomer, as it seems preferred by the nuclear structure
calculations and applying the Gallagher rule, but rather suggest a K$^{\pi}$\,=\,9$^{-}$. 
This statement, however, is quite vague and further detailed investigations are required to clarify the
situation.\\

\begin{figure*}
	\centering
	\resizebox{0.90\textwidth}{!}{
		\includegraphics{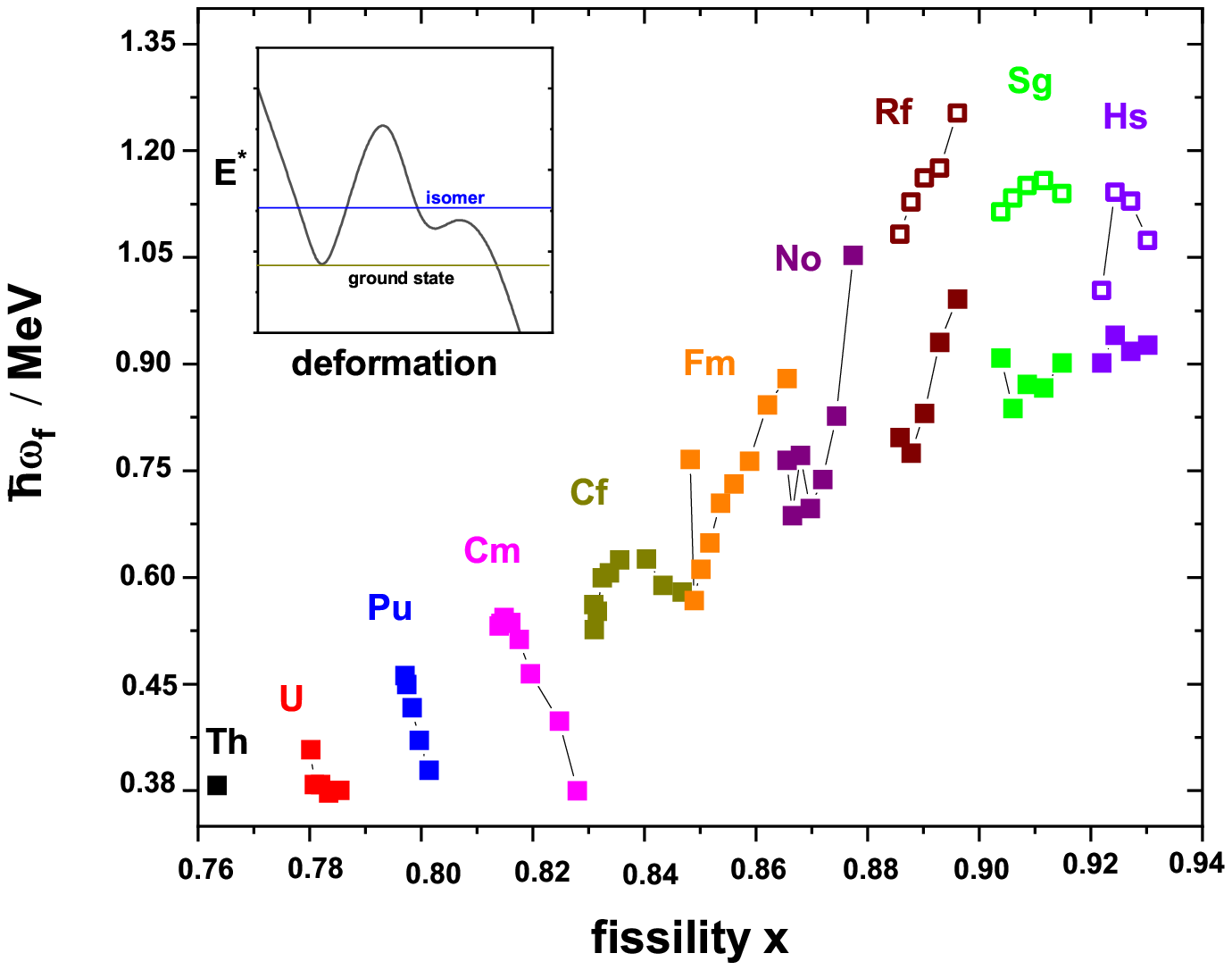}
	} 
	\caption{Barrier curvature energy for even - even nuclei.}
	
	\label{fig:24}       
\end{figure*}

\subsection{\bf 7.4 Stability of K isomers against fission}
It is well known that spontaneous fission of nuclei with odd numbers of protons and/or neutrons is hindered compared to even - even nuclei. This feature qualitativly
can be understood by the conservation of angular momentum and parity. While a pair of nuclei coupling their angular momenta to L\,=\,0 may change the nuclear level
at crossing points during
deformation towards the fission configuration, thus following the energetically most favourable path \cite{Joha59}, for an unpaired nucleon this is normally not possible
as it has to keep its angular momentum, which leads to an effective increase of the fission barrier, in literature denoted as 'specalization energy' and thus to an increase of the half-life \cite{Rand73}. Quantitatively
this increase can be expressed by a hindrance factor HF = T$_{sf}$(exp)/T$_{ee}$, where T$_{sf}$(exp) is the experimental half-life and T$_{ee}$ the 
'unhindered' half-life, usually taken as the geometric mean of the half-lives of the neighbouring even-even nuclei. There is no general rule for estimation of
such hindrance factors, since the increase of the effective fission barrier depends of the nuclear structure of the considered nuclei and on the increase in
energy of the occupied single particle level at deformation. Neverteless a correlation between the steepness of increase or decrease in energy at deformation 
seems evident \cite{Hess17}. In odd-odd nuclei with an unpaired proton and an unpaired neutron, both nucleons have to keep their angular momenta, leading
to higher hindrance factors than obtained for odd-mass nuclei. Consequently only very few cases of spontaneous fission of odd-odd nuclei are known. 
It should be mentioned that detection of spontaneous fission is complicated by the fact, that the odd-odd nucleus may decay by $\beta^{-}$-, $\beta^{+}$- or electron capture decay into an even-even nucleus of much lower half-life that undergoes spontaneous fission. So techniques to discriminate between 'direct' spontaneous fission and spontaneous fission after $\beta$- or electron capture decay have to be applied.\\
In K isomers, where pairs of nucleons are broken and the nucleons are excited into different levels, the situation resembles that in odd-odd nuclei, i.e.
a strong hindrance of spontaneous fission can be expected. So far, only for two cases, $^{256m}$Fm \cite{Hall89} and $^{254m1}$No (see sect. 6.2) spontaneous fission was observed. It should be mentioned that here one is faced with similar situation as in odd-odd nuclei. In some nuclei ($^{253,254}$Rf, $^{250}$No) two fission activities were observed. In these three cases, however, it was shown, that the isomer decays by internal transitions into the ground-state, which then undergoes spontaneous fission.\\
To obtain information about fission hindrance of K isomers F.P. He\ss berger et al. \cite{Hes10} performed for the case of $^{254m1,m2}$No some basic
calculations to estimate fission halflives of these isomers. The calculations were based on the empirical description of fission half-lives
suggested by V.E. Viola and B.D. Wilkins \cite{Viola66}. Similar to description of $\alpha$ decay spontaneous fission was estimated as a tunneling
through a one-dimensional fission barrier. The fission half-life thus was expressed as\\
 \\
 (7.4.1) T$_{sf}$ = ln2 x (nP)$^{-1}$ \\
 \\
  where 'n' denotes the frequency of barrier assaults, 'P' the barrier penetration probability.\\
The fission barrier was approximated by an inverted parabola, and the barrier transmission probability was calculated using the 
Hill-Wheeler approximation \cite{Hill53}, resulting in\\
 \\
 (7.4.2) P = [1+exp(2$\pi$/$\hbar\omega_{f}$)x(B$_{f}$-E))]$^{-1}$ \\
 \\
 with $\hbar\omega_{f}$ representing the barrier curvature energy, B$_{f}$ the fission barrier and E the excitation energy above the ground state.\\
 The fission half-life for the ground state (E$^{*}$\,=\,0) can thus be written as\\
  \\
  (7.4.3) lg(T$_{sf}$/years) = (2.73/$\hbar\omega_{f}$ x B$_{sf}$) - 28.04 \\
 \\
 It should be noted, however, that is approximation is quite crude, as in the region of the actinides the fission barrier is known to be 
 double humped \cite{Bjor80} as sketched in fig. 24 in the inset.
 On the basis of the above relation for the fission half-lives F.P. Heßberger et al. \cite{Hess86} analyzed the barrier curvature energies for 
 the known even-even nuclei using the fission barrier as the sum B$_{sf}$ = B$^{LD}_{sf}$ - $\delta$U$_{gs}$ with 
 the liquid drop fission barrier B$^{LD}_{sf}$ as suggested 
 by M. Dahlinger et al. \cite{Dahl82} and $\delta$U$_{gs}$ being the ground-state shell correction energy as presented by P. M\"oller and J.R. Nix
 \cite{MoN80}. On this basis F.P. He\ss berger et al. \cite{Hess86} obtained quite smooth behavior of $\hbar\omega_{f}$ around 0.4 MeV as function of the 
 fissility parameter x\footnote{ x = Z$^{2}$ / (49.22 x A (1-0.3803 x I$^{2}$ - 20.489 x I$^{4}$)) and I = (A-2Z)/A} for the isotopes of uranium to
 californium (region 1), $\hbar\omega_{f}$ around 0.8 MeV, for the isotopes of nobelium, rutherfordium and seaborgium, where the outer fission barrier
 was predicted to lie in energy below the ground state \cite{Moll77} (region 3) and a strong variation with x for the fermium and nobelium isotopes
 (region 2), which was explained by quite broad barriers in region 1, narrow barriers in region 3 and a 'transition region' (region 2). On this basis 
 F.P. He\ss berger et al. \cite{Hes10} calculated, however using liquid drops fissions barriers according to A. Sierk \cite{Sierk86} and ground state shell correction energies from R. Smolanczuk and A. Sobiczewski \cite{SmS95}, fission half-lives for $^{254m1,254m2}$No of 1 s ($^{254m1}$No) and 
 $<$0.5 $\mu$s ($^{254m2}$No) and hence hindrance factors of $\approx$1375 ($^{254m1}$No) and $>$3.3 x 10$^{6}$ ($^{254m2}$No). 
 The hindrance factor HF = 1375 seems quite low, but one has to consider that it is very sensitive to the value of the barrier curvature $\hbar\omega_{f}$ 
 and the effective fission barrier (B$_{sf}$-E$^{*}$). In \cite{Hes10} quite conservative values of (B$_{sf}$-E$^{*}$) = 5.643 MeV and  $\hbar\omega_{f}$ = 0.75 MeV 
 were used; using a slighty higher value $\hbar\omega_{f}$ = 0.8 MeV and (B$_{sf}$-E$^{*}$) = 5.47 MeV using the fission barrier from \cite{Moll15}, one obtains
 T$_{sf}$ = 0.013 s, hence a two orders of magnitude lower value (see table 11).\\
 In fig. 24 the results for the barrier curvature energies $\hbar\omega_{f}$ according to relation 7.4.3.
 using, however, fission barriers as published by P. M\"oller et al. \cite{Moll15}
 (full squares) for the elements thorium to hassium are presented as function of the fissility parameter x; for comparison results for the isotopes of the elements rutherfordium, seaborgium, and hassium
 using the dynamical fission barriers B$^{dyn}_{f}$ as published by R. Smolanczuk et al. \cite{SmS95} are shown as open squares in fig. 24.
 Evidently an increase of the barrier curvature energies at increasing fissility is observed, with partly strong variations within an isotope line,
 which is seen a signature for a change of the shape of the fission barrier. The values using fission barriers of \cite{SmS95} are higher than those using fission barriers of \cite{Moll15} since \cite{SmS95} predict higher fission barriers.\\ 
 For the case of $^{254m1}$No one obtaines a fission half-live of T$_{sf}$\,=\,0.0339 s using the value of $\hbar\omega_{f}$ for the ground state. Respecting, however, that the excitation energy of the isomer may be already above the height of the outer barrier, thus taking the barrier curvature obtained for
 $^{250}$No, which is expected to have a narrow single-humped barrier \cite{Moll77} one obtains  T$_{sf}$\,=\,0.48 $\mu$s, and hence hindrance factors of 
 HF\,=\,40560 and HF\,=\,2.86$\times$10$^{9}$. \\
 Recently the idea of parametrizing the barrier curvature energy was resumed by J. Khuyagbaatar \cite{Khu22}. He obtained somewhat higher values of 
 $\hbar\omega_{f}$ and a fission half-life of 9.4$\times$10$^{-4}$ for $^{254m1}$No. The results are summarized in table 11; evidently the hindrance factor reported in \cite{Hes10} seems unrealistic low, seemingly using a too low value for the barrier curvature energy. But it also seen that the hindrance factors resulting from the other analyses also straggle in a wide range of more than four orders of magnitude, which can be ascribed due to the rather simple form
 of calculating the fission half-lives.\\
 In the case of $^{256}$Fm Khuyagbaatar report a calculated unhindert fission half-life of 2.2$\times$10$^{-7}$ s, while the analysis presented here delivers a
 value T$_{sf}$(calc) = 3.6$\times$10$^{-8}$ s and hence hindrance factors of 3.6$\times$10$^{3}$ \cite{Khu22} and 2.2$\times$10$^{5}$.\\
 A comparison for $^{254m1}$No is shown in table 11. Evidently the cases 3 and 4 exhibit hindrance factors of $>$1$\times$10$^{6}$ which is higher than hindrance factor for odd-mass nuclei which are typically (except a few cases) $<$$\times$10$^{6}$, i.e. one can conclude that fission of $^{254m1}$No is more hindered than
 typically for nuclei with one unpaired nucleon. A comparison with odd-odd nuclei is hardly possible, as there are only two cases of odd-odd nuclei where direct 
 fission is reported, $^{262}$Db (a less certain case) and $^{260}$Md. 
 In the case of $^{262}$Db a quite low hindrance factor (HF$\approx$3250) is obtained, while for $^{260}$Md a quite high value (HF$\approx$9.2$\times$10$^{8}$) is obtained (see \cite{Hess17}).
 Thus one may conclude that fission hindrance of 2- quasiparticle - K isomers (with two unpaired nucleons) indeed rather resembles the case of odd-odd nuclei with each an unpaired proton and neutron. At present state, such a conclusion is indeed still 
 somewhat speculative, more information of spontaneous fission of K isomers and odd-odd nuclei is required as well as efforts from theory to calculate 
 fission half-lives of K isomers is required.

 	\begin{table}
 		\centering
 		\caption{Comparison of fission hindrances for $^{254m1}$No for different barrier curvature energies $\hbar\omega_{f}$ values. 
 		T$_{sf}$ = 1375 s for all cases.}
 		\label{tab:1}       
 		\begin{tabular}{ccc}
 			\hline\noalign{\smallskip}
 		reference	& T$_{sf}$(calc)  & HF\\
 			\hline\noalign{\smallskip}     
 			\cite{Hes10} & 1 s  & 1375 \\
 			\cite{Hes10}, $\hbar\omega_{f}$=0.8 MeV & 0.013 s & 1.06$\times$10$^{5}$ \\
 		    this work, $\hbar\omega_{f}$=0.783 MeV & 0.034 s & 4.1$\times$10$^{4}$ \\
 	        this work, $\hbar\omega_{f}$=1.05 MeV & 0.048 $\mu$s & 2.9$\times$10$^{9}$ \\
            \cite{Khu22} & 0.94 ms s & 1.4$\times$10$^{6}$ \\
 		\end{tabular}
 	\end{table}

 	Finally one other item should be mentioned; besides $\hbar\omega_{f}$ fission half-lives calculated using the above mentioned method are also quite
 	sensitive to the fission barrier. To show the situation fission half-lives of $^{256}$Rf for the ground-state and at an excitation energy of 1 MeV
 	were calculated using once the fission barriers from \cite{Moll15} and the analyzed barrier curvatures and once the dynamical fission barriers from \cite{SmS95}.
 	Using in both cases the values of $\hbar\omega_{f}$  extracted for the ground-state fission half-lives of T$_{sf}$(E$^{*}$=1 MeV)\,=\,7.6 $\mu$s (fission barriers from \cite{Moll15}) and  T$_{sf}$(E$^{*}$= 1 MeV)\,=\,30 $\mu$s (fission barriers from \cite{SmS95}) and hence half-life ratios 
 	T$_{sf}$(gs)/T$_{sf}$(E$^{*}$=1 MeV) = 860 and T$_{sf}$(gs)/T$_{sf}$(E$^{*}$=1 MeV) = 210 are obtained; although these ratios vary within a factor of four, this difference is not of significant relevance for qualitative discussion on fission hindrance of K isomers.

\section{8. Outlook}

The occurence of K isomers in strongly prolate deformed nuclei is a physically interesting phenomenom.
Thanks to improved experimental techniques investigation of K isomers became a key aspect in 
spectroscopy of transfermium nuclei during the past two decades. Although about thirty new K isomers were
discovered in that region of nuclei, information on the structure and decay properties is 
scarce for most of the cases due to low production rates for most of the isomers which do not allow
for detailed $\gamma$ spectroscopic investigations.\\
A specific feature pointed out quite often is the longer half-life of some isomers compared to the ground - state,
which was also discussed in context with the stability of 'superheavy' nuclei in vicinity of the
closed proton and neutron shells expected at Z\,=\,114, 120 or 126 and N\,=\,172 or 184. Here one has to
keep in mind, however, that K isomerism is a phenomenom in strongly deformed nuclei, while the nuclei around the
closed proton an neutron shells are spherical. So K isomers are not expected in the region of spherical
superheavy elements.\\
Stability and decay properties of K isomers (also with respect to those of the ground-state) are strongly related
to the properties of the radioactive decay modes, internal transitions, $\alpha$ decay, spontaneous fission, relevant
for the decay of the isomers.\\
The ratios (partial half-lives) for internal transitions are essentially defined by the K hindrance, those for $\alpha$
decay by the Q-value, the 'angular momentum hindrance' and by hindrance due to changes in nuclear structure and the parity.
Spontaneous fission half-lives are essentially defined by hight and shape of the fission barrier and a fission hindrance
due to unpaired nucleons. It is thus the interplay between probabilities of these three decay modes which finally defines 
the decay of the K isomer. As long as partial half-lives for $\alpha$ decay and spontaneous fission are significantly larger
than those for internal transitions, the latter will dominate and $\alpha$ decay and spontaneous fission will be exotic decay
modes with low intensities as in the case of $^{254m1}$No. Half-lives of the K isomers are definitely shorter than those of
the ground-states in such cases. The situation changes, when partial half-lives for $\alpha$ decay or spontaneous fission come into the same
order of magnitude as the half-lives for internal transitions. In those cases $\alpha$ decay or spontaneous fission can become the essential
decay modes of K isomers and also half-lives of the isomeric state may become longer than
that of the ground state due to strong hindrance of the isomeric decays while $\alpha$ decay or spontaneous fission of the
ground-state is unhindered or at least less hindered. 
An example is the decay of $^{270,270m}$Ds by
$\alpha$ emission, where $\alpha$ decay of the isomeric state is strongly hindered, and thus half-life is longer than that of the
ground-state despite of a higher Q$_{\alpha}$ - value of the isomer. Another phenomenom is the observation of two fission activities
for $^{250}$No and $^{254}$Rf ($^{254}$Rf, T$_{1/2}$ = 247 $\mu$s), with the shorter one attributed to the ground-state and the longer one attributed to the isomeric
state. Here it was shown that direct fission of the isomer is at best a decay mode of very low intensity due to the strong hindrance
caused by the unpaired nucleons despite a significantly ($\le$1 MeV) lower fission barrier of the isomer. The longer-lived
fission activity attributed to the isomer rather represents spontaneous fission of the ground-state delayed by internal transitions.
Thus we can conclude, although details of the decays are theoretically not well understood, that decay of the latter K isomers can be understood 
on the basis of the more or less well known properties of internal transitions, $\alpha$ decay and spontaneous fission and
certainly do not contain some kind of 'new physics'.\\
Although properties of K isomers in the transfermium region will not have direct impact on the expected properties of 'spherical' superheavy
nuclei, study of their properties will deliver valuable information on energies and ordering of single particle levels 
(Nilsson levels) which certainly will have feedback on the prediction and location of spherical proton and
neutron shells in the range Z\,=\,114-126 and N\,=\,172-184.\\ 

%
%

\end{document}